\documentclass[12pt,preprint]{aastex}
\usepackage{epsf}

\def\linebreak{\hfil\break}

\def\
{\hfil\linebreak}

\def\orch{{\it Orchestra}}
\def\deg{\ifmmode {^\circ}\else {$^\circ$}\fi}
\def\degree{\ifmmode {^\circ}\else {$^\circ$}\fi}
\def\mum{\ifmmode {\rm \mu {\rm m}}\else $\rm \mu {\rm m}$\fi}
\def\arcsec{\ifmmode ^{\prime \prime}\else $^{\prime \prime}$\fi}

\def\inch{\ifmmode ^{\prime \prime}\else $^{\prime \prime}$\fi}
\def\arcmin{\ifmmode ^{\prime}\else $^{\prime}$\fi}
\def\qprime{\ifmmode q^{\prime}\else $q^{\prime}$\fi}

\def\degree{\ifmmode {^\circ}\else {$^\circ$}\fi}
\def\arcsec{\ifmmode ^{\prime \prime}\else $^{\prime \prime}$\fi}

\def\inch{\ifmmode ^{\prime \prime}\else $^{\prime \prime}$\fi}
\def\arcmin{\ifmmode ^{\prime}\else $^{\prime}$\fi}

\def\mjup{\ifmmode { M_J}\else $ M_J$\fi}
\def\rjup{\ifmmode { R_J}\else $ R_J$\fi}
\def\mearth{\ifmmode { M_{\oplus}}\else $ M_{\oplus}$\fi}
\def\rearth{\ifmmode { R_{\oplus}}\else $ R_{\oplus}$\fi}
\def\lstar{\ifmmode { L_d / L_{\star}}\else $ L_d / L_{\star}$\fi}
\def\ldlstar{\ifmmode { L_{\star}}\else $ L_{\star}$\fi}
\def\lsun{\ifmmode { L_{\odot}}\else $ L_{\odot}$\fi}
\def\mstar{\ifmmode M_{\star}\else $ M_{\star}$\fi}
\def\msun{\ifmmode M_{\odot}\else $ M_{\odot}$\fi}
\def\tstar{\ifmmode T_{\star}\else $ T_{\star}$\fi}
\def\rstar{\ifmmode R_{\star}\else $ R_{\star}$\fi}
\def\rsun{\ifmmode R_{\odot}\else $ R_{\odot}$\fi}
\def\mjup{\ifmmode M_{J}\else $ M_{J}$\fi}
\def\rjup{\ifmmode R_{J}\else $ R_{J}$\fi}
\def\mjupyr{\ifmmode { M_J~yr^{-1}}\else $ M_J~yr^{-1}$\fi}
\def\msunyr{\ifmmode { M_{\odot}~yr^{-1}}\else $ M_{\odot}~yr^{-1}$\fi}
\def\gyr{\ifmmode {\rm g~yr^{-1}}\else $\rm g~yr^{-1}$\fi}
\def\kms{\ifmmode {\rm km~s^{-1}}\else $\rm km~s^{-1}$\fi}
\def\ms{\ifmmode {\rm m~s^{-1}}\else $\rm m~s^{-1}$\fi}
\def\rhill{\ifmmode R_H\else $R_H$\fi}
\def\rfast{\ifmmode R_{fast}\else $R_{fast}$\fi}
\def\rgap{\ifmmode R_{gap}\else $R_{gap}$\fi}
\def\vhill{\ifmmode v_H\else $v_H$\fi}
\def\qdstar{\ifmmode Q_D^\star\else $Q_D^\star$\fi}
\def\resc{\ifmmode r_{esc}\else $r_{esc}$\fi}
\def\mesc{\ifmmode m_{esc}\else $m_{esc}$\fi}
\def\rmin{\ifmmode r_{min}\else $r_{min}$\fi}
\def\rmax{\ifmmode r_{max}\else $r_{max}$\fi}
\def\mmax{\ifmmode m_{max}\else $m_{max}$\fi}
\def\rmind{\ifmmode r_{min,d}\else $r_{min,d}$\fi}
\def\rmaxd{\ifmmode r_{max,d}\else $r_{max,d}$\fi}
\def\mmaxd{\ifmmode m_{max,d}\else $m_{max,d}$\fi}
\def\qz{\ifmmode q_{0}\else $q_{0}$\fi}
\def\qi{\ifmmode q_{i}\else $q_{i}$\fi}
\def\ql{\ifmmode q_{l}\else $q_{l}$\fi}
\def\qs{\ifmmode q_{s}\else $q_{s}$\fi}
\def\r0{\ifmmode r_{0}\else $r_{0}$\fi}
\def\m0{\ifmmode m_{0}\else $m_{0}$\fi}
\def\M0{\ifmmode M_{0}\else $M_{0}$\fi}
\def\xm{\ifmmode x_{m}\else $x_{m}$\fi}
\def\gyr{\ifmmode {\rm g~yr^{-1}}\else ${\rm g~yr^{-1}}$\fi}
\def\cms{\ifmmode {\rm cm~s^{-1}}\else ${\rm cm~s^{-1}}$\fi}
\def\gcms{\ifmmode {\rm g~cm^{-2}}\else $\rm g~cm^{-2}$\fi}
\def\gcmc{\ifmmode {\rm g~cm^{-3}}\else $\rm g~cm^{-3}$\fi}
\def\VP{2012 VP$_{113}$}
\def\ALMA{{\it ALMA}}

\def\2470{[24]--[70]}

\newbox\grsign \setbox\grsign=\hbox{$>$} \newdimen\grdimen \grdimen=\ht\grsign
\newbox\simlessbox \newbox\simgreatbox
\setbox\simgreatbox=\hbox{\raise.5ex\hbox{$>$}\llap
     {\lower.5ex\hbox{$\sim$}}}\ht1=\grdimen\dp1=0pt
\setbox\simlessbox=\hbox{\raise.5ex\hbox{$<$}\llap
     {\lower.5ex\hbox{$\sim$}}}\ht2=\grdimen\dp2=0pt

\begin{document}

\title{Formation of Super-Earth Mass Planets at 125--250 AU from a Solar-type Star}
\vskip 7ex
\author{Scott J. Kenyon}
\affil{Smithsonian Astrophysical Observatory,
60 Garden Street, Cambridge, MA 02138} 
\email{e-mail: skenyon@cfa.harvard.edu}

\author{Benjamin C. Bromley}
\affil{Department of Physics, University of Utah, 
201 JFB, Salt Lake City, UT 84112} 
\email{e-mail: bromley@physics.utah.edu}
%
%

\begin{abstract}

We investigate pathways for the formation of icy super-Earth mass planets 
orbiting at 125--250~AU around a 1~\msun\ star. An extensive suite of coagulation 
calculations demonstrates that swarms of 1~cm to 10~m planetesimals can form
super-Earth mass planets on time scales of 1--3~Gyr. Collisional damping of
$10^{-2} - 10^2$~cm particles during oligarchic growth is a highlight of these
simulations.  In some situations, damping initiates a second runaway growth phase
where 1000--3000~km protoplanets grow to super-Earth sizes. Our results establish 
the initial conditions and physical processes required for {\it in situ} formation 
of super-Earth planets at large distances from the host star.  For nearby dusty
disks in HD~107146, HD~202628, and HD~207129, ongoing super-Earth formation at 
80--150~AU could produce gaps and other structures in the debris.  In the solar
system, forming a putative planet X at $a \lesssim$ 300~AU ($a \gtrsim$ 1000~AU)
requires a modest (very massive) protosolar nebula.

\end{abstract}


\section{INTRODUCTION}
\label{sec: intro}

For nearly a century, planet X has haunted the outer Solar System. 
Originally envisioned as a way to explain irregularities in the 
orbits of Uranus and Neptune \citep[e.g.,][]{lowell1915,bower1931,
brouwer1955,rawlins1970,rawlins1972,harrington1988}, 
planet X -- along with the variants Nemesis and Tyche -- has since 
been invoked to explain the orbital properties of some comets and 
trans-Neptunian objects \citep[e.g.,][]{matese1999,gomes2006,lykaw2008}
and apparent periodicities in mass extinction events on Earth
\citep{davis1984,whitmire1984}. Generating these phenomena requires a
massive planet X, $\sim$ a few Earth masses to tens of Jupiter masses,
with a semimajor axis $a \sim$ 200--20,000~AU.

Observational and theoretical constraints severely limit possibilities 
for planet X \citep[e.g.,][]{hogg1991,zakam2005,iorio2009,fern2011}. 
Along with improved mass measurements for Uranus and Neptune from 
{\it Voyager}, better analyses of positional data eliminate the need 
for a massive planet to explain any `irregularities' in the orbits 
of Uranus and Neptune \citep[e.g.,][]{standish1993}.
By excluding the possibility of a brown dwarf anywhere in the
Solar System, a Jupiter-mass planet inside $\sim$ 25,000~AU, and
a Saturn-mass planet inside $\sim$ 10,000~AU, observations from 
the {\it IRAS}, {\it 2MASS}, and {\it WISE} all-sky infrared 
surveys specifically rule out Nemesis and some other variants 
of planet X \citep[e.g.,][]{luhman2014}. 

The recent discovery of \VP\ \citep{trujillo2014} in the 
`inner Oort cloud' has renewed 
interest in planet X. With orbital parameters 
similar to those of Sedna\footnote{Sedna: semimajor axis $a$ = 542~AU, 
eccentricity $e$ = 0.86, and perihelion distance $q_p$ = $a (1-e)$ = 
76.23~AU; \VP: $a$ = 266~AU, $e$ = 0.698~AU, and $q_p$ = 80~AU}, 
\VP\ joins a group of a dozen relatively large 
($R \approx$ 200--1000~km) distant objects ($q_p \ge$ 30~AU, $e \gtrsim$ 0.7) 
with arguments of perihelion $\omega \approx$ 0\deg.  The observed 
distribution of $\omega$ for these objects is statistically unlikely 
\citep[see also][]{delafm2014}. Although dynamical interactions 
between the young Sun and a passing star can produce objects on 
highly eccentric orbits like Sedna and 
\VP\ \citep[e.g.,][]{morby2004a,kb2004d}, the known planets 
randomize $\omega$ on 0.1--1~Gyr time scales 
\citep{gomes2006,lykaw2008,trujillo2014}.
Super-Earth mass planets at $a \approx$ 200--300~AU can maintain
the observed distribution of $\omega$ for the age of the Solar
System \citep{trujillo2014,delafm2014,iorio2014}.

In addition to explaining the orbital dynamics of objects beyond 40--50 AU,
identifying one or more Earth-mass planets with $a \sim$ 200~AU has
interesting implications for the architecture of the Solar System
beyond Neptune. Despite a lack of massive planets with $a \gtrsim$ 
35~AU, there is a diverse set of trans-Neptunian objects with radii 
$R \lesssim$ 1000--2000~km on orbits with a broad range of $e$ and $i$
\citep[e.g.,][]{gladman2008,petit2011}. Aside from objects like Sedna 
and \VP, the distribution of semimajor axes for these objects suggests 
a clear `edge' at roughly 48~AU. Confirming a massive planet at $\sim$ 
200~AU might imply a very large reservoir of material beyond this edge.

Establishing the existence of a massive planet beyond 100~AU has 
profound consequences for the origin of the Solar System.  In current 
theory, planets grow out of material orbiting within a circumsolar 
disk of gas and dust. As the gaseous disk dissipates, these planets 
(i) clear out material along their orbits and (ii) migrate radially 
to a more stable orbit \citep[e.g.,][]{youdin2013,morbi2013}.  
If an extra super-Earth forms with the known gas giants at 
$a \approx$ 10--20~AU before the gaseous disk disappears, this 
planet could migrate radially to much larger semimajor axes as in 
models for Fomalhaut b \citep{crida2009}. 

Planet-planet scattering is a promising alternative to radial migration
\citep{bk2014}. In this picture, several massive planets form out of
disk material at 5--10~AU. Close orbital encounters among these 
planets then scatter lower mass planets to large semimajor axes
\citep[e.g.,][]{rasio1996,lev1998,juric2008,bk2011a}. If the gaseous
component of the disk is massive and extends to 125--250~AU, gas drag
and dynamical friction can circularize the orbits of super-Earth-mass
planets on time scales comparable to the disk lifetime. Because lower
mass planets are left on eccentric orbits, measuring the orbital
elements of any planet with $a \gtrsim$ 100~AU constrains the mass
and outer radius of the solar nebula \citep{bk2014}. 

Establishing the orbital eccentricity of a super-Earth at 200~AU provides
a way to distinguish between these two possibilities. Migrating 
planets have nearly circular orbits 
\citep[e.g.,][]{ward1997,masset2003,ida2008a,crida2009}.  Although 
gas drag can produce nearly circular orbits among scattered massive 
planets, lower mass planets have more eccentric orbits \citep{bk2014}.

To provide a distinct set of predictions for massive planets at 200~AU, 
we consider models where one or more super-Earths form directly out of 
solid material well beyond the outer edge of the Kuiper belt
\citep[see also][]{stern2005,kb2008,kb2010,kb2012}.  After summarizing
several issues of growth at semimajor axes $a \approx$ 125--250 AU 
(\S\ref{sec: back}), we describe numerical simulations of planet 
formation for a variety of initial particle sizes and input parameters 
(\S\ref{sec: calcs}--\ref{sec: evol}). Within a ring of width 
$\delta a \approx 0.2 a$ and total mass \M0\ $\approx$ 15~\mearth, 
ensembles of 1~cm to 10~m particles grow into super-Earth mass planets 
on time scales of 1--2~Gyr at 125--250~AU.  After considering the 
implications of these results (\S\ref{sec: disc}), our discussion 
concludes with a brief summary (\S\ref{sec: summary}).

\section{BACKGROUND}
\label{sec: back}

In the current core accretion theory, planets form in circumstellar
disks of gas and dust \citep[e.g.,][]{gold2004,chiang2010,youdin2013}. 
After micron-sized grains within the disk agglomerate into cm-sized 
objects \citep[e.g,][]{birn2010}, continued coagulation 
\citep[e.g.,][]{windmark2012,garaud2013} or some type of instability 
\citep[e.g.,][]{youdin2005,johan2012} concentrates solids into 
km-sized or larger planetesimals. When most of the solids end up in 
planetesimals, these objects collide, merge, and grow into planets 
\citep[e.g.,][]{kb2006,chambers2008,kb2008,koba2010b,raymond2011}. 
In a popular variant of this picture, coagulation or instabilities 
produce only a few large planetesimals, which then rapidly accrete 
cm-sized pebbles \citep{ormel2010a,bk2011a,lamb2012,lamb2014,chambers2014}.

Here, we focus on the growth of large swarms of planetesimals into
super-Earth mass planets. When planetesimals form, their relative 
velocities are small. Physical collisions among planetesimals yield
larger, merged objects.  As planetesimals merge into larger and 
larger protoplanets, dynamical friction and viscous stirring excite 
the orbits of all the solids 
\citep[e.g.,][]{weth1989,ida1992a,ida1992b,weth1993,gold2004}. 
Relative collision velocities grow in step with the escape velocity
of the largest protoplanet.  Eventually, collisions between small 
planetesimals produce debris instead of mergers. Collisions between
debris particles make even more debris.  The resulting collisional 
cascade grinds the debris into small dust grains; radiation pressure
ejects these grains from the nascent planetary system 
\citep[e.g.,][]{dohn1969,burns1979,will1994}. Together, the collisional 
cascade and radiation pressure typically remove 80\% to 95\% of the 
initial mass in solids from the outer disk of a planetary system 
\citep{kb2002b,kb2008,kb2010,koba2010b}. As a result, the largest planets 
have maximum radii of $\sim$ 3000--7000~km and maximum masses of 
0.1--0.5~\mearth\ \citep[e.g.,][]{kenyon2002,kb2002b,inaba2003b,kb2008,kb2010,koba2010b}.

Within this evolutionary sequence, there are several ways to limit the 
loss of small objects and to promote the formation of super-Earth mass 
planets at 125--250~AU. 

\begin{itemize}

\item Increase the binding energy \qdstar\ of small planetesimals. Larger 
binding energies delay the collisional cascade, allowing 
protoplanets to grow to larger sizes.

\item Change the size distribution of the fragments produced in a collision. 
If the fragments have a steeper size distribution, it takes longer for the 
collisional cascade to grind them to dust. Longer collisional cascades enable
larger protoplanets.

\item Include collisional damping of the smallest objects. If collisional 
damping can reduce the velocities of small collision fragments, collisional 
grinding proceeds more slowly and protoplanets grow larger. 

\end{itemize}

Of these options, modifying \qdstar\ seems the least realistic. Modest 
changes to \qdstar\ consistent with numerical simulations of the binding 
energy for small objects \citep[e.g.,][]{benz1999,lein2009,lein2012} 
lead to factor of $\sim$ 2 changes in the mass of the largest object 
\citep[e.g.,][]{kb2010,koba2010b,kb2012}. Achieving much larger 
protoplanets requires factor of $\gtrsim$ 10 increases in the binding
energy of 0.1--10~km objects. Calculations of the likely structure of
these objects appear to preclude such large increases in \qdstar.
Thus, we eliminate this possibility.

Modifying the size distribution of the fragments is more promising. 
In all numerical simulations of planet formation, the size distribution
of the debris is a power law $n(r) \propto r^{-q}$ with a maximum size 
\rmaxd\ and a minimum size 
\rmind\ \citep[e.g.,][]{weth1993,kl1999a,obrien2003,kb2004a,chambers2008,
morby2009b,koba2010b,weid2010}.
If \mesc\ is the mass in fragments, defining \rmaxd\ and $q$
establishes the mass distribution uniquely \citep[e.g.,][]{kcb2014}.
Many investigators set \rmaxd\ $\approx \gamma \resc$ where \resc\ is 
the radius of a particle with mass \mesc\ and $\gamma \approx$ 0.1--0.3. 
Because the equilibrium size distribution for a collisional cascade has 
$q \approx$ 3.5--3.7 \citep{dohn1969,will1994,obrien2003}, many 
investigators fix $q$ = 3.5.  Although others employ more complicated 
formulae for \rmaxd\ and $q$ \citep{morby2009b,weid2010}, these also 
require $q \lesssim$ 4.0 at small sizes. 


In contrast to these restrictive approaches, detailed numerical 
simulations of binary collisions between planetesimals often yield 
debris with steep size distributions 
\citep[e.g.,][]{durda2004,durda2007,lein2012}. For planetesimal
radii $R_0 \approx$ 10--100~km, the debris has typical 
\rmaxd\ $\approx$ 0.1--1.0 $R_0$. The smallest debris particles
have sizes equal to the smallest size resolved in the calculation,
$\rmind \sim$ 0.1--1~km. Over this range of sizes, the slope of
the size distribution depends on the ratio $r_c$ of the collision
energy to the binding energy. Collisions with larger $r_c$ have
larger $q$. For $r_c \gtrsim$ 1,  extreme values for the slope of 
the size distribution are $q \approx$ 4--6 \citep{lein2012} or 
$q \approx$ 8--10 \citep{durda2004,morby2009b}. 

In addition to exploring calculations with a range in $q$, it is 
important to consider the impact of collisional damping on the
cascade \citep{gold2004}. For typical surface densities of solid
particles, collisional damping is effective for particle sizes of
1--10~mm \citep[see also][]{kb2002a,kb2014}. Numerical simulations 
which follow such small particles generally do not include collisional 
damping \citep[e.g.,][] {stark2009,theb2012b,gaspar2013}. Calculations 
which include collisional damping rarely follow particles with sizes 
smaller than 1~cm \citep[e.g.,][]{weth1993,kb2008,morby2009b,koba2010b,
weid2010,koba2014}. Thus, it is possible that a full evolutionary 
calculation with particle sizes smaller than 1~cm will yield new 
phenomena.

In the following section, we outline the tools we use to calculate 
the evolution of a swarm of particles with sizes of 1~\mum\ to
$\gtrsim 10^4$~km. The Appendix summarizes several tests of our
algorithms and comparisons with results of other approaches. Once
the approach is set, \S\ref{sec: evol} describes results for 
evolutionary calculations at 125--250~AU.

\section{PLANET FORMATION CALCULATIONS}
\label{sec: calcs}

Over the past decade, we have developed \orch, an ensemble of computer codes 
for the formation and evolution of planetary systems.  \orch\ consists of a 
radial diffusion code which derives the time evolution of a gaseous or a 
particulate disk \citep{bk2011a}, a multiannulus coagulation code which infers 
the time evolution of a swarm of planetesimals \citep{kb2004a,kb2008,kb2012}, 
and an $n$-body code which follows the orbits of gravitationally interacting 
protoplanets \citep{bk2006,bk2011b,bk2013}.  Within the coagulation code, 
\orch\ includes algorithms for treating interactions between small particles 
and the gaseous disk \citep[e.g.,][]{ada76,weiden1977a} and between coagulation 
mass bins and $n$-bodies \citep{bk2006}. To treat interactions between small 
particles and the $n$-bodies more rigorously, the $n$-body calculations include 
tracer particles. Massless tracer particles allow us to calculate the evolution 
of the orbits of small particles in response to the motions of massive protoplanets. 
Massless and massive tracer particles enable calculations of the response of 
$n$-bodies to the changing gravitational potential of small particles 
\citep{bk2011a, bk2011b, bk2013}.

\subsection{Numerical Grid}
\label{sec: calcs-grid}

To explore a reasonable parameter space with good statistics, we perform 
coagulation calculations within a single annulus with semimajor axis $a$
and width $\delta a$ around a star with mass \mstar\ = 1~\msun. Within
this annulus, there are $M$ mass batches with characteristic mass $m_k$ 
and radius $r_k$ \citep{weth1993,kl1998}. Batches are logarithmically
spaced in mass, with mass ratio $\delta \equiv m_{k+1} / m_{k}$.  Each 
mass batch contains $N_k$ particles with total mass $M_k$ and average 
mass $\bar{m}_k = M_k / N_k$. Particle numbers $N_k < 10^{15}$
are always integers.  Throughout the calculation, the average mass is 
used to calculate the average physical radius $\bar{r}_k$, collision 
cross-section, collision energy, and other necessary physical variables. 
As mass is added and removed from each batch, the average mass changes 
\citep{weth1993}.

For any $\delta$, numerical calculations lag the result of an ideal calculation
with infinite mass resolution (see the Appendix). At $a$ = 125--250~AU, 
calculations with $\delta$ = 1.05--1.19 yield better solutions to the 
growth of 100+~km objects than calculations with $\delta$ = 1.41--2.00.
Although simulations with $\delta$ = 1.05--1.10 allow somewhat better 
tracking of the late stages of planet growth, the improvement over results
with $\delta$ = 1.19 does not compensate for the factor of 2--4 extra cpu
time required to complete these calculations. Thus, we consider a suite of
calculations with $\delta$ = 1.19 ($ = 2^{1/4}$).

In this suite of calculations, we follow particles with sizes ranging from
a minimum size \rmin\ = 1~\mum\ to the maximum size in the annulus \rmax. 
The algorithm for assigning material to the mass bins extends the maximum
size as needed to accommodate the largest particles. Typically, the number 
of large objects with \rmax\ $\gtrsim$ 1000~km in each calculation is small 
$N_k \lesssim$ 10; each protoplanet generally lies in its own mass bin. To
conserve cpu time, we do not promote these objects into the $n$-body code.

\subsection{Initial Conditions}
\label{sec: calcs-init}

All calculations begin with a mono-disperse swarm of planetesimals with 
initial size \r0\ and mass density $\rho_p$ = 1.5 g cm$^{-3}$.  These particles
have initial surface density $\Sigma_0$, total mass $M_0$, and horizontal and 
vertical velocities $h_0$ and $v_0$ relative to a circular orbit.  The horizontal 
velocity  is related to the orbital eccentricity, $e$ = 1.6 $(h/V_K)^2$, where 
$V_K$ is the circular orbital velocity.  The orbital inclination is
${\rm sin}~i$ = $\sqrt{2} v/V_K$.

For conditions at 125--250~AU, typical growth times are longer than the 1--5~Myr 
lifetime of the gaseous disk \citep{haisch2001,will2011,clout2014}. Thus, we set 
the initial surface density of the gas to zero and ignore gas drag on small solids 
\citep{ada76,weiden1977a,raf2004}. Compared to calculations which include gas drag 
\citep[e.g.,][]{weth1993,chambers2008,kb2008,koba2010b}, our simulations retain a 
larger fraction of small particles which might experience significant radial drift 
during the likely lifetime of the gaseous disk. However, our goal is to isolate 
initial conditions and physical processes which might allow formation of super-Earths
at 125--250~AU. We plan to consider the impact of gas drag in future papers.

\subsection{Evolution}
\label{sec: calcs-evol}

The mass and velocity distributions of the planetesimals evolve in time due to
inelastic collisions, drag forces, and gravitational encounters.  As summarized 
in \citet{kb2004a,kb2008}, we solve a coupled set of coagulation equations which
treats the outcomes of mutual collisions between all particles in all mass bins.
We adopt the particle-in-a-box algorithm, where the physical collision rate is 
$n \sigma v f_g$, $n$ is the number density of objects, $\sigma$ is the geometric 
cross-section, $v$ is the relative velocity, and $f_g$ is the gravitational focusing 
factor \citep{weth1993,kl1998}.  Depending on physical conditions in the disk, 
we derive $f_g$ in the dispersion or the shear regime \citep{kl1998,kb2012}.  
For a specific mass bin, the solutions include terms for (i) loss of mass from 
mergers with other objects and (ii) gain of mass from collisional debris and 
mergers of smaller objects.

As discussed in \citet{weth1993}, the most massive protoplanets on roughly 
circular orbits are `isolated' from one another. Isolated protoplanets can
accrete smaller objects but cannot collide with other isolated protoplanets.
To identify isolated protoplanets, we find an ensemble of the largest objects 
whose combined gravitational range is smaller than the width of the annulus. 
For an individual protoplanet, the gravitational range is
\begin{equation}
R_{g,k} = K a R_{H,kk} + 2 a e_k
\label{eq: range}
\end{equation}
Here, $K = 2 \sqrt{3}$ and $R_{H,kk} = [(m_k + m_k)/3 \msun]^{1/3}$ is the
mutual Hill radius.  The gravitational ranges of the isolated protoplanets 
satisfy the summation,
\begin{equation}
\sum_{k_{min}}^{k_{max}} ~ n_k R_{g,k} \le \delta a ~ ,
\label{eq: iso}
\end{equation}
where $n_k$ is the number of protoplanets in mass bin $k$.

Collision outcomes depend on the ratio $Q_c/Q_D^*$, where $Q_D^*$ is the 
collision energy needed to eject half the mass of a pair of colliding 
planetesimals to infinity and $Q_c$ is the center of mass collision energy 
\citep[see also][]{weth1993,will1994,tanaka1996b,stcol1997a,kl1999a,obrien2003,koba2010a}. 
Following \citet{weth1993}, two colliding planetesimals with horizontal velocity 
$h_1$, $h_2$ and vertical velocity $v_1$, $v_2$ have relative horizontal and
vertical velocities $h_c = (h_1^2 + h_2^2)^{1/2}$ and 
$v_c = (v_1^2 + v_2^2)^{1/2}$ \citep[see also][]{kl1998}. The escape velocity 
of the colliding pair is $v_{esc} = (2 G m_c / r_c)^{1/2}$, where 
$m_c = m_1 + m_2$ is the combined mass and $r_c = r_1 + r_2$ is the combined 
radius. The center of mass collision energy is then
\begin{equation}
Q_c = 0.5 \mu (h_c^2 + v_c^2 + v_{esc}^2) / m_c
\label{eq: Qc}
\end{equation}
where $\mu = m_1 m_2 / m_c $ is the reduced mass.

Consistent with N-body simulations of collision outcomes 
\citep[e.g.,][]{benz1999,lein2008,lein2009}, we set
\begin{equation}
\qdstar = Q_b r^{\beta_b} + Q_g \rho_p r^{\beta_g}
\label{eq: Qd}
\end{equation}
where $Q_b r^{\beta_b}$ is the bulk component of the binding energy,
$Q_g \rho_g r^{\beta_g}$ is the gravity component of the binding energy,
and $r$ is the radius of a planetesimal.

In this study, planetesimals have $Q_b$ = $2 \times 10^5$ erg g$^{-1}$ cm$^{0.4}$, 
$\beta_b = -0.40$, $Q_g$ = 0.22 erg g$^{-2}$ cm$^{1.7}$, and $\beta_g$ = 1.30.
These parameters are broadly consistent with published analytic and numerical
simulations \citep[e.g.,][]{davis1985,hols1994,love1996,housen1999}.  At small 
sizes, they agree with results from laboratory 
\citep[e.g.,][]{ryan1999,arakawa2002,giblin2004,burchell2005}
and numerical \citep[e.g.,][]{lein2009} experiments of impacts between icy o
bjects.  For the weakest objects with $r \approx$ 10--100~m, our adopted 
\qdstar\ is a factor of 3--10 smaller than in other studies 
\citep[e.g.,][]{bottke2010}. Thus, our small planetesimals are relatively
weak and easy to break.

For two colliding planetesimals with masses $m_1$ and $m_2$, the mass of the 
merged planetesimal is
\begin{equation}
m = m_1 + m_2 - \mesc ~ ,
\label{eq: msum}
\end{equation}
where the mass of debris ejected in a collision is
\begin{equation}
\mesc = 0.5 ~ (m_1 + m_2) \left ( \frac{Q_c}{Q_D^*} \right)^b ~ .
\label{eq: mej}
\end{equation}
The exponent $b$ is a constant of order unity 
\citep[e.g.,][]{davis1985,weth1993,kl1999a,benz1999,obrien2003,lein2012}. 
In previous calculations, we adopted $b$ = 9/8; here, we also
consider $b$ = 1 \citep[see also][]{koba2010a,koba2010b}. 

To place the debris in the grid of mass bins, we set the mass of the 
largest collision fragment as $\mmaxd = 0.2 ~ \mesc$ and adopt a 
differential size distribution $n(r) \propto r^{-q}$. After placing
a single object with mass \mmaxd\ in the grid, we place material in 
successively smaller mass bins until (i) the mass is exhausted or
(ii) mass is placed in the smallest mass bin. Any material left over
is removed from the grid.  To explore a broad range of $q$ 
(\S\ref{sec: back}), we derive results for $q$ = 3.5, 4.0, 4.5, 4.5, 
5.0, and 5.5.

To illustrate the general impact of $b$ and $q$ on the evolution, we
consider two simple examples.  When $Q_c/\qdstar$ = 0.5 (early in the 
evolution), our approach yields (\mesc, \mmaxd) =  (0.25, 0.05) $m_c$ 
when $b$ = 1 and (0.23, 0.045) $m_c$ when $b$ = 9/8.  The larger exponent 
results in less debris which is placed in bins with smaller average mass.  
When $Q_c/\qdstar$ = 1.5 (late in the evolution), (\mesc, \mmaxd) = 
(0.75, 0.15) $m_c$ when $b$ = 1 and (0.79, 0.16) $m_c$ when $b$ = 9/8.  
The larger exponent produces more debris which is placed in bins with 
larger average mass.  More debris makes collisional damping and dynamical 
friction more effective, which enhances the growth of the largest objects.  
Thus, the smaller (larger) exponent tends to enhance (retard) growth 
early in the evolution at the expense of slower (more rapid) growth 
later in the evolution. 

The broad range of $q$ we consider has a similar impact on the timing
and magnitude of runaway growth. When the size distribution is shallow
($q$ = 3.5-4.0), debris is placed in a broad range of mass bins, limiting
the impact of collisional damping and dynamical friction. When $q$ is
larger, debris is placed in a narrower range of mass bins, enhancing
the impact of collisional damping and dynamical friction. Overall, we
then expect calculations with larger $q$ and larger $b$ to have faster
growth; calculations with smaller $q$ and smaller $b$ have slower growth.
In \S4, we discuss the impact of these choices in more detail.

As we place the debris in specific mass bins, we also redistribute the
kinetic energy per unit mass of each colliding pair of planetesimals. 
As in \citet{kl1998}, we assume all collisions between mass batches 
conserve the horizontal and vertical components of kinetic energy. For
an initial kinetic energy, $m_1 (h_1^2 + v_1^2) + m_2 (h_2^2 + v_2^2)$, 
any merged planetesimal with mass $m$ receives a fraction $m / (m_1 + m_2)$
of this kinetic energy; any fragment with mass $m_f$ receives a fraction
$m_f / (m_1 + m_2)$.  Recalling the center of mass collision energy from 
eq. (\ref{eq: Qc}), this approach assumes that the escape velocity
component of the collision energy is equal to the energy required to 
disperse the fragments to infinity. 

To compute the evolution of the velocity distribution, we also include 
collisional damping from inelastic collisions and gravitational 
interactions.  For inelastic and elastic collisions, we follow the 
statistical, Fokker-Planck approaches of \citet{oht1999} and 
\citet{oht2002}, which treat pairwise interactions (e.g., dynamical 
friction and viscous stirring) between all objects. For evaluating 
these interactions within a single annulus, we eliminate terms to
calculate the probability that objects in one annulus interact with 
objects in other annuli \citep{kb2001,kb2004b,kb2008}. We also compute 
long-range stirring from distant oligarchs \citep{weiden1989}. The 
Appendix describes several tests of these algorithms.

\section{EVOLUTION OF THE LARGEST OBJECTS}
\label{sec: evol}

To evolve a sea of mono-disperse planetesimals in a single annulus, we set the
size \r0, the surface density $\Sigma_0$, and the orbital elements $e_0$ and $i_0$.
For these simulations, \r0\ = $10^n$~cm where $n$ is an integer between 0 and 8
inclusive. Beyond the snow line, the standard minimum mass solar nebula has 
$\Sigma \approx$ 30~g~cm$^{-2}~(a/{\rm 1~AU})^{-3/2}$ \citep{kb2008}, which implies 
$\Sigma$ = 0.0215 g~cm$^{-2}$ at $a$ = 125~AU. In an annulus with $\delta a = 0.2 a$, 
the total mass is $M_0$ = 15.8~\mearth. For simplicity, we adopt the same mass in
an annulus at 250~AU. In this annulus, $\Sigma$ = $5.37 \times 10^{-3}$~g~cm$^{-2}$,
roughly 70\% of the surface density of a minimum mass solar nebula extrapolated to 
250~AU.  Previous calculations suggest the growth time for large objects scales 
inversely with the mass in solid objects \citep{kl1999a,kb2002b,kb2008,kb2010}.  
Thus, we consider only one surface density.

In most published simulations, the initial orbital elements of planetesimals are set 
to match the escape velocity of the largest objects \citep[e.g.,][]{koba2010b,weid2010}.  
For particles with \r0\ $\gtrsim$ 10--100~m at 125--250~AU, the time scale to reach 
equilibrium is longer than the 1--5~Myr lifetime of the gaseous disk (Appendix). 
Here, we set $i_0$ = $e_0$/2 and adopt $e_0 = 10^{-4}$ for \r0\ = 1~cm to 1~km, 
$e_0 = 10^{-3}$ for \r0\ = 10~km, and $e_0 = 10^{-2}$ for \r0\ = 100--1000~km.  
For the adopted $\Sigma$ at 125--250~AU and these $e_0$, swarms of particles are
gravitationally stable and have $e_0$ intermediate between the likely $e$ for 
particles within a turbulent gaseous disk and an equilibrium $e$ set by a balance 
between stirring and damping (Appendix).  After describing results for these initial 
conditions, we consider how different choices impact the results.

In the next two sub-sections, we focus on the evolution of the largest objects at
125~AU (\S\ref{sec: evol-125}) and at 250~AU (\S\ref{sec: evol-250}). We then conclude 
this section with a brief discussion of the limitations of our calculations.

\subsection{Results at 125~AU}
\label{sec: evol-125}

\subsubsection{Super-Earth Formation}
\label{sec: evol-125-se}

Without gas drag, all calculations follow a similar pattern
\citep[e.g.,][]{kenyon2002,kb2002b,kb2004a,kb2008,kb2010}. Starting from a
mono-disperse set of planetesimals, collisions produce mergers and negligible 
debris.  Growth of larger planetesimals allows dynamical friction to circularize 
the orbits of the largest particles at the expense of raising the $e$ and $i$ 
of the smallest particles. Collisional damping gradually reduces the velocities 
of the smallest objects.  Eventually, significant gravitational focusing factors 
enable runaway growth, where the largest planetesimals grow rapidly by factors 
of 10--100. As these protoplanets grow, viscous stirring raises the velocities 
of smaller planetesimals. Gravitational focusing factors diminish considerably.
Runaway growth ends; oligarchic growth begins.  
Throughout oligarchic growth, collisions produce more and more debris. Once 
protoplanets reach radii of 1000~km or larger, the collisional cascade tries to
remove nearly all of the small particles remaining in the annulus.  As the cascade 
proceeds, protoplanets reach a maximum radius which remains fixed for the rest 
of the calculation.

When $b$ = 9/8 and $q$ = 3.5, protoplanets reach radii of 1000~km or larger in
roughly 1 Gyr (Fig.~\ref{fig: rmax-125a}). In all calculations, growth starts slowly. 
Ensembles of 1~cm particles take $\sim 10^5$~yr to reach 10~cm sizes and another
$\sim 10^6$~yr to reach 1~m sizes. After a very brief period of runaway growth,
the largest objects have \rmax\ $\approx$ 0.1--0.3~km and make the transition 
to oligarchic growth. Large protoplanets then reach 100~km sizes in 200--300~Myr. 
During a second period of runaway growth at 300--500~Myr, protoplanets grow 
past 1000~km. Some reach sizes of 5000--7000~km. We address the origin of this 
second phase of runaway growth in \S4.1.2.

Calculations starting with larger planetesimals follow the same steps. 
For \r0\ = 1~cm to 1~km, the time scale to reach the first runaway growth 
phase increases roughly linearly with the initial particle size. When 
\r0\ = 10--1000~km, the largest planetesimals endure a long phase of 
slow growth, make a smooth transition to oligarchic growth, and then 
gradually accumulate enough material to reach sizes of 500--3000~km.

This evolution is fairly independent of $b$ (Fig.~\ref{fig: rmax-125b}). At the 
smallest sizes (\r0\ $\approx$ 1~cm to 1~km), the evolution of $\rmax(t)$ 
for $b$ = 1 is nearly identical to the evolution for $b$ = 9/8. Because
planetesimals with \r0\ = 100--1000~km hardly grow in 10~Gyr, this evolution
is also insensitive of $b$. For 10~km objects, however, growth is somewhat 
faster when $b$ = 1 than when $b$ = 9/8. In our suite of calculations, 
it takes 
150--175~Myr (225--275~Myr) to produce 100~km objects and
1000--1250~Gyr (1250--1250~Myr) to produce 1000~km objects
when $b$ = 1 ($b$ = 9/8).  The relative abundance of lower mass
fragments causes this difference. During slow growth, calculations with $b$ 
= 1 produce somewhat more fragments than calculations with $b$ = 9/8. The
larger population of fragments circularizes the orbits of the largest
protoplanets more effectively, enabling a stronger runaway growth phase.
Despite this difference, the largest protoplanets reach sizes of roughly
5000~km for all $b$.

To test the sensitivity of these results to stochastic variations within each
simulation, we perform 5--10 calculations for $q$ = 3.5 and each combination of 
$b$ and \r0. For $b$ = 9/8 and \r0\ = 1~cm to 0.1~km, the largest objects have 
radii \rmax\ $\approx$ 3000--4500~km, with a median size \rmax\ $\approx$ 3500~km. 
With \r0\ = 1~km, the median \rmax\ $\approx$ 6900~km; the median \rmax\ drops 
to 4700~km for \r0\ = 10~km. Ensembles of larger planetesimals grow very little: 
\rmax\ = 500~km for \r0\ = 100~km and \rmax\ = 1500~km for \r0\ = 1000~km. 

When $b$ = 1, protoplanets typically reach smaller maximum sizes.  Among the 
suite of calculations, variations in \rmax\ are similar to those for $b$ = 9/8, 
8\% to 10\% for sets of 5--10 calculations with identical starting conditions.  
However, the median \rmax\ is 10\% to 15\% smaller. When $b$ = 1, two factors 
conspire to produce smaller protoplanets. During runaway and the early stages of
oligarchic growth, collisions produce more debris when $b$ = 1. At this time,
collisional damping is ineffective. Thus, calculations with $b$ = 1 lose somewhat
more mass from the grid than those with $b$ = 9/8. Throughout the late stages of 
oligarchic growth and the second phase of runaway growth, relative particle 
velocities are much larger.  Collisions then produce more debris when $b$ = 9/8. 
With so much debris, collisional damping becomes more effective for calculations 
with $b$ = 9/8 than those with $b$ = 1. Damping promotes the growth of super-Earths, 
which are therefore more common when $b$ = 9/8.

Although protoplanets with \rmax\ = 3000--7000~km are characteristic of growth 
with $q$ = 3.5, the largest protoplanets in several calculations reach sizes
exceeding $10^4$~km ($m \gtrsim$ 1~\mearth). Each of these simulations yields
2--3 super-Earths.  For \r0\ = 1~cm to 10~km, 3\% (2\%) of simulations with 
$b$ = 9/8 (1) produce super-Earth mass planets. Each of these examples began 
with small planetesimals, \r0\ $\le$ 100~cm; however, small number statistics
precludes correlating occasional super-Earth formation with \r0.

Super-Earth formation is much more common in calculations with $q \gtrsim$ 4.5
(Fig.~\ref{fig: rmax-125c}). For all $q$, protoplanets grow at similar rates for
100~Myr to 1~Gyr. At later times, protoplanets in calculations with $q \gtrsim$ 
4.5 grow much more rapidly.  In simulations with \r0\ = 1~cm to 1~km and $b$ 
= 9/8, protoplanets reach median maximum sizes \rmax\ $\approx$ 9000--12000~km 
($m \approx$ 0.75--1.8~\mearth).  The variations about the medians are much 
larger, $\sim$ 20\% to 25\% instead of the 10\% range for $q$ = 3.5--4.0.  When 
\r0\ = 10~km, maximum sizes are roughly 20\% smaller.  None of the simulations 
with larger planetesimals (\r0 $\gtrsim$ 10~km) yields a super-Earth.

Results for $b$ = 1 are similar. Suites of simulations
with identical starting conditions have typical protoplanet radii of 
7500--10000~km, roughly 15\% smaller than calculations with $b$ = 9/8. Because
the variations about the median size are large, $\sim$ 25\%, many simulations
with $b$ = 1 and \r0\ = 1~cm to 1~km produce super-Earth mass planets. Ensembles 
of larger protoplanets with \r0\ $\gtrsim$ 10~km never produce super-Earths.

\begin{figure}
\includegraphics[width=6.5in]{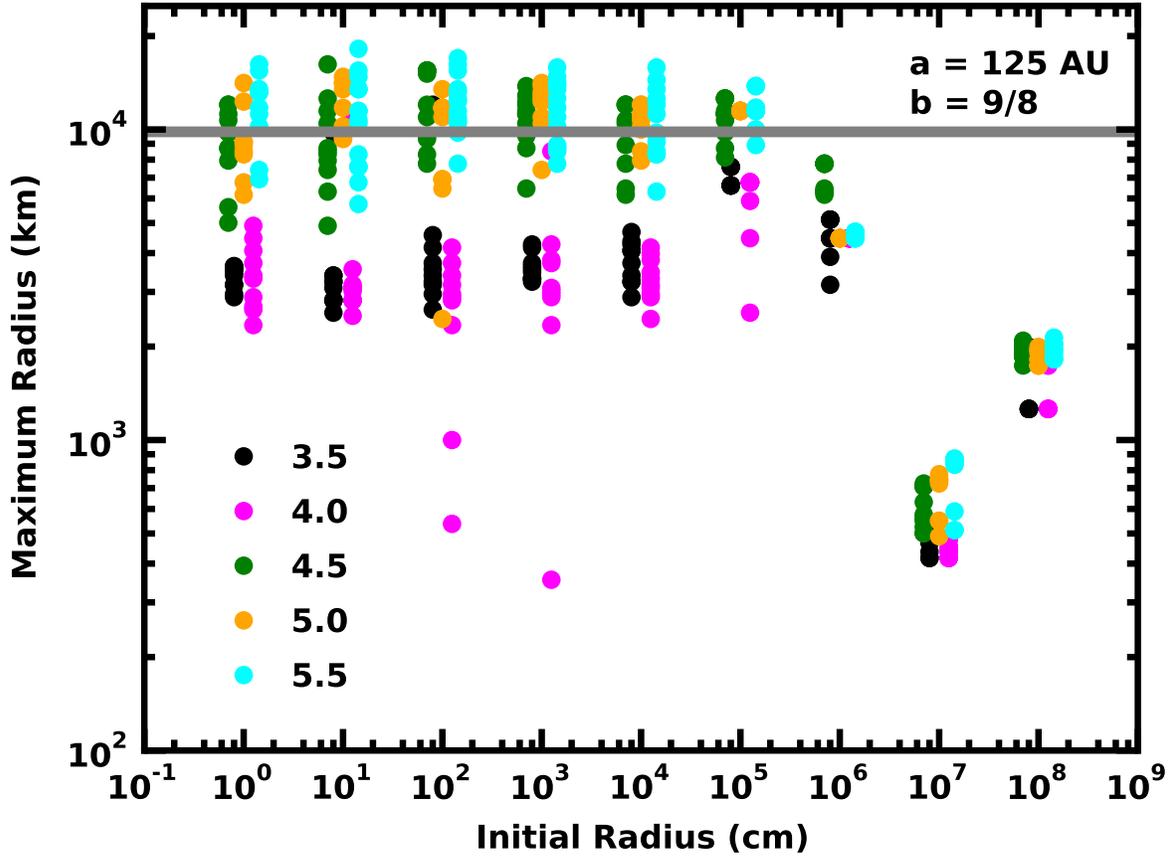}
\vskip 3ex
\caption{%
Maximum planet radius at $a$ = 125~AU as a function of \r0\ and $q$
for $b$ = 9/8.  To improve clarity, points are slightly offset from
the nominal \r0. The horizontal grey bar indicates the radius of an
Earth-mass planet with $\rho_s$ = 1.5 \gcmc. Calculations with 
\r0\ = 1~cm to 1~km and $q \gtrsim$ 4.5 produce super-Earth mass planets.
\label{fig: rmax-125q1}
}
\end{figure}

\begin{figure}
\includegraphics[width=6.5in]{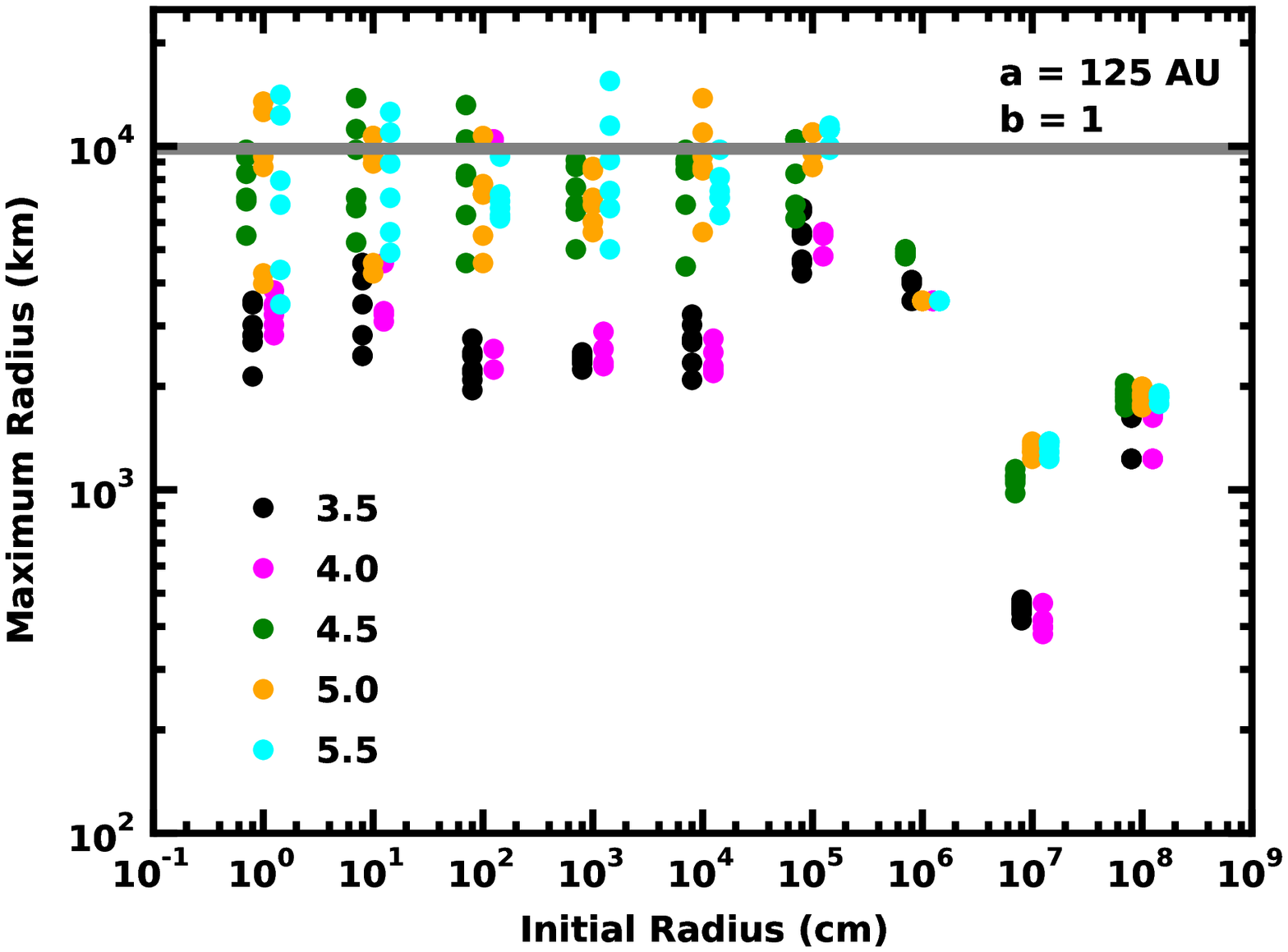}
\vskip 3ex
\caption{%
As in Fig. \ref{fig: rmax-125q1} for calculations with $b$ = 1.
\label{fig: rmax-125q2}
}
\end{figure}

Figs.~\ref{fig: rmax-125q1}--\ref{fig: rmax-125q2} summarize results for \rmax\ as 
a function of $b$, \r0, and $q$. When \r0\ = 1~cm to 0.1~km, $b$ = 9/8, and
$q$ = 3.5 or 4, fewer than 5\% of all simulations produce super-Earth mass
planets (Fig.~\ref{fig: rmax-125q1}).  Although rings composed of larger 
planetesimals (\r0\ = 1~km) usually yield more massive planets when $q$ = 
3.5 or 4, super-Earths are still rare. In contrast, super-Earths form in 
more than half of the simulations with \r0\ = 1~cm to 1~km and $q$ = 4.5, 
5, or 5.5.  When $b$ = 1 and $q$ = 4.5, 5, or 5.5, super-Earth formation is 
somewhat less common (Fig.~\ref{fig: rmax-125q2}). With \r0\ $\lesssim$ 1~km, 
super-Earths form in roughly 25\% of the simulations with $q \gtrsim$ 4.5.  
Among simulations with $q \lesssim$ 4, super-Earth formation is rare.

\subsubsection{Collisional Damping when $q$ = 3.5--4.0}
\label{sec: evol-125-damp1}

Compared to previous calculations of planet formation at 100--150~AU
with similar starting conditions and fragmentation parameters, our 
new calculations yield fairly similar results.  Published multiannulus 
calculations at 100--150 AU with identical fragmentation parameters and 
$q$ = 3.4 have a median \rmax\ $\approx$ 1250~km \citep{kb2010}. 
In the calculations for this paper with $q$ = 3.4 (3.5), the median 
\rmax\ is 1500--2000~km (2000--3000~km).  By neglecting gas drag and 
including the evolution of small particles with radii smaller than 1~m, 
the calculations in this paper produce somewhat more massive large objects. 

Although the new calculations allow the growth of larger objects when
$q$ = 3.4--3.5, the collisional cascade still converts most of the 
initial solid mass into particles with sizes smaller than 1~\mum. 
Radiation pressure rapidly ejects these particles. In \citet{kb2010},
the cascade removes from 0.1\% (\r0\ = 100~km) to 90\% (\r0\ = 1~km)
of the initial mass.  Our new results suggest a weak dependence of 
mass loss on the initial planetesimal radius for small planetesimals; 
the cascade removes 80\% of the initial mass when \r0\ = 1~cm and more 
than 90\% when \r0\ = 1~km.  When \r0\ $\gtrsim$ 10~km, the cascade 
removes very little mass as in the \citet{kb2010} simulations; the 
slow evolution of swarms of large planetesimals then prevents the 
formation of large planets \citep[see also][]{kb2010}.

Despite the rough similarity in \rmax\ and the ejected mass among these 
simulations, the time evolution of the size distribution is very 
different. In previous multiannulus simulations, the cumulative
size distribution can be split into three power-laws 
\citep[e.g.,][]{kb2004c}:
\begin{equation}
\label{eq: sd}
n_c(r) = \left\{ \begin{array}{l l l}
        n_s r^{-\qs} & \hspace{5mm} & r \le r_2 \\
\\
        n_i r^{-\qi} & \hspace{5mm} & r_2 \le r < r_1 \\
\\
        n_l r^{-\ql} & \hspace{5mm} & r \ge r_1 \\
         \end{array}
         \right .
\end{equation}
Here, $r_1 \approx$ 10--100~km, $r_2 \approx$ 0.1--1~km,
\qs\ $\approx$ 2.5--2.7, \qi\ $\approx$ 0--2, and 
\ql\ $\approx$ 3--4 \citep[see also][]{kb2012}.  
Growth sets the slope at large sizes, $r \ge r_1$; collisional debris 
is responsible for the slope at small sizes, $r \le r_2$
\citep{kb2004c,kb2008,kb2012,pan2005,schlicht2011,schlicht2013}.  
At intermediate sizes, competition between collisional destruction 
and debris production sets the slope.

To visualize the evolution of the size distribution in the new
simulations, we normalize the cumulative size distribution to 
the expected size distribution at small sizes, 
$n_{c, rel} (r) = n_c(r) / n_s r^{-q_s}$. With this normalization,
the expected cumulative size distribution is:
\begin{equation}
\label{eq: sd-rel}
n_{c, rel} (r) = \left\{ \begin{array}{l l l}
        1 & \hspace{5mm} & r \le r_2 \\
\\
        n_2 r^{\qs-\qi}  & \hspace{5mm} & r_2 \le r < r_1 \\
\\
        n_1 r^{\qs-\ql} & \hspace{5mm} & r \ge r_1 \\
         \end{array}
         \right .
\end{equation}
For this expression, we set $n_1 = n_l/n_s$ and $n_2 = n_i/n_s$.

\begin{figure}
\includegraphics[width=6.5in]{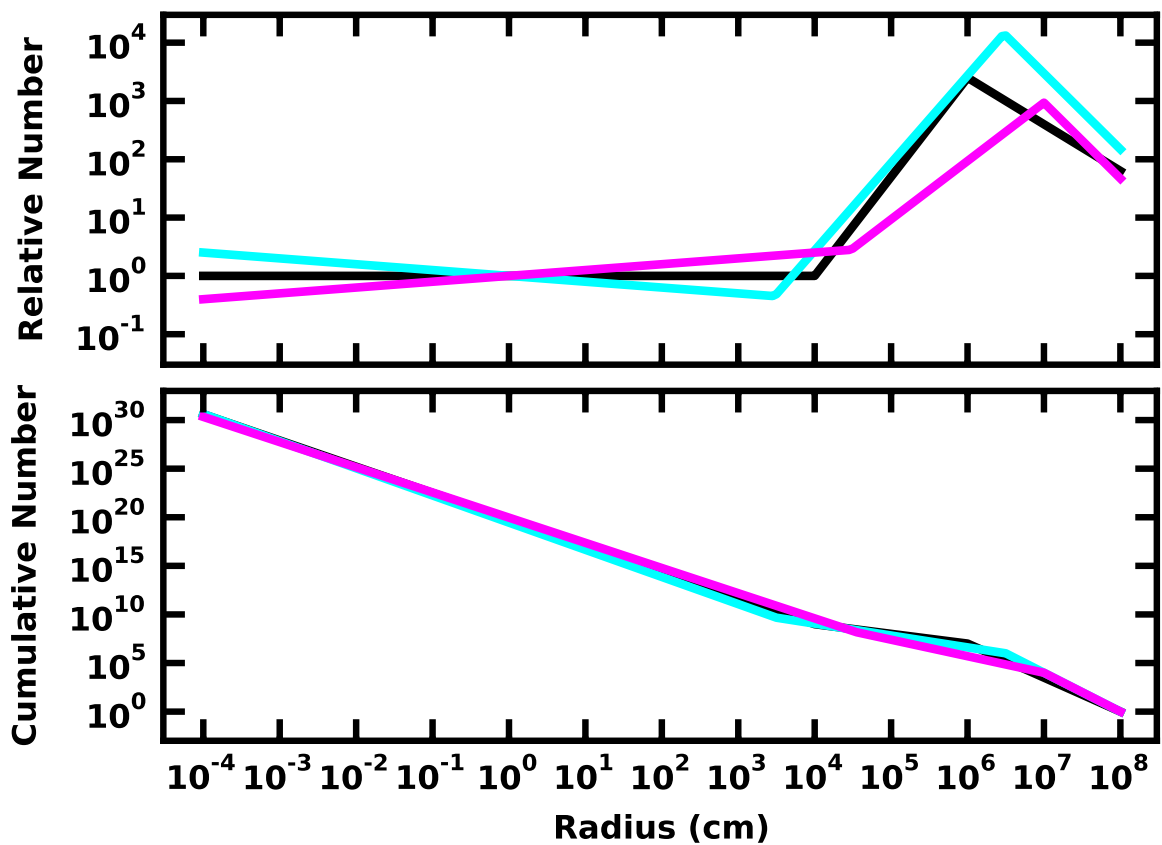}
\vskip 3ex
\caption{%
Comparison of cumulative (lower panel) and relative cumulative
(upper panel) size distributions for three idealized power law
size distributions. 
To construct the relative size distribution, we adopt
a normalization with $q_s$ = 2.7 which yields $n_{c,rel} \equiv$ 
1 at $r$ = 1~cm.
{\it Black curves:} $n_l$ = 1, \qs\ = 2.7, \qi\ = 1.0, \ql\ = 3.5,
$r_1$ = 0.1~km, $r_2$ = 10~km.
{\it Cyan curves:} $n_l$ = 1, \qs\ = 2.8, \qi\ = 1.2, \ql\ = 4.0,
$r_1$ = 0.03~km, $r_2$ = 30~km.
{\it Magenta curves:} $n_l$ = 1, \qs\ = 2.6, \qi\ = 1.7, \ql\ = 4.0.
$r_1$ = 0.3~km, $r_2$ = 100~km.
Relative size distributions allow an easier assessment of the 
differences between power-law components.
\label{fig: sd0}
}
\end{figure}

Fig.~\ref{fig: sd0} illustrates some of the advantages of the 
relative size distribution. In the lower panel, three cumulative
size distributions with different power-law components look fairly
similar. Shifting to the relative size distribution in the upper 
panel, the three curves are very distinct.
With $\qi < \qs < \ql$, the relative cumulative size distribution
rises from $r_2$ to $r_1$ and then falls from $r_1$ to \rmax. This
normalization makes it fairly easy to infer $r_1$ and $r_2$ from
the breaks in the relative size distribution. 

\begin{figure}
\includegraphics[width=6.5in]{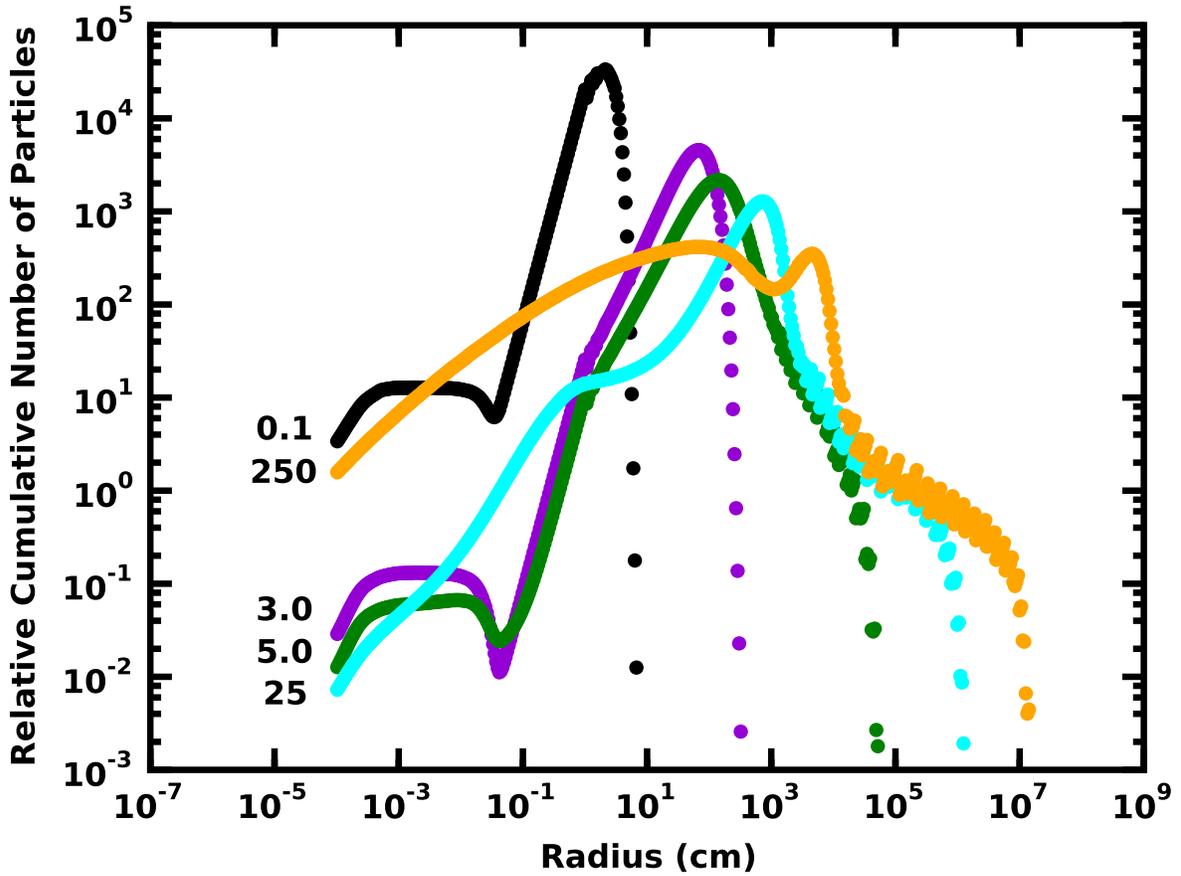}
\vskip 3ex
\caption{%
Evolution of the relative cumulative size distribution $n_{c,rel}$
for a calculation with \r0\ = 1~cm, B = 9/8, and $q$ = 3.5. Here,
$n_{c,rel} = n_c / n_r$ where $n_c$ is the cumulative number 
distribution from the calculation, $n_r = n_0 r^{-2.5}$, and 
$n_0 = 10^{23}$.  Numbers to the left of each curve indicate 
time in Myr. 
\label{fig: sd1}
}
\end{figure}

Fig.~\ref{fig: sd1} shows the evolution of the cumulative relative 
size distribution for the first 300~Myr of a calculation with 
\r0\ = 1~cm, $b$ = 9/8, and $q$ = 3.5. At 0.1--1.0~Myr, the size 
distribution approximately follows the expected shape, with
(i) a flat debris tail at $r \approx$ 1--100~\mum; 
(ii) a sharp rise at $r \approx$ 100~\mum\ to $r \approx$ 1--10~cm; and
(iii) a steep fall for $r \gtrsim$ 1--10~cm. 
The small dip at $r \approx$ 100~\mum\ establishes particle sizes where
removal of particles by merger with much larger particles dominates 
debris production by destructive collisions.  The peak of the curves 
corresponds to particle sizes which contain most of the mass.

As the calculation proceeds, larger and larger particles contain more and
more of the mass. It takes $\sim$ 30~Myr for the peak in the relative size 
distribution to reach 10~m and another 270~Myr to reach 100~m. The growth 
of the largest particles follows this evolution, with 10~m particles in 
1~Myr, 1~km particles in 3~Myr, and 10~km particles in 30 Myr. By 300~Myr, 
the largest particles have reached sizes of 200~km. 

Throughout this evolution, the size distribution develops several distinctive 
features. At small sizes ($r \approx$ 10--100~\mum\ to 1--10~cm), the debris 
tail contains more particles and evolves towards a single power-law. Among
larger particles with $r \approx$ 1--100~km, there is a steeper power-law 
composed of large objects which grow by accreting much smaller objects
\citep[see also][]{kenyon2002,gold2004,kb2008,kb2010,kb2012,schlicht2011}.

\begin{figure}
\includegraphics[width=6.5in]{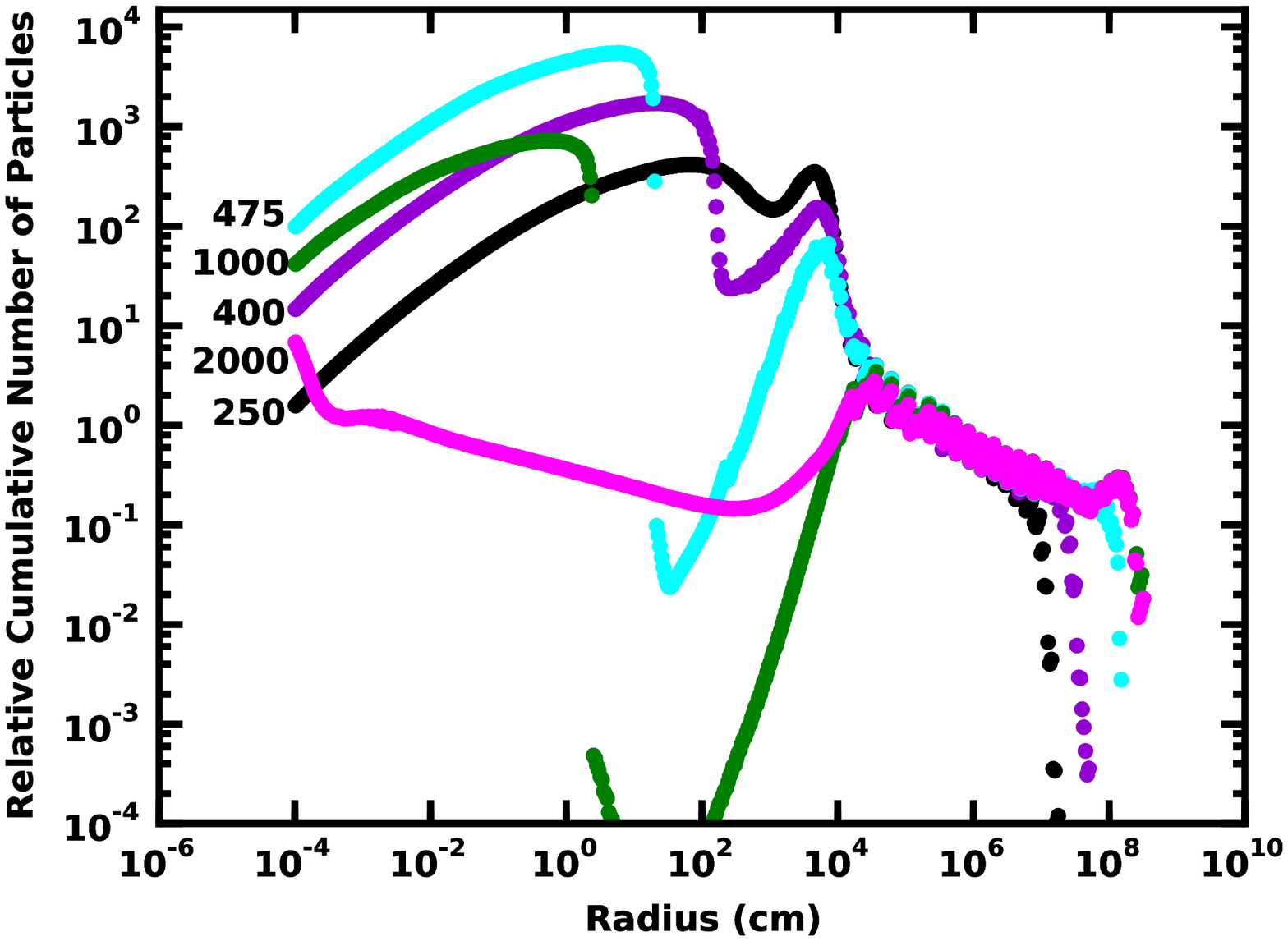}
\vskip 3ex
\caption{%
As in Fig.~\ref{fig: sd1} for later epochs.
\label{fig: sd2}
}
\end{figure}

Once the largest particles have sizes exceeding 500--1000~km, the size 
distribution changes dramatically (Fig.~\ref{fig: sd2}). From 300~Myr to 
500~Myr, (i) the dip at sizes of 10~m shifts to smaller and smaller 
particles and becomes progressively deeper; 
(ii) the peak at sizes of 100~m shifts to larger and larger sizes; and
(iii) the fraction of mass in the debris tail grows significantly. These
changes herald a short (second) runaway growth phase, where the largest 
particles with $r \approx$ 100~km to 1000~km begin to contain most of the
mass and achieve a size distribution with a standard slope $q_l \approx$ 
3 \citep[see also][]{kb2012}. By $\sim$ 1.5~Gyr, collisions have removed
most of the mass in small particles, effectively ending growth of the 
largest particles.

To understand the interesting evolution of the size distribution, it is
helpful to consider the behavior of the velocity distribution. To simplify
this discussion, we calculate the eccentricity $e_k$ of each mass bin 
relative to the Hill eccentricity $e_h = (\mmax / 3 \msun)^{1/3}$,
\begin{equation}
e_{k, rel} = e_k / e_h ~ ,
\label{eq: erel}
\end{equation}
where \mmax\ is the mass of the largest object in the annulus.

\begin{figure}
\includegraphics[width=6.5in]{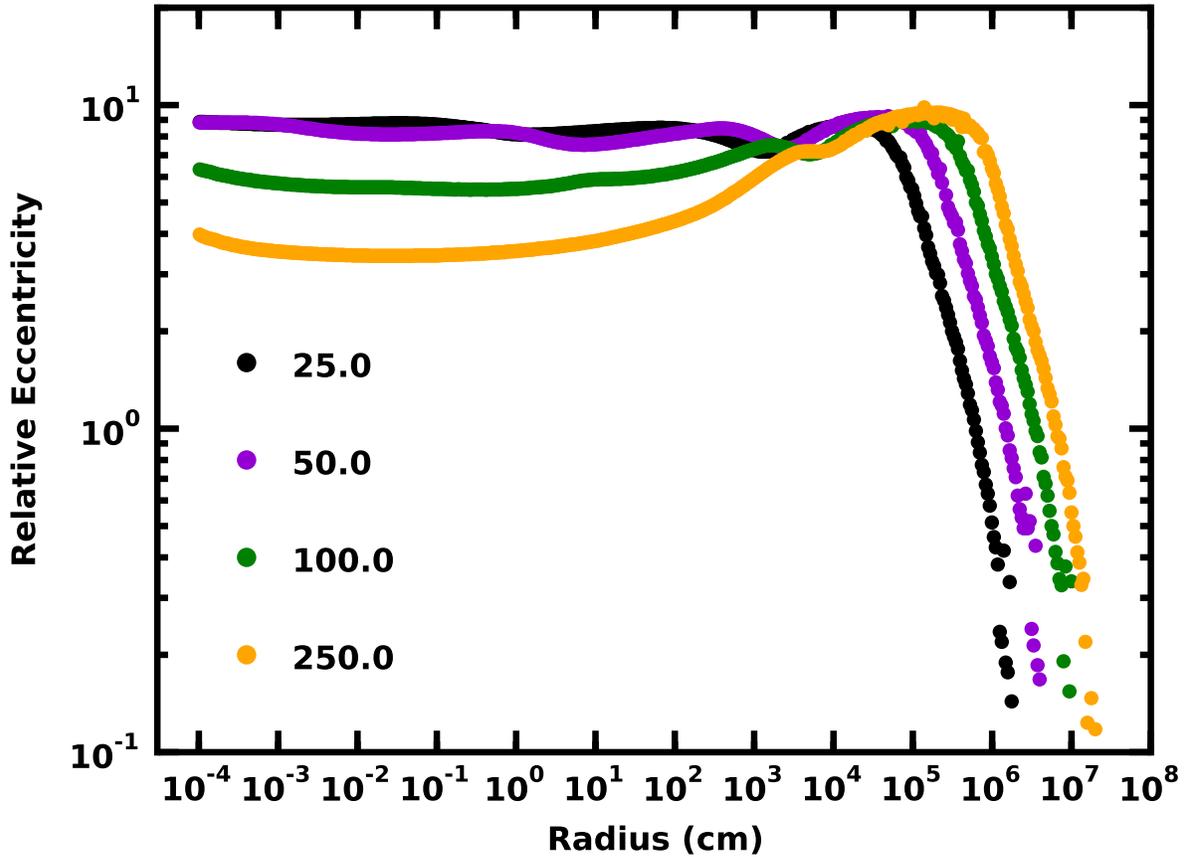}
\vskip 3ex
\caption{%
Evolution of the relative eccentricity $e_H$ in Hill units for
a calculation with \r0\ = 1~cm, B = 9/8, and $q$ = 3.5 at times 
(in Myr) indicated in the legend. As the maximum particle size
grows, small particles maintain $e_H \approx$ 10. At larger 
sizes ($r \gtrsim$ 0.03 \rmax), the eccentricity follows a power 
law $e_H \approx 0.2 (r / \rmax)^{-1}$.
\label{fig: vd1}
}
\end{figure}

During the first period of runaway growth and oligarchic growth,
the relative eccentricity distribution follows a standard pattern
\citep[Fig.~\ref{fig: vd1}; 
see also][]{weth1993,kl1999a,oht2002,kb2002b,gold2004,kb2004a,kb2008}. 
When the largest objects reach sizes of 1~km, collisional damping 
and viscous stirring have established an equilibrium where the 
random velocities of small particles ($r \lesssim$ 0.1~km) are 
$\sim$ ten times the Hill velocity of the largest particles, a factor 
of 2--4 smaller than in our previous calculations where \rmin\ = 1~m 
\citep[e.g.,][]{kb2008,kb2010,kb2012}.
At the largest sizes, dynamical friction reduces the eccentricities 
to a minimum $e_{k,rel} \approx$ 0.1--0.2. In between, dynamical 
friction maintains an approximate power-law, with 
$e_{k,rel} \propto r^{-n}$ and $n \approx$ 0.5--1.
Despite the larger stirring as \rmax\ reaches 30~km, a rough balance 
between collisional damping and viscous stirring preserves 
(i) the constant relative eccentricity, $e_{k, rel} = e_k / e_H \approx$ 10, 
among the small particles and (ii) the power law decline among the 
larger particles.

As the evolution proceeds past 100~Myr, it is convenient to consider
the behavior of large, intermediate, and small particles.  For our 
adopted \qdstar, large objects with $r \gtrsim$ 10~km do not shatter; 
they gain mass and energy from collisions with smaller particles in 
the grid.  Growing protoplanets also try to stir all other particles 
to larger and larger velocities. The magnitude of the stirring grows 
with the size of the protoplanet; the impact on smaller particles is 
fairly independent of the size of the small particles. Dynamical 
friction with small objects circularizes the orbits of the protoplanets.

Intermediate mass particles with $r \approx$ 10--100~m contain most of 
the mass but are easily shattered.  Collisions among these objects place 
mass and energy into smaller mass bins.  Compared to stirring by large 
particles, damping is relatively inefficient. Thus, intermediate mass 
bins lose mass and gain velocity with time.

Small objects with $r \lesssim$ 1~cm contain little mass but do not 
shatter. These particles gain mass from the debris of collisions among
intermediate mass particles and lose mass from collisions with large
objects. Slow growth among these particles gradually shifts material
to larger bins. As this material moves into the intermediate mass bins,
destructive collisions cycle it back into smaller mass bins. With most
of the mass in intermediate mass particles, the small particles gain 
mass with time.

The evolution of $e_{k,rel}$ for small particles depends on the
relative importance of collisional damping and stirring by large
particles. Without any gain in mass from destructive collisions,
stirring by large particles dominates damping. However, destructive
collisions rapidly cycle material from 10--100~m particles into 
smaller particles. Collisions among stronger small particles are
not destructive. Thus, the net mass in small particles grows 
(Figs.~\ref{fig: sd1}--\ref{fig: sd2}). During this evolution, 
collisional damping gradually overcomes stirring by larger particles;
$e_{k,rel}$ gradually declines.

Ultimately, the shape of the \qdstar\ relation drives this evolution.
In alternative calculations where $\qdstar \propto r^{-\beta_b}$ and
$\beta_b \le$ 0 at small radii, dramatic reductions in $e_{k, rel}$ 
among small particles cannot occur. When the largest particles reach 
10--100~km sizes, the smallest particles in the grid shatter first. 
As large objects continue to grow, shattering impacts larger and 
larger particles, but the mass in the smallest particles always 
decreases. This evolution severely limits collisional damping and 
allows a collisional cascade to destroy all objects below some size 
limit. 

With $\beta_b > 0$, intermediate mass particles always shatter first.  
For our adopted parameters, particles with $r \approx$ 30~m have the 
smallest \qdstar.  As the random velocities of all particles slowly 
increase, debris from collisions of these weak planetesimals populates
mass bins where particles are stronger and do not shatter. Collisional
damping can then reduce $e_{k,rel}$ for objects in these mass bins. 

The drop in $e_{k, rel}$ from collisional damping is responsible for a
second runaway in the growth of the largest objects. With $e_{k, rel}$
$\approx$ 0.5--2 at 300--500~Myr, gravitational focusing factors rise
dramatically. As the largest protoplanets grow from 500~km to 2000~km,
viscous stirring overcomes collisional damping for smaller and smaller
objects. Thus, the `collisional damping front' moves from roughly 10~m
at 300~Myr to 10~cm at 500~Myr to 0.1~cm at 1.5~Gyr; the relative
eccentricities for the larger objects rise from $e_{k, rel} \approx$ 
5--6 at 500~Myr to $e_{k, rel} \approx$ 20 at 1.5~Gyr to 
$e_{k, rel} \approx$ 30--40 at 10~Gyr.

In all simulations with $q$ = 3.5--4.0, the ultimate sizes of the largest 
objects depend on the timing of the dramatic reduction in $e_{k,rel}$ for 
the smallest objects. Usually, oligarchic growth produces large objects with
$e_{k,rel} \gtrsim$ 3--10 before collisional damping is able to reduce the
relative eccentricities for the smallest objects. Collisions between small 
objects and oligarchs then occur in the dispersion regime, where gravitational 
focusing factors are small and growth is slow. Despite damping, collisions 
among small objects are destructive. The cascade then slowly removes the
mass in small objects before oligarchs reach super-Earth sizes.

Sometimes, conditions conspire to enable the production of several super-Earth
mass planets. When oligarchs and small objects have $e_{k,rel} \lesssim$ 1--3,
collisions are in the shear regime where gravitational focusing factors are
large. Growth is then very rapid. For simulations with $q$ = 3.5--4.0, 
the debris from every destructive collision is spread out over several orders 
of magnitude in radius. The large radius spread in the debris (i) limits the 
impact of dynamical friction on the oligarchs (enabling higher orbital 
eccentricities) and (ii) reduces the ability of collisional damping. Thus, 
super-Earth formation is very rare. 

\subsubsection{Collisional Damping when $q$ = 4.5--5.5}
\label{sec: evol-125-damp2}

When $q \gtrsim$ 4.5, the second phase of runaway growth becomes more
explosive (Figs.~\ref{fig: rmax-125c}). At 300~Myr to
1~Gyr, the largest particles rapidly grow from \rmax\ = 2000--4000~km
to super-Earth sizes with \rmax\ $\gtrsim 10^4$~km. Although super-Earth
production is fairly insensitive to $q$ for $q \gtrsim$ 4.5, maximum
planet sizes depend on $b$. Calculations with $b$ = 9/8 almost
always yield larger planets than calculations with $b$ = 1.

Two factors enable routine super-Earth formation when $q \gtrsim$ 4.5.
In all calculations, the collisional cascade begins after the formation
of large objects with \rmax\ $\gtrsim$ 500--1000~km. Debris production
grows dramatically. Because radiation pressure ejects particles smaller 
than 1~\mum, the mass loss rate from the grid also grows.  Although the 
debris production rate is insensitive to $q$, calculations with larger 
$q$ concentrate the debris into a smaller range of radii and eject fewer
particles with radii smaller than 1~\mum. At similar stages of evolution,
more mass remains in the grid. Thus, the largest protoplanets tend to
grow to larger masses.

\begin{figure}
\includegraphics[width=6.5in]{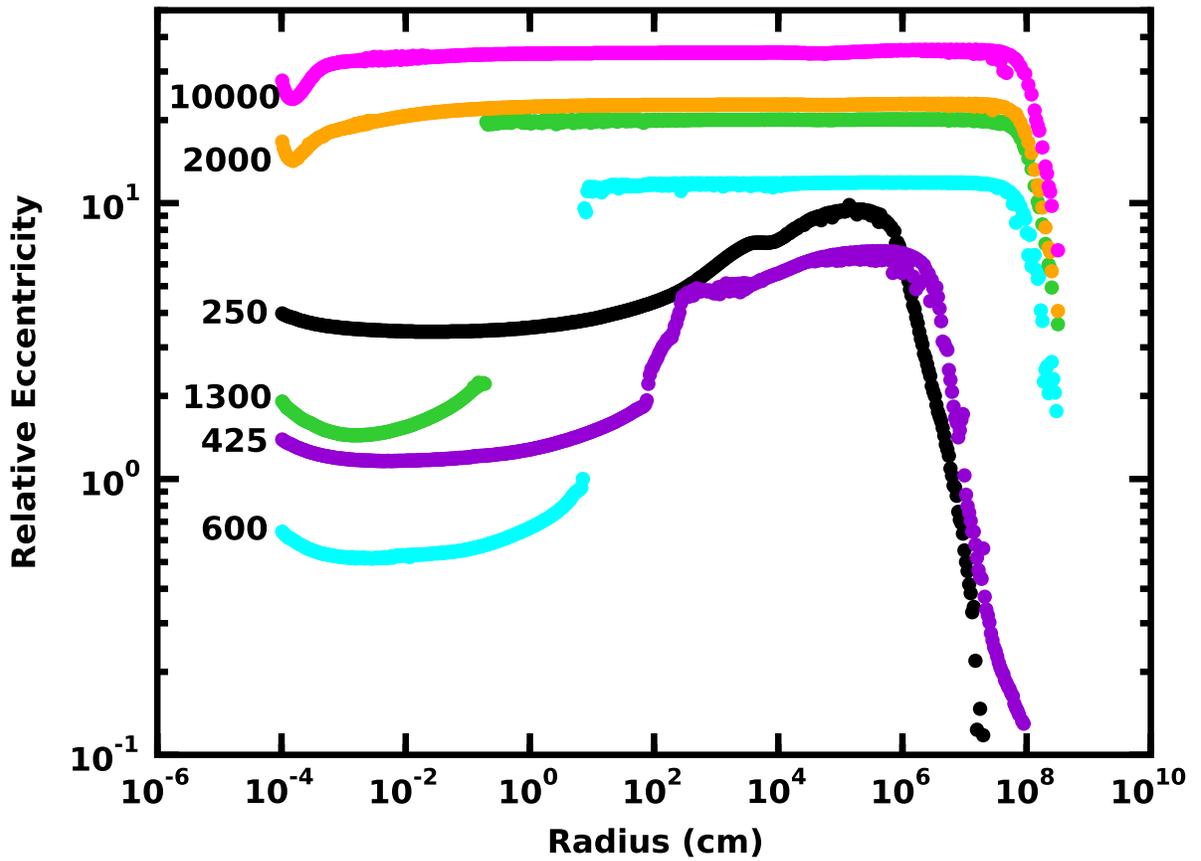}
\vskip 3ex
\caption{%
As in Fig.~\ref{fig: vd1} for later epochs. As the evolution 
proceeds, collisional damping depresses the eccentricity for
small particles. Once these particles have been swept up or 
destroyed, the eccentricity of small particles rises.
\label{fig: vd2}
}
\end{figure}

Calculations with larger $q$ also enhance the impact of collisional 
damping. When $q$ is small (3.5--4.0), debris is spread out over a 
broad range of sizes. Collisional damping -- which depends on the
collision rate -- then reduces the eccentricities of small particles
more efficiently than those of larger particles (Fig.~\ref{fig: vd2}).
With less mass in smaller objects than in larger objects, the growth of
the largest protoplanets is fairly slow. When $q$ is large (4.5--5.5),
debris is concentrated in the largest particles. Collisional damping
is then more effective for larger particles, which now contain much 
more mass than the smaller particles. Once relative eccentricities 
fall below the Hill velocity, protoplanets rapidly accrete the more 
massive particles and reach super-Earth sizes.

\begin{figure}
\includegraphics[width=6.5in]{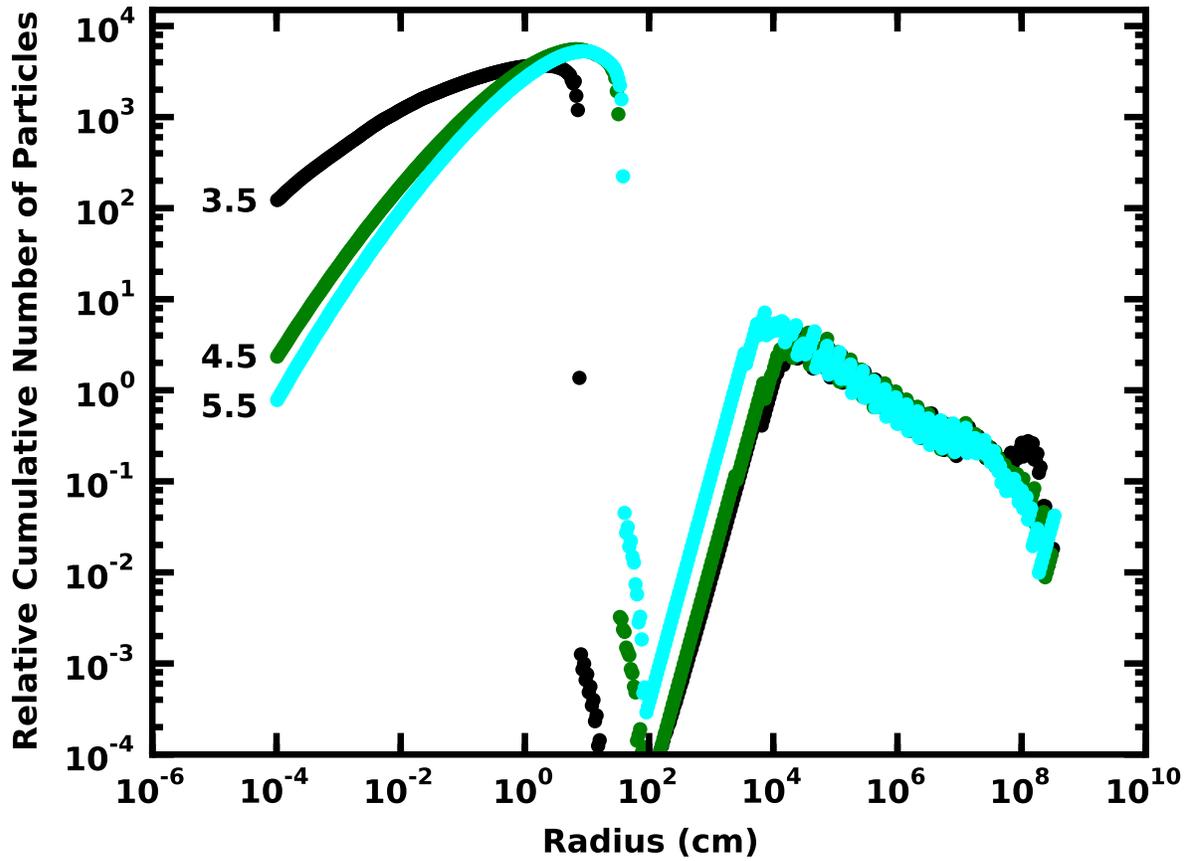}
\vskip 3ex
\caption{%
Relative cumulative size distributions when \rmax\ = 3000~km
for calculations with $q$ = 3.5 (black points), 4.5 (green
points), and 5.5 (cyan points). Calculations with larger $q$
have more material in particles with $r$ = 1~cm to 0.1~km and
less material in particles with $r \lesssim$ 1~cm.
\label{fig: sd3}
}
\end{figure}

Fig.~\ref{fig: sd3} compares the relative size distributions for 
calculations with various $q$. To illustrate the differences clearly,
we focus on evolution times where the largest objects have \rmax\ =
2000--3000~km and are just starting their second phase of runaway 
growth. When $q$ = 3.5, the slope of the size distribution for small
particles with $r \lesssim$ 1--10~cm is shallow. Calculations with
larger $q$ have steeper size distributions with more mass in the 
largest objects. Among larger particles with $r \gtrsim 10^4$~cm,
the size distributions are nearly independent of $q$. At intermediate
sizes (10~cm $\lesssim r \lesssim$ 10--100~m), the concentration of
mass in larger particles is striking.

\begin{figure}
\includegraphics[width=6.5in]{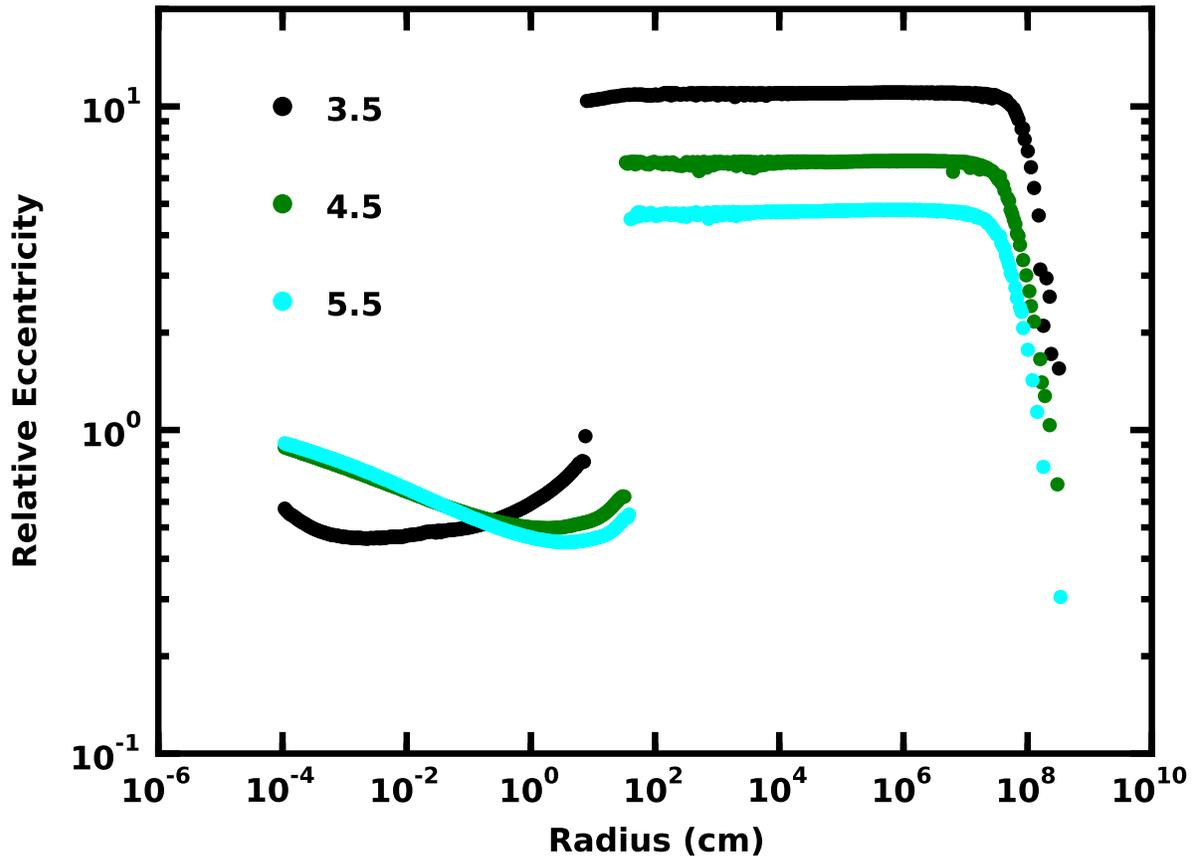}
\vskip 3ex
\caption{%
As in Fig.~\ref{fig: sd3} for the relative eccentricity.
In calculations with larger $q$, particles with 
$r \gtrsim$ 1~cm have smaller $e_{k, rel}$; smaller 
particles have larger $e_{k, rel}$.
\label{fig: vd3}
}
\end{figure}

Fig.~\ref{fig: vd3} compares the relative eccentricity distributions for 
the same epochs as Fig.~\ref{fig: sd3}. At small sizes ($r \lesssim$
1--10~cm), the changing impact of collisional damping with increasing
$q$ is apparent: damping is more effective at small sizes when $q$ = 3.5
and at large sizes when $q$ = 4.5 and 5.5. With more debris in larger
particles, the damping front -- which marks the transition between particles 
with large $e$ and small $e$ -- extends to larger $r$ for calculations 
with larger $q$. For all particles with $r \gtrsim$ 1--10~cm, calculations 
with larger $q$ have smaller relative eccentricities. When $q$ = 3.5, 
the largest protoplanets have $e_{k, rel} >$ 1; despite the small 
$e_{k, rel}$ for the smallest particles, these planets grow relatively 
slowly during the second runaway growth phase. When $q$ = 4.5--4.5, the 
largest protoplanets have $e_{k, rel} <$ 1; these planets can grow rapidly. 

\subsubsection{Evolution with Different Starting Conditions}
\label{sec: evol-125-einit}

To examine how the results of these simulations depend on initial conditions,
we consider suites of simulations where the initial eccentricity and inclination
are a factor of three larger or smaller than our nominal $e_0$ and $i_0$. For
each \r0, we perform another 3--5 calculations with the different starting values
of $e$ and $i$. Within each suite of simulations, the median \rmax\ is nearly
identical; there is no obvious variation of \rmax\ with $e_0$. The range in 
\rmax\ for these simulations is also similar, $\sim$ 20\% to 25\%.

Within the full suite of simulations, there are modest differences in the
time evolution of \rmax\ and the size distribution as a function of starting
$e$ and $i$. In rings composed of small objects with \r0\ $\lesssim$ 10--100~m,
gravitational focusing is negligible at the start of the simulation. Thus, 
growth rates depend only on the geometric cross-section and the relative numbers
of small and large objects. Although collisions produce more debris when the 
initial $e$ and $i$ are large, the debris contains a negligible amount of mass.
Thus, the initial slow growth of small particles is independent of $e_0$.

As the evolution proceeds, the rapid growth of the largest particles has two 
regimes. When the largest objects make the transition from slow growth to 
runaway growth, gravitational focusing factors increase dramatically. Ensembles
of particles with smaller initial $e$ reach this transition earlier than 
ensembles of particles with larger initial $e$. Defining $t_R$ as the time 
of this transition, our results suggest $t_R (e_0/3) \approx 0.25 t_R (3 e_0) $.

When the largest objects make the transition from runaway to oligarchic growth, 
the growth rate again depends on the initial eccentricity. Throughout runaway
growth, collisions in simulations with the largest $e_0$ produce the most 
debris. More debris leads to more effective collisional damping and somewhat
more effective dynamical friction between the largest and smallest objects.
As a result, oligarchs grow faster in calculations with larger $e_0$. Despite
starting out more slowly, these oligarchs overtake the oligarchs in calculations
with smaller $e_0$. Thus, calculations with larger $e_0$ produce super-Earths
somewhat more rapidly than calculations with smaller $e_0$.

Among swarms of large particles with \r0\ $\gtrsim$ 1~km, the evolution 
seems to depend little on $e_0$.  Over typical growth times of 10--100~Myr
for 1~km particles, collisional damping, dynamical friction, and viscous 
stirring produce similar size distributions for calculations with different 
$e_0$. At the start of runaway growth, all calculations then appear to follow 
a similar sequence. Within this sequence, there is a large dispersion of 
outcomes due to stochastic variations in the collision rate. This dispersion
dominates small differences in outcomes due to the starting conditions.

\begin{figure}
\includegraphics[width=6.5in]{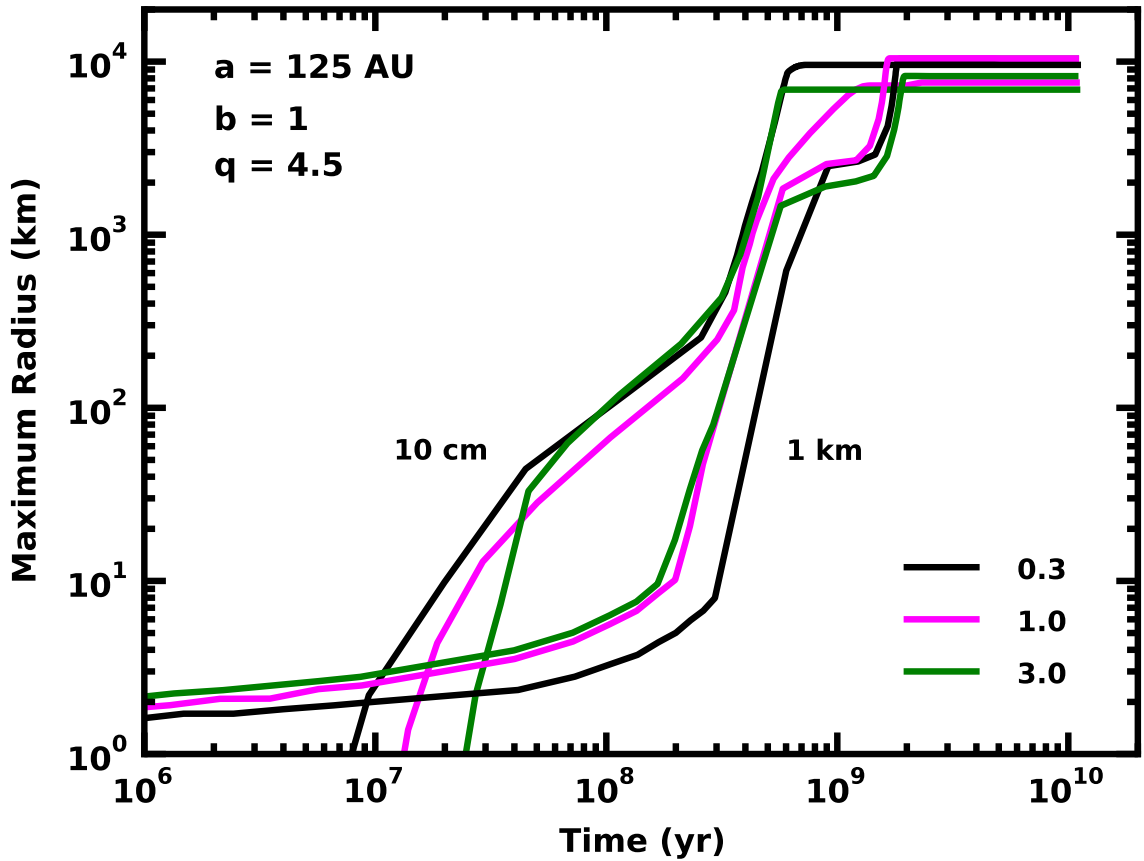}
\vskip 3ex
\caption{%
Growth of the largest object as a function of initial 
eccentricity. For the two particle sizes considered
(\r0\ = 10~cm and 1~km), the legend in the lower right
corner indicates the initial eccentricity relative to
the nominal $e_0$.
Overall, growth is fairly independent of the initial
$e$ and $i$.
\label{fig: rmax-e}
}
\end{figure}

Fig.~\ref{fig: rmax-e} illustrates some of these points for calculations
with $b$ = 1 and $q$ = 4.5. When \r0\ = 10~cm, particles with smaller initial
$e$ grow somewhat faster. Once \rmax\ exceeds 10~km, oligarchs in calculations
with larger $e$ grow faster and eventually overtake oligarchs with smaller
initial $e$. At late times, the largest objects enter a second runaway growth
phase, where they accrete material rapidly at rates independent of $e_0$. 
Eventually, the largest objects reach similar (super-Earth) sizes. When
\r0\ = 1~km, growth is more stochastic. Sometimes, small $e$ calculations
lag those with larger $e$ (as in the Figure). In other simulations, large $e$
calculations lag. All eventually produce 1--2 protoplanets with large radii.

\subsection{Results at 250~AU}
\label{sec: evol-250}

At 250~AU, the evolution from small planetesimals into super-Earth mass planets
follows the behavior described for calculations at 125~AU.  To avoid repeating 
the discussion of \S\ref{sec: evol-125}, we summarize differences between the
two sets of simulations in the rest of this section.

The time scale for planet formation is the simplest difference between sets
of calculations at 125~AU and at 250~AU.  At any semimajor axis, the time scale 
for planets to reach identical sizes scales inversely with the initial mass in 
planetesimals and linearly with the orbital period. With identical initial 
planetesimal masses in our two sets of calculations, the formation time scales
with orbital period. Thus, planets at 250~AU take roughly 3 times longer to 
form than those at 125~AU. For calculations with \r0\ = 1~cm, this difference 
is not a major issue: 1~Gyr formation times at 125~AU become 3~Gyr time scales at
250~AU. Both time scales are shorter than the age of the Solar System. For 
calculations with \r0\ $\approx$ 10--100~m, this difference is crucial: 2--3~Gyr
formation time scales at 125~AU become 6--10~Gyr time scales at 250~AU. Thus,
large particles at 250~AU do not form super-Earth mass planets over the age of 
the Solar System.

\begin{figure}
\includegraphics[width=6.5in]{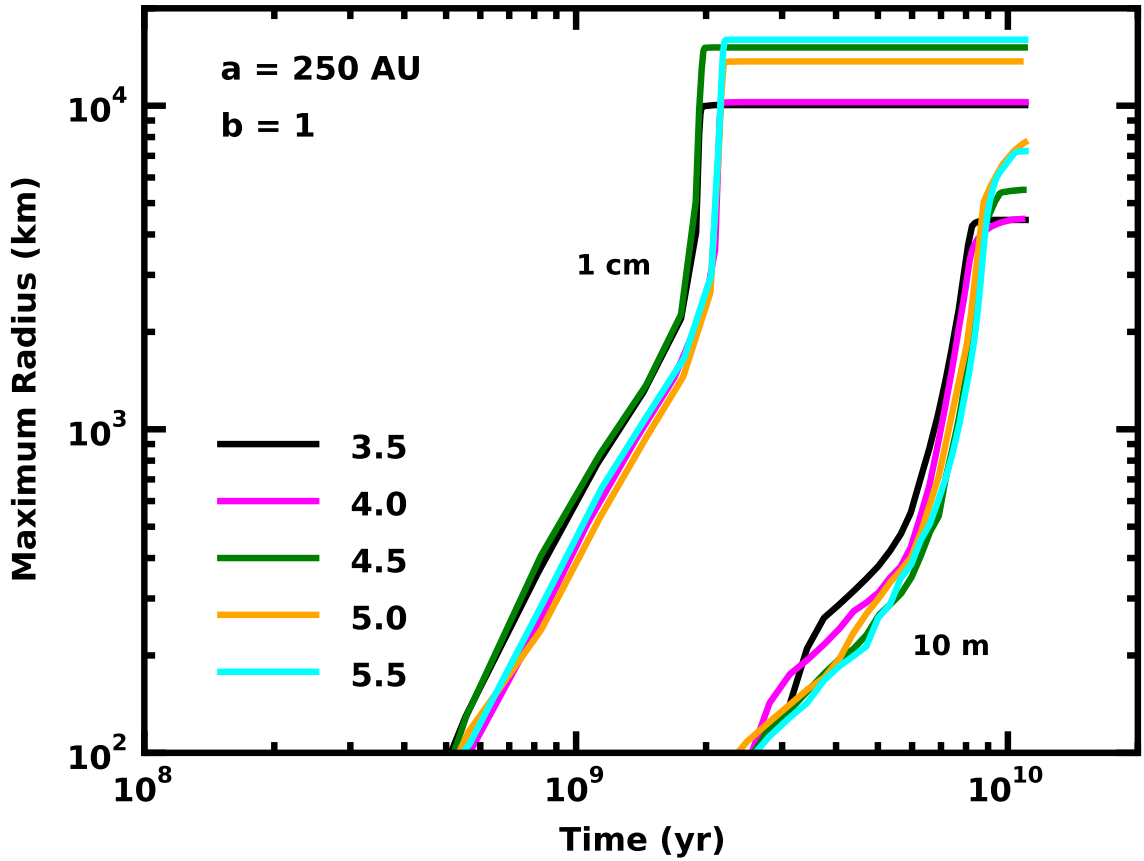}
\vskip 3ex
\caption{%
Growth of the largest object at 250~AU as a function of 
$q$ for $b$ = 1 and \r0\ = 1~cm and 10~m. For rings of
small particles with \r0\ = 1--10~cm, coagulation produces
super-Earth mass planets for all $q$ in 2--4~Gyr. Larger
particles with \r0\ $\gtrsim$ 10~m cannot grow into 
super-Earths over the age of the Solar System.
\label{fig: rmax-250}
}
\end{figure}

Fig.~\ref{fig: rmax-250} illustrates this point for the growth of 1~cm and 10~m
particles.  When \r0\ = 1--10~cm, growth follows the same stages outlined in 
\S\ref{sec: evol-125}.  After 1--10~Myr of slow growth, periods of runaway growth, 
oligarchic growth, and runaway growth produce super-Earth mass planets in 2--4~Gyr. 
The growth time is roughly 3 times longer than the time scale at 125~AU (see 
Figs.~\ref{fig: rmax-125a}--\ref{fig: rmax-125b}).  When \r0\ $\gtrsim$ 10~m, 
every phase of growth takes longer.  Thus, the first runaway growth phase begins 
at 100~Myr to 1~Gyr, instead of 1--10~Myr.  Although collisional damping still 
promotes rapid growth during the second runaway growth phase, rapid growth begins 
at 5--10~Gyr -- too late to allow super-Earth formation within the age of the
Solar System. 

Although growth at 250~AU takes longer, small particles reach super-Earth sizes 
independently of $q$ (Fig.~\ref{fig: rmax-250}). For \r0\ = 1--10~cm and all $q$, 
collisional damping drives the second period of runaway growth, where 
2000--3000~km protoplanets rapidly grow into super-Earths. The time scale to 
produce super-Earths is also insensitive to $q$, 2--4~Gyr.  Although super-Earths
form for all $q$, calculations with $q$ = 4.5--5.5 typically yield planets roughly 
twice as massive as calculations with $q$ = 3.5--4.0.

At larger sizes ($\r0 \gtrsim$ 1--10~m), final planet masses depend on $q$.
Calculations with $q$ = 3.5--4.5 stall at typical sizes of 4000--5000~km; 
simulations with $q$ = 4.5--5.5 reach radii of 7000--8000~km at 10~Gyr. On
longer time scales, calculations with large $q$ eventually reach super-Earth
masses. Models with smaller $q$ never produce massive planets.

Results at 250~AU seem fairly insensitive to $b$. For \r0\ = 1--10~cm,
all calculations yield super-Earths. The largest planets have masses of
2--3~\mearth\ for $b$ = 1 and $b$ = 9/8. For \r0\ $\gtrsim$ 10~m, planets
also have maximum sizes that do not depend on $b$. When \r0\ = 1~m, calculations
with $b$ = 9/8 and $q$ = 4.5--5.5 routinely yield super-Earth mass planets
on time scales of 2--4~Gyr. Simulations with $b$ = 9/8 (1) and $q$ = 3.5--4.0
(3.5--5.5) cannot produce super-Earths in 10~Gyr.

\begin{figure}
\includegraphics[width=6.5in]{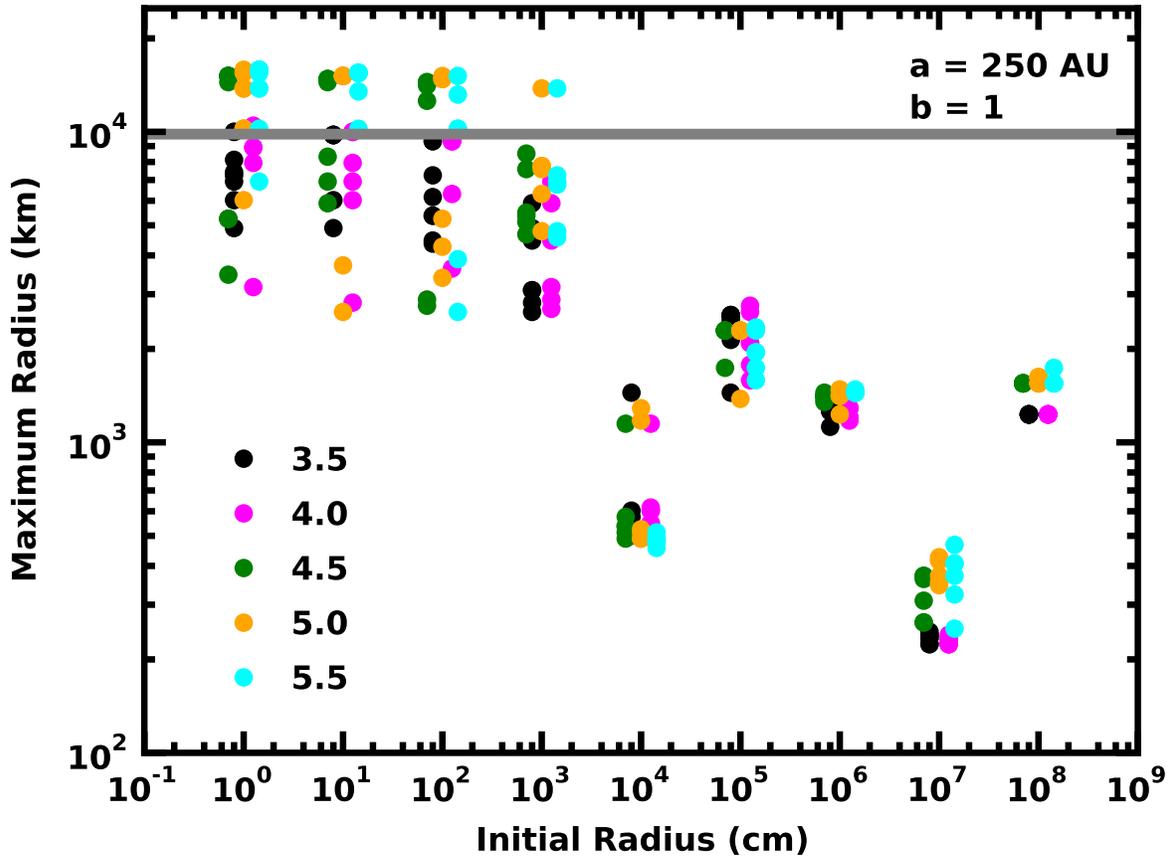}
\vskip 3ex
\caption{%
Maximum planet radius at $a$ = 250~AU as a function of \r0\ and $q$
for $b$ = 1.  To improve clarity, points are slightly offset from
the nominal \r0. The horizontal grey bar indicates the radius of an
Earth-mass planet with $\rho_s$ = 1.5 \gcmc. Calculations with \r0\ = 
1~cm to 1~m produce super-Earth mass planets. Larger planetesimals
evolve into much smaller planets on 1--10~Gyr time scales.
\label{fig: rmax-250q1}
}
\end{figure}

\begin{figure}
\includegraphics[width=6.5in]{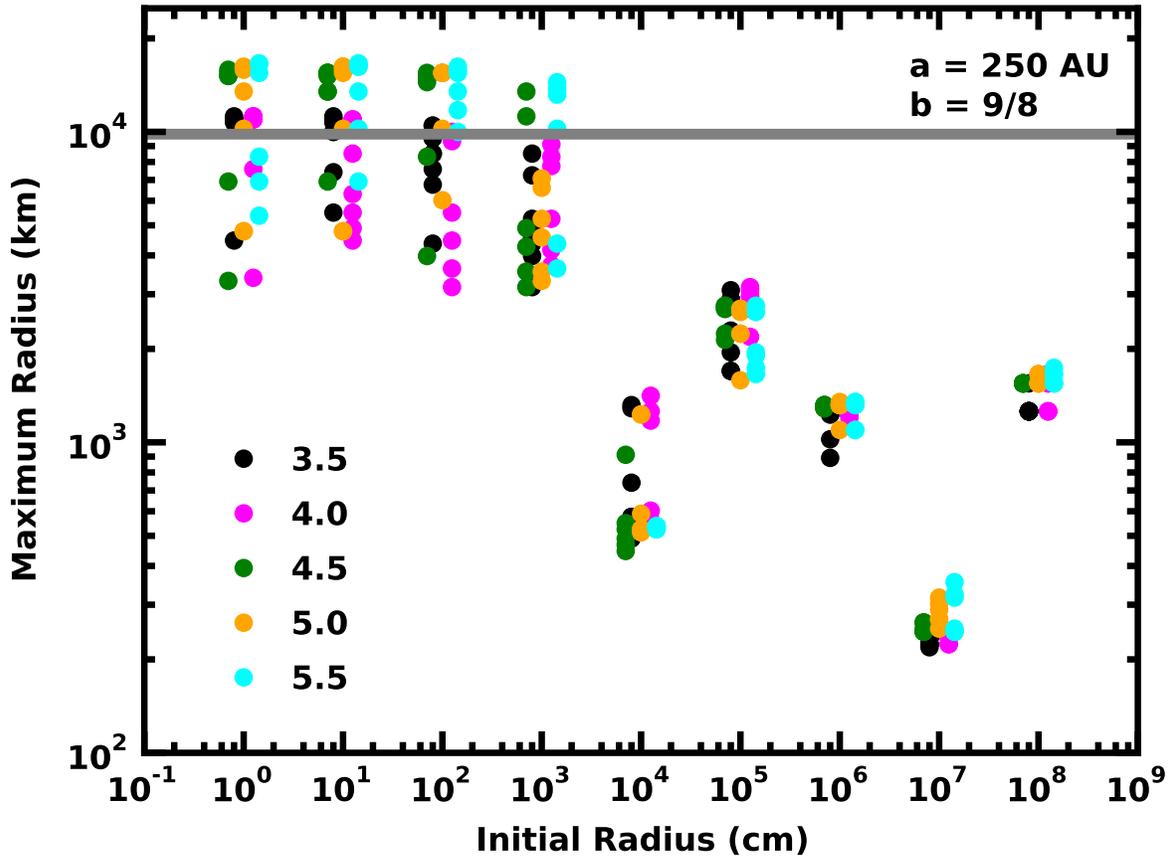}
\vskip 3ex
\caption{%
As in Fig. \ref{fig: rmax-250q1} for calculations with $b$ = 9/8.
On 1--4~Gyr time scales, super-Earths form from planetesimals
with \r0\ = 1~cm to 1~m. Occasionally, ensembles of 10~m 
planetesimals can grow into 1--2 super-Earths if $q \gtrsim$ 4.5.
Larger planetesimals produce planets with radii ranging from
a few hundred km to a few thousand km.
\label{fig: rmax-250q2}
}
\end{figure}

Figs.~\ref{fig: rmax-250q1}--\ref{fig: rmax-250q2} summarize \rmax\ for the 
complete suite of calculations at 250~AU. When $b$ = 1, nearly all calculations
with \r0\ = 1--10~cm yields a super-Earth mass planet. Although roughly half 
of simulations with \r0\ = 1~m yield a super-Earth, many produce smaller planets 
with \r0\ = 3000--7000~km. These results are fairly insensitive to $q$. With 
\r0\ = 10~m, only a few calculations yield super-Earths. Ensembles of larger 
planetesimals never produce super-Earths.  
When $b$ = 9/8 and \r0\ = 1~cm to 1~m, super-Earth formation is also common.
Larger planetesimals almost never produce a super-Earth. These results are also
fairly independent of $q$.

\subsection{Limitations of the Models}
\label{sec: evol-lim}

In previous papers, we have summarized various limitations in coagulation
calculations for planet formation \citep[e.g.,][]{kl1999a,kb2004a,kb2008}. 
Here, we focus on how choices in methods and initial conditions impact our
results.

\subsubsection{Resolution}

Finite resolution -- in the mass spacing parameter $\delta$ and the width 
of the annulus $\delta a$ -- is characteristic of all numerical coagulation 
calculations.
As noted in the Appendix, simulations with finite $\delta$ lag the results of
a calculation with `perfect'  mass resolution 
\citep[see also][]{weth1990,kl1998}. For $\delta$ = 1.2, this lag is 
negligible relative to simulations with $\delta$ = 1.05--1.10 (Appendix). 
Thus, uncertainties due to finite $\delta$ are small.

Adopting a single annulus for our calculations limits our ability to follow
the evolution of a planetesimal swarm as a function of semimajor axis 
\citep[e.g.,][]{spaute1991,weiden1997b,kb2004a,kb2008}. Despite this 
disadvantage, single annulus calculations yield fairly accurate estimates 
for the growth times and outcomes of coagulation calculations. For the initial 
conditions adopted here, the formation time depends only on the orbital 
period.  Across the width of our single annulus, $\delta a / a$ = 0.2, the 
orbital period changes by 35\%. If the surface density is constant across
the annulus, planetesimals at the inner edge of the annulus grow 35\% faster 
than planetesimals at the outer edge of the grid. With typical formation
times of $\sim$ 1~Gyr and a factor of two range in the formation time,
uncertainties due to the size of the annulus are small. Thus, our results
yield good first estimates for the growth of super-Earths at 125--250~AU.

Although we allow any maximum size in the coagulation grid, setting a finite
lower limit on particle size \rmin\ changes the evolution of larger particles. 
During slow and runaway growth, particles with $r \sim \rmin$ have negligible 
mass and little impact on the evolution of larger particles. Once the collisional 
cascade begins, debris rapidly fills the smallest mass bins. Particles with sizes 
$r \sim$ 10--100~\rmin\ are swept up by much larger protoplanets and destroyed by
particles with somewhat smaller sizes $r \sim$ 1--10~\rmin.  With no particles 
smaller than \rmin\ in the grid, our calculations have fewer destructive 
collisions among particles with $r \lesssim \rmin$ compared to a calculation
with an infinitesimally small \rmin. Thus, our calculations produce an `excess'
of small particles relative to an `ideal' calculation 
\citep[see also][and references therein]{campo1994,obrien2003,kriv2006,kriv2008}. 

The standard method to address this issue is to extend the size distribution
below \rmin, assume these particles have relative velocities similar to 
particles with $r \approx \rmin$, derive collision rates and outcomes with 
particles in the grid, and correct the excess number of particles with 
$r \approx$ \rmin\ \citep[e.g.,][]{campo1994,obrien2003}. Although this 
approach is attractive when \rmin\ $\approx$ 1--100~cm, it is problematic
when \rmin\ $\approx$ 1~\mum. For solar-type stars,
radiation pressure likely places smaller particles on highly eccentric or
hyperbolic orbits as long as the ensemble of solids is optically thin 
\citep[e.g.,][]{burns1979}. Relative collision velocities are then much
larger, but collision rates are much smaller. 

We plan to perform a set of multiannulus calculations to quantify outcomes 
with realistic orbital geometry for small particles. To provide an initial
estimate of the importance of the small size cut-off, we perform several
calculations with \rmin\ = 0.1~\mum. Aside from shifting the excess of
small particles to smaller sizes, this change has a modest impact on
the overall size distribution or the growth of small objects. Thus, this
uncertainty is minimal.

\subsubsection{Fragmentation}

The fragmentation algorithm is another uncertainty. We use a standard energy
scaling approach which mimics the results of more detailed SPH and $n$-body
calculations of binary collisions. Although our fragmentation parameters -- 
$Q_b$, $Q_g$, $\beta_b$, and $\beta_g$ -- are consistent with the detailed
calculations, other choices are possible 
\citep[e.g.,][]{morby2009b,bottke2010,kb2010,koba2014}. Based on published 
results, it is straightforward to assess how alternative choices impact our 
results.  For simplicity, we separate the discussion into the gravity regime 
for large objects ($r \gtrsim$ 0.1~km) and the strength regime for small objects
($r \lesssim$ 0.1~km).

In the gravity regime, there is general agreement among $n$-body and SPH 
calculations for $\beta_g$
\citep{benz1999,lein2012}. Small variations about the adopted value have 
little impact on collision outcomes. Current simulations suggest at least 
a factor of 2--3 uncertainty in $Q_g$ \citep{morby2009b,bottke2010}. 
Although the ultimate sizes of protoplanets grow with increasing $Q_g$
\citep[e.g.,][]{kb2010}, efficient collisional damping probably limits the 
importance of this parameter on our conclusions. Reducing $Q_g$ by a factor
of 100 or more probably prevents the formation of 1000~km or larger
protoplanets. More modest changes, however, probably have little impact
on our results.

In the strength regime, different choices for $Q_b$ and $\beta_b$ do 
not affect runaway and oligarchic growth \citep{kb2010}.  Making small 
particles harder to break (by increasing $Q_b$ and decreasing $\beta_b$) 
increases the population of small objects and the effectiveness of 
collisional damping, enabling a broader set of initial conditions to 
produce super-Earths.  

For smaller particles, likely outcomes depend on the relative importance
of shattering and collisional damping.
As an example, we consider the impact of reducing \qdstar\ by a factor of 
ten for particles with $r \lesssim$ 100~m in the strength regime. In our
nominal calculations, collisions with impact velocities $v \gtrsim$ 
400~\cms\ shatter 30~m particles. As the largest particles grow past 
100~km, collisional damping maintains collision velocities for these
particles just above this limit (Fig.~\ref{fig: vd1}). Although these 
particles shatter, particles much smaller than 10~m and much larger than 
100~m do not shatter, which allows collisional damping to initiate the 
second phase of runaway growth. When \qdstar\ is a factor of ten smaller, 
impact velocities $v \gtrsim$ 125~\cms\ shatter small particles, 
distributing debris into smaller mass bins. If the collisional damping 
rate remains identical to that in our nominal calculations, continued 
shattering removes the debris rapidly. Collisions of much more massive 
particles cannot produce debris fast enough to replenish material lost 
from smaller mass bins. The collisional cascade then stalls the growth 
of the largest objects, which never reach super-Earth masses.

However, increased debris production associated with a smaller \qdstar\ also
increases the effectiveness of collisional damping. A simple estimate suggests
that damping counters the impact of a larger rate of shattering. When 
\qdstar\ is a factor of ten smaller, debris production for fixed $Q_c$ is a
factor of ten larger (eq.~\ref{eq: mej}). With a factor of ten more mass in
small particles, the damping rate is a factor of ten larger. Setting the
approximate rates for viscous stirring and dynamical friction from large 
particles to this damping rate \citep[e.g.,][]{gold2004}, the likely 
equilibrium velocity of small particles, 75--100~\cms, is smaller than the
125~\cms\ shattering velocity. Thus, collisional damping appears capable of
countering any large reduction in \qdstar\ and maintaining the evolution on
the path to super-Earth formation. 

As noted in \S4.1.2, setting $\beta_g \le$ 0 limits the impact of collisional 
damping on the velocity evolution of small particles. When $\beta_g \le$ 0, 
the smallest particles in the grid shatter first. Debris from these collisions
permanently leaves the grid. As successively larger particles begin to shatter,
smaller particles continue to shatter. More debris leaves the grid. With too
little mass among smaller particles, collisional damping is ineffective.
Velocities for small particles remain large, preventing a second runaway growth
phase leading to the formation of super-Earth mass planets. Maximum protoplanet
radii are then small, $\sim$ 2000~km.  Repeating our calculations with different 
parameters for \qdstar\ is necessary to confirm this conclusion.

Aside from choosing $q$ and the four parameters in the \qdstar\ relation,
we also set the mass of the largest fragment in the debris, $m_L = 0.2 \mesc$,
where \mesc\ is the total mass of the debris 
\citep[see also][]{weth1993,kl1999a}. This approach differs from 
\citet{morby2009b}, $m_L \approx 0.004 e^{(Q_c / 4 \qdstar)^2} \mesc$;
\citet{koba2010a}, $m_L \approx \epsilon (Q_c / \qdstar)^{-1} \mesc$ 
and $\epsilon \approx$ 0.01-0.5; and
\citet{weid2010}, $m_L = 0.5 (Q_c / \qdstar)^{-0.78} \mesc $.
For moderate collision energies ($Q_c \approx \qdstar$), 
our algorithm places a somewhat smaller fraction \citep{weid2010}, 
a similar fraction for $\epsilon$ = 0.2 \citep{koba2010a}, 
or a much larger fraction \citep{morby2009b} of the ejected 
mass in the largest particle within the debris. 
At low collision energies ($Q_c \ll \qdstar$), 
\citet{koba2010a} and \cite{weid2010} have more massive fragments;
\citet{morby2009b} has less massive fragments. At high collision
energies ($Q_c \gg \qdstar$), all other algorithms place more 
material in smaller fragments.

Although accurately assessing the impact of these choices requires a 
broad set of additional simulations, we can outline the probable role 
of $m_L$ in our calculations at 125--250~AU. The \citet{morby2009b} 
algorithm always places more mass in less massive particles within 
the debris.  Per collision, this approach thus assigns more mass to
particles with $r < \rmin$, leading to less mass among debris with 
$r > \rmin$. Fewer debris particles results in less efficient 
collisional damping and limited growth during any second phase of
runaway growth. Super-Earth formation seems unlikely. 

In the \citet{koba2010a} and \citet{weid2010} approaches, super-Earth
formation probably follows a similar path as in our calculations.  At 
the start of the second runaway growth phase, most of the mass is in 
particles with $r \gtrsim$ 1~km and $Q_c \approx \qdstar$. For identical
$q$, these algorithms place the debris in similar mass bins. Among small 
objects ($r \lesssim$ 1~cm) with $Q_c \lesssim \qdstar$, our algorithm
removes mass from the grid more rapidly than these two approaches and
thus underestimates the growth rate for super-Earths. At intermediate 
sizes (10~cm to 0.1~km), our approach places mass in larger fragments,
removing mass less rapidly and thus overestimating the growth rate for
super-Earths. For collisions among particles with these sizes, however, 
producing smaller fragments probably enhances collisional damping. If
this enhanced collisional damping is sufficient to retain these small 
particles before low velocity destructive collisions remove them,
super-Earth formation is likely.

To summarize, fragmentation is a critical issue in the growth of
super-Earths at 125--250~AU. For reasonable choices of parameters
in \qdstar, likely outcomes are similar to the results discussed
in \S\ref{sec: evol-125}--\ref{sec: evol-250}. Plausible choices
for $m_L$ have a broader range of consequences. In our assessment,
the \citet{morby2009b} algorithm probably prevents super-Earth
formation; other algorithms \citep[e.g.,][]{koba2010a,weid2010} 
probably allow super-Earth formation. Verifying these conclusions
require additional simulations which are beyond the scope of the
present effort.

\subsubsection{Initial Conditions}

Aside from the initial radius for planetesimals \r0, our choices for the
initial conditions have little impact on our results. Our adopted set of
\r0\ spans a broad range of plausible sizes. Calculations with smaller
\r0\ will behave almost identically to simulations with \r0\ = 1~cm.
Because collision times scale with initial radius, outcomes for swarms of 
planetesimals with \r0\ $>$ 1000~km will follow those for \r0\ = 1000~km
but on longer time scales. Thus, calculations with much smaller (larger)
\r0\ yield (do not yield) super-Earth mass planets.

Another initial condition -- the size distribution of planetesimals -- is
probably also not important. When planetesimals initially have a broad range 
of sizes, dynamical friction starts to modify the velocity distribution
at the start of the 
calculation. Growth of large particles (\r0\ = 1--100~km) then usually 
proceeds as much as a factor of two faster than growth from a mono-disperse 
set of particles \citep{kb2010,kb2012}.  We expect similar differences at 
125--250~AU. Thus, starting with a broad range of sizes might enable some 
initial conditions to yield super-Earths on 2--3~Gyr time scales instead of 
4--6~Gyr time scales.

A broad range of sizes probably has much less impact on the outcomes
of calculations with small particles.  When \r0\ is small, dynamical 
friction and gravitational focusing are less important. Independent 
of the initial conditions, short collision times and slow growth 
rapidly produce a broad range of sizes. Thus, our conclusions for the 
growth of small particles into super-Earths are fairly independent of
the initial range of sizes.

Our results indicate that the starting eccentricity also has little
impact on the outcomes of the calculations. For simulations with 
\r0\ $\gtrsim$ 1~km, our choices for the initial eccentricity span 
the likely range of possibilities.  At smaller sizes, the equilibrium 
$e$ and $i$ are much smaller than our adopted $e_0$ and $i_0$. If 
interactions with the gas circularize the orbits of small particles 
effectively, the initial $e$ and $i$ might be much smaller.  
Gravitational instability of the swarm is then very likely 
\citep[e.g.,][]{michi2007,michi2009}, leading to much more 
rapid growth of large particles.

\subsubsection{Dynamics}

The approximations used in coagulation codes begin to break down 
when (i) most of the mass is in a few planets and (ii) the dynamical 
interactions between these objects dominates collective interactions 
between small and large objects. To quantify these limits, we first
consider dynamical constraints. At 125--250~AU, the escape velocity
from the Solar System is roughly 2.5--4~\kms. With $\rho_s$ = 1.5~\gcmc, 
planets have escape velocity
\begin{equation}
v_{esc} \approx 9 \left ( R_p \over 10^4~{\rm km} \right ) ~ \kms\ ~ .
\label{eq: v-esc}
\end{equation}
Gravitational interactions raise the relative velocities of massless
particles to roughly 60\% of the escape velocity. Thus, planets with $R_p \gtrsim$ 
5000~km can eject small particles from the solar system. 

At the same time, collisional damping opposes gravitational scattering.
Our calculations suggest damping overcomes scattering until planets 
reach super-Earth sizes. Thus, dynamical ejection of small particles 
is probably unimportant throughout most of each calculation.

When protoplanets are large enough, their dynamical interactions are
stronger and more chaotic than suggested by our Fokker-Planck 
approximation.  Using the isolation criterion of eq.~(\ref{eq: iso}), 
two (four) equal-mass oligarchs on roughly circular orbits are isolated 
from one another when they have \rhill\ $\approx$ 0.028 $a$ (0.014 $a$). 
These limits correspond to planets with masses of roughly 
20~\mearth\ (3~\mearth).  With typical masses of 2--3~\mearth, the 
two super-Earths 
produced in many of our calculations are clearly isolated. Large groups 
(10--20) of much lower mass planets are also stable.  Although four 
super-Earths are nominally isolated, these planets are close to the 
dynamical stability limits and could be dynamically unstable 
\citep[e.g.,][and references therein]{kb2006,bk2006}. 

Deriving the number of massive protoplanets in each calculations allows us 
to quantify the importance of strong dynamical interactions between protoplanets.
Among 200 (150) calculations at 125~AU (250~AU), 45\% (0\%) produce one
super-Earth, 47\% (78\%) produce two, 6\% (12\%) produce three, and
2\% (10\%) produce four.  In most of these calculations -- 98\% at 125~AU 
and 90\% at 250~AU -- dynamical interactions among super-Earths are very 
unlikely. In others, dynamical interactions are probably common. 

During the other 800 (450) calculations at 125~AU (250~AU) which do not
yield a super-Earth, protoplanets almost never interact dynamically. 
Because calculating these interactions is very cpu intensive 
\citep[e.g.,][]{kb2006}, our neglect of these interactions is reasonable.
Once we consider multiannulus calculations of super-Earth formation at
125--250~AU, we will revisit this issue.

Although protoplanets eventually contain most of the mass remaining in 
each simulation, the approximations in our coagulation code rarely break
down. Most simulations do not produce one or more super-Earths. In these
situations, most of the mass is in protoplanets with sizes ranging from
a few hundred km to a few thousand km. Our code accurately handles these
situations. When 1--2 super-Earths form, these objects begin as 2000--3000~km
objects which accrete material from the swarm of small particles with
low eccentricity. Because these two objects do not interact dynamically,
our statistical approach accurately derives the accretion rate of small
particles.

In a few cases, 3--4 oligarchs accrete small particles and grow to
super-Earth masses. Our approach handles the growth of these oligarchs
accurately. Dynamical interactions among these protoplanets become 
important near the end of the accretion process, when their masses approach 
an Earth mass. At this point our approximations begin to break down.
Usually, we promote these objects into an $n$-body code and follow their 
dynamical interactions precisely \citep[e.g.,][]{kb2006,bk2006,bk2011a}.
Here, we simply note the formation of 3--4 objects with near super-Earth
masses and save the more cpu intensive calculation for another study.

\section{DISCUSSION}
\label{sec: disc}

Our analysis indicates a plausible path for {\it in situ} production of 
super-Earth mass planets at 125--250~AU around solar-type stars. This
path begins with a massive ring, $M_0 \gtrsim$ 15~\mearth, composed
of small particles, $r_0 \approx$ 1~cm to 10~m. Successive mergers of
these objects lead to larger and larger protoplanets.  On a time scale 
of $\sim$ 1~Gyr, coagulation yields several planets with masses
$m \approx$ 1--4~\mearth. 

Although we did not consider a range of initial masses for rings at 
125--250~AU, previous studies demonstrate how the growth times and
final masses of protoplanets depend on the initial mass and semimajor
axis of a swarm of solid objects \citep[e.g.,][]{kb2008, kb2010}. 
For all initial disk masses, the time scale for protoplanets to reach 
a fiducial radius (e.g., 1000~km) depends inversely on the initial disk 
mass. When $M_0 \approx$ 15~\mearth\ at 125~AU, it takes protoplanets 
400~Myr to 1~Gyr to reach radii of 1000~km. Doubling (halving) the 
initial disk mass halves (doubles) these time scales to 200--500~Myr 
(0.8--2~Gyr).  Although the total masses of the most massive protoplanets
also scale roughly linearly with mass, gravitational interactions among 
the protoplanets set the ultimate masses of the most massive protoplanets.
For the conditions we consider, halving the initial disk mass reduces 
this ultimate mass by roughly 50\%; however, doubling the initial mass
may simply double the number of super-Earths rather than double the
masses of the most massive super-Earths.  Finally, the time scale for
planet formation depends linearly on the orbital period.  Doubling the 
initial semimajor axis nearly triples the formation times.

Our conclusions raise several interesting issues concerning planet 
formation theory. Super-Earth formation at 100--250~AU also has direct 
observational consequences for the structure of debris disks around 
solar-type stars and for the architecture of the outer Solar System. 
After a brief discussion of several theoretical points, we describe how 
the new calculations impact our understanding of debris disks and the 
outer Solar System. We conclude this section with several observational 
tests which can help to constrain plausible models.

\subsection{Theoretical Issues: Initial Conditions}
\label{sec: disc-theory-init}

Our model for {\it in situ} super-Earth formation at 125--250~AU requires 
a massive reservoir of solids in the outer regions of the protosolar nebula.  
Observations of nearby star-forming regions suggest these initial conditions 
are plausible.  In the $\rho$ Oph and Taurus-Auriga molecular clouds, disks 
with radii of 100--300~AU are fairly common around pre-main sequence stars 
with ages of roughly 1~Myr 
\citep[e.g.,][and references therein]{and2005,and2009,and2010,will2011}. 
Of the twelve $\rho$ Oph disks with outer radii larger than 100~AU, five
have measured surface densities at least as large as the 0.02~\gcms\ (125~AU)
or the 0.005~\gcms\ (250~AU) required by our simulations. Within most disks 
around young stars, observations require particles at least as large as 
1--10~mm \citep[e.g.,][]{will2011}. In many cases, observations also reveal 
that these particles have scale heights much smaller than the local scale 
height of the gas \citep[e.g.,][]{furl2006,furl2009}.  Thus, current evidence 
implies that these disks have particle sizes and total masses similar to those 
required by our calculations.

Once cm-sized `pebbles' form and settle towards the disk midplane, various
processes shape their evolution. Collisions can slowly merge pebbles into
larger objects or shatter them into fine dust grains
\citep[e.g.,][]{,birn2010,birn2012, windmark2012,garaud2013}. 
At 125--250~AU, viscous evolution slowly drives the gas (and any solids coupled 
to the gas) to larger distances \citep[e.g.,][]{lbp1974,step1998,cha2009}. 
Large solids feel a headwind from the gas, which drags them towards (and 
perhaps into) the Sun \citep[e.g.,][]{ada76,weiden1977a,raf2004}. However, 
the gas can also concentrate pebbles into much larger solid objects 
\citep[e.g.,][]{youdin2005,johansen2007}. Nearly all published studies of 
these phenomena focus on outcomes inside 30~AU, where planet formation is
a common process. Scaling these results to conditions at 125--250~AU, it 
appears that 1~cm and larger particles can survive for the 1--3~Myr lifetime
of the gaseous disk.  Detailed studies of grain evolution are necessary to
consider whether this conclusion is robust. If it is, our calculations suggest
super-Earth formation is a likely outcome in large protostellar disks. 

Observations of solid objects beyond 30~AU might constrain the outer radius 
of the gaseous component of the protosolar nebula.  \citet{kretke2012} show 
that the gaseous disk excites Kozai oscillations in planetesimal leftovers 
from the formation of gas giant planets at 5--20~AU.  Some of these 
planetesimals end up on orbits with low eccentricity and high inclination. 
The current lack of these objects suggests an outer edge of $\sim$ 80~AU 
for the protosolar nebula. Observations over the next decade should allow 
significant refinement in this estimate. 

\subsection{Theoretical Issues: Fragmentation}
\label{sec: disc-theory-frag}

The calculations in \S4 illustrate the complex interplay between 
collisional damping, fragmentation, and the growth of the largest
protoplanets in a gas-free environment. When protoplanets reach sizes of
500--1000~km, they initiate a collisional cascade which gradually grinds 
km-sized and smaller objects into micron-sized dust grains.  Radiation 
pressure then ejects these grains from the planetary system. In our
calculations, ejected grains comprise $\sim$ 25\% to 80\% of the initial 
mass in the ring. 

Throughout the collisional cascade, the largest protoplanets try to accrete 
smaller planetesimals and collision fragments. The rate these protoplanets 
grow depends on the magnitude and timing of collisional damping
\citep[see also][]{kb2002a,kb2004a,gold2004}.  When the largest protoplanets 
{\it and} small fragments have relative eccentricity, $e_{k,rel} < 1$ 
(see eq. \ref{eq: erel}), protoplanets rapidly accrete the fragments. 
Super-Earth formation is likely.  When either of these conditions is {\it not} 
met, protoplanets cannot accrete fragments as rapidly as the cascade grinds the 
fragments to dust. Protoplanets grow slowly and almost never reach super-Earth 
masses. 

In our calculations, the two conditions for super-Earth formation depend on 
the slope $q$ of the power-law size distribution for collisional debris and 
the slope $b$ in the relation for the mass of the debris produced in a collision
($\mesc \propto (Q_c / \qdstar)^b$; see eq. \ref{eq: mej}).  At 125~AU, 
calculations with 
$q \approx$ 4.5--5.5 and $b \approx$ 9/8 are much more likely to meet these 
constraints on $e_{k,rel}$ than calculations with $q \approx$ 3.5-4.0 and
$b \approx$ 1. At 250~AU, super-Earth formation is fairly independent of 
$q$ and weakly dependent on $b$.

Collisional damping is the physical mechanism that sets the final masses
of protoplanets as a function of $(q, b)$. When $q$ is large, 
more fragments are concentrated into a smaller range of fragment masses.
These fragments collide more rapidly, enabling collisional damping to
reduce collision velocities more effectively. Smaller velocities initiate
a second runaway growth phase where protoplanets reach super-Earth masses.
When $q$ is small, fragments are spread out among a much larger range of 
masses. Damping is less effective; protoplanets grow more slowly. The
$b$ parameter magnifies this behavior. When $b$ is large (small), high 
velocity collisions produce more (fewer) fragments, enhancing (reducing) 
collisional damping and the second runaway growth phase.

Choosing the `best' $b$ and $q$ for coagulation calculations is not
straightforward.  Despite recent advances in numerical simulations of 
debris production during binary collisions 
\citep[e.g.,][]{durda2004,durda2007,lein2012,benavidez2012},
theory offers mixed guidance on several aspects of the algorithms used
to model fragmentation.  Currently, there is reasonable agreement on 
the binding energy \qdstar\ in the gravity and strength regimes
\citep[see also][]{housen1999,housen2011}. Modest changes consistent 
with the range of results from numerical simulations have modest impact 
on the growth of protoplanets \citep[e.g.,][]{kb2010,kb2012}.  

There is also reasonable agreement on the magnitude of $b$. For a broad
variety of approaches, $b$ = 1--1.25 \citep[e.g.,][]{davis1985,housen1990,
benz1999,housen1999,koba2010a,lein2012}. For high velocity collisions 
between roughly equal mass objects, our approach is consistent with
analytic derivations and numerical simulations. For cratering collisions
where objects have very unequal masses, analytic fits to experimental
data infer the ejected mass as a function of the impact velocity and the 
escape velocity of the more massive target \citep[e.g.,][]{housen2011}.
When $b$ = 1 or 9/8, our algorithm for the ejected mass generally agrees 
with these detailed fits \citep[e.g.,][]{kcb2014}.

However, there is less agreement on \rmaxd, the size of the largest fragment
within the debris. For example, the recommended procedure of \citet{lein2012} 
results in two large fragments with total mass of roughly 10\% of the total 
mass in the debris.  From the results of the \citet{durda2007} calculations, 
\citet{morby2009b} derive a factor of 5--10 smaller mass than \citet{lein2012}.
Our approach -- where we place 20\% of the total debris mass into the largest
fragment -- is similar to \citet{lein2012} but very different from 
\citet{morby2009b}. In future calculations, we plan to see if these differences
have a clear impact on the overall outcomes of coagulation calculations.

Moreover, there is little agreement on the size distribution for the debris. 
\citet{lein2012} propose a single power-law with $q$ = 3.85 which extends from
the largest remnant to the smallest particle allowed in the calculation. In
contrast, \citet{morby2009b} adopt $q \approx$ 9--10 for the largest particles
in the debris and $q$ = 3.5 for the smallest particles. In both approaches,
the largest particles contain most of the mass; thus, \citet{morby2009b} have
a much larger effective $q$ than \citet{lein2012}. Our calculations show that
the magnitude of $q$ has a dramatic impact on the size of the largest 
protoplanets. Using SPH/$n$-body simulations of binary collisions to derive a 
better appreciation of the behavior of $q$ as a function of the collision
velocity, mass ratio, and composition would enable a better understanding 
of the outcomes of planet formation simulations.

Finally, there has been little discussion of \rmind\ the minimum size of
fragments produced during destructive collisions. Usually, \rmind\ is zero; 
sometimes $\rmind \ll \rmaxd$.  Together with the binding energy of 
planetesimals and larger protoplanets, \rmind, \rmaxd, and $q$ establish 
the importance of collisional damping and the likelihood of super-Earth 
formation in swarms of icy planetesimals. Robust predictions for the outcomes 
of planet formation calculations therefore require a broader discussion of 
these fragmentation parameters.

In an interesting approach to these issues, \citet{krijt2014} infer a 
minimum size for debris fragments \rmind\ using constraints on the 
{\it surface energy} of colliding particles. In their model, a fraction $f$ 
of the impact kinetic energy makes new surface area. When $f$ = 0, the surface 
energy of colliding particles is conserved. When $f >$ 0, the collision
produces additional surface area. For size distributions with 3 $<q<$ 4, 
they derive \rmind\ as a function of the bulk properties of the colliding 
planetesimals. In high velocity collisions with fixed $q$, their picture 
produces size distributions similar to our algorithm.  At lower velocities, 
collisions produce only a few (sometimes one) particles rather than a broad 
size distribution. 

Although the \citet{krijt2014} analytic model provides an interesting 
way to choose some aspects of the size distribution, it is valid only
for 3 $< q <$ 4 and does not directly constrain $f$ or $q$. Expanding 
on their methods might yield a more general analytic prescription for 
the size distribution of the debris and narrow the allowed range for $q$
\citep[see also][]{housen1999}. 

Without a useful analytic model, numerical collision calculations with
a larger dynamic range in the sizes of debris particles would provide
useful guidance on collision outcomes and the properties of the debris. 
Larger simulations would place better limits on plausible values for 
\rmind\ and $q$ in coagulation calculations.

\subsection{Applications: Debris Disks}
\label{sec: disc-obs1}

Aside from these theoretical issues, our results have clear consequences
for the origin of the Solar System and debris disks at large distances
from their host stars.
Among known debris disks around solar-type stars, several have material 
at $a \approx$ 100--200~AU. The disk in HD~107146 (G2~V spectral type, 
age $\sim$ 100~Myr) extends from $\sim$ 30~AU to $\sim$ 150~AU
\citep{corder2009,ricci2014}. Recent \ALMA\ data indicate a reduction in
dust surface density at 70--80~AU \citep{ricci2014}. The large ring 
observed in HD~202628 (G2~V, 2~Gyr) lies at $\sim$ 200~AU and has a
structure similar to the ring in Fomalhaut \citep{krist2012}. In 
HD~207129 \citep[G2~V, 2--3~Gyr;][]{krist2010,marshall2011}, a ring 
with roughly similar geometry is somewhat smaller ($\sim$ 160~AU).

Published analyses suggest fairly large masses of solids in HD~107146 
and HD~207129. For HD~107146, fits to the dust distribution from \ALMA\ data
indicate a mass of roughly 0.2~\mearth\ in cm-sized and smaller particles 
\citep{ricci2014}. Deriving the total mass in solids requires an adopted
size distribution. If $n(r) \propto r^{-q}$ for $r$ = 1~cm to 100~km, 
likely values for $q$ yield total masses ranging from 2--3~\mearth\ for 
$q \approx$ 4 to 100~\mearth\ for $q \approx$ 3.5 
\citep[see also][]{ricci2014}. 

In HD~207129, detailed analysis of {\it Herschel} images and the spectral
energy distribution implies a total mass of 0.007~\mearth\ in mm and smaller 
particles \citep{lohne2012}. Adopting a power-law size distribution for 1~mm 
to 100~km objects yields masses ranging from 0.1~\mearth\ ($q \approx$ 4) 
to 70~\mearth\ ($q \approx$ 3.5). 

Together with our numerical simulations, the large solid masses derived for 
these systems show that super-Earth mass planets are reasonable outcomes of 
the planet formation process. The
geometries of the rings in HD~202628 and HD~207129 are similar to the
geometry adopted for our rings at 125~AU and at 250~AU. With ages of 
1--3~Gyr, our models suggest several super-Earths might be growing within
the dusty debris of each system.

In HD~107146, \citet{ricci2014} infer a reduced surface density of dust at
70--80~AU within the extended disk of debris. If the lower surface density
is a `gap' in the disk produced by a planet, the likely mass of the planet
is a few Earth masses. Based on our calculations, super-Earth formation at
70--80~AU around a 100~Myr old star requires a larger initial surface density
than we considered for this study \citep[see also][]{kb2008,kb2010}. With a 
factor of 3--5 increases in surface density, super-Earth formation in 100~Myr 
is a plausible outcome of our simulations \citep[see also][]{ricci2014}.

\subsection{Applications: Solar System}
\label{sec: disc-obs2}

The calculations in this study provide interesting constraints on the 
formation of a suspected planet X beyond 100~AU in the Solar System. 
Scaling the results at 125~AU and 250~AU, the time scale to grow 
Earth-mass or larger planets from 1--10~cm pebbles is
\begin{equation}
t_{SE} \approx 1 \left ( { M_0 \over 15~\mearth\ } \right )^{-1} \left ( { a \over {\rm 125~AU} } \right )^{3/2} {\rm Gyr} ~ ,
\end{equation}
where $M_0$ is the initial mass in an annulus of width $\delta a \approx 0.2 a$ 
centered at semimajor axis $a$. Requiring planet formation to conclude within
3--4~Gyr allows super-Earths in annuli with $M_0 \approx$ 15~\mearth\ at
$a \lesssim$ 300~AU. More massive rings enable super-Earths outside 300~AU.

Together with \citet{bk2014}, this study outlines plausible alternatives
to standard migration models for super-Earths located at large distances
from their host stars. In the migration hypothesis, a planet forms close to 
the host star and migrates out through a disk of gas or small planetesimals
\citep[e.g.,][]{crida2009}. In the scattering picture, a gas giant planet
dynamically scatters a much smaller planet from 5--10~AU to 100--200~AU;
the gaseous disk then circularizes the orbit of the smaller planet at large 
$a$ \citep{bk2014}.
The {\it in situ} model envisions a massive swarm of solids in a ring at
large $a$; coagulation within the ring allows the solid to grow into one
or more super-Earth mass planets.

In each of these scenarios, it is easier to understand a planet X in orbit
at $a \approx$ 100--250~AU than at $a \gtrsim$ 1000~AU. For the {\it in situ}
picture, formation at $a \approx$ 1000~AU requires a large reservoir of 
solids, $\gtrsim$ 300~\mearth. Fewer than 1\% of the protostellar disks 
around Taurus-Auriga pre-main sequence stars and only 10\% to 15\% of the 
disks around Taurus-Auriga embedded protostars contain this much mass
\citep[see the discussion in][and references therein]{najita2014}. In
contrast, 20\% to 60\% of the disks around young stars contain enough mass 
to produce a super-Earth at 100--200~AU. Thus, {\it in situ} formation
well beyond 200--300~AU seems unlikely. 

Observations of young stars similarly limit the migration and scattering
scenarios. In both mechanisms, the planet ends up on a roughly circular 
orbit within the gaseous disk surrounding the newly-formed central star. 
With typical characteristic radii of 20--200~AU \citep[e.g.,][]{and2010},
disks around young stars rarely extend to 500--1000~AU. Thus, formation
at 5--10~AU followed by migration/scattering to 500--1000~AU is also
very unlikely.

\subsection{Tests}
\label{sec: disc-tests}

Testing theories for super-Earths beyond 100~AU requires additional 
observations. Although recent observations identify Jovian mass planets 
associated with several circumstellar disks 
\citep[e.g.,][]{lagrange2010,currie2014b}, direct imaging observations 
of isolated super-Earths with $a \gtrsim$ 100~AU around other stars 
is unlikely with current or planned facilities 
\citep[e.g.,][]{hugot2012,jova2013,mac2014}. However, detecting super-Earths 
with surrounding debris clouds might allow direct tests as in Fomalhaut b 
\citep[e.g.,][]{kalas2005,chiang2009,kw2011a,currie2012,galich2013,kalas2013,kcb2014}.
Unless planet X has a moderate albedo, direct detection within the
Solar System is also very challenging \citep{trujillo2014}. Observations
of debris disks and Sedna-like objects provide alternate ways to 
test the various theoretical possibilities.

Among debris disks, comparisons between observations and the debris
predicted from collisional cascade or planet formation calculations
allow indirect tests \citep[e.g.,][]{kb2002b,kb2008,kb2010,lohne2012,koba2014}. 
Although discrepancies between predictions and observations remain,
these results generally confirm the cascade models 
\citep[see also][]{gaspar2013}. However, previous analyses did not 
include collisional damping of small particles within the cascade.
In our calculations, damping enhances the population of 
$10^{-2} - 10^2$~cm particles and thus probably enhances the predicted emission
from small particles. As we learn more about the importance of damping 
as a function of the initial disk mass and semimajor axis, we will be
able to compare the predicted emission with observations of complete
ensembles of debris disks observed with {\it IRAS}, {\it Spitzer}, and
{\it Herschel}.

For nearby debris disks, \ALMA\ observations offer the best hope of
placing better limits on super-Earths at large distances from the 
parent star. When super-Earths form within a ring of small solid
particles, they should clear out a gap with a half-width of roughly 
3 Hill radii \citep[e.g.,][]{ida2000b,kirsh2009,bk2011b,bk2013}.
If \ALMA\ detects narrow regions of low dust surface density in other
disks and rings (as in HD~107146), comparisons with the predicted 
surface density within a gap might yield constraints on super-Earth 
formation in debris disks \citep[e.g.,][]{ricci2014}.

Within the Solar System, current evidence for a massive planet beyond 
100~AU in the Solar System is limited and indirect. In the discovery 
paper for \VP, \citet{trujillo2014} suggest the need for a planet at 
200--250~AU to explain the clustering of $\omega$ among a dozen objects 
with perihelion distance $q_p \gtrsim$ 30~AU and orbital eccentricity 
$e \gtrsim$ 0.7. In addition to confirming the clustering, 
\citet{delafm2014} note other clusterings in the orbital parameters. 
They propose two super-Earth mass planets at roughly 200~AU and 250~AU. 
However, constraints on the precession of the inner planets appear to 
constrain any planet X to much larger $a \gtrsim$ 1000~AU 
\citep[e.g.,][and references therein]{iorio2009,iorio2012,iorio2014}.
Future studies need to reconcile the distribution of orbital parameters
for Sedna-like objects with the precession rates for the gas giants
and terrestrial planets.

New calculations can place better constraints on the formation of planet X 
and lower mass objects like Sedna and \VP. For the {\it in situ} picture,
ensembles of Earth-mass planets formed at 200~AU will eventually overcome
collisional damping and scatter smaller objects (and possibly themselves) 
throughout the Solar System. Comparing the orbital elements of known objects 
on Sedna-like orbits with orbits derived from detailed numerical simulations 
of planet formation will allow good tests of this formation model 
\citep[see also][]{kretke2012,bk2014}. 

Larger samples of Sedna-like objects can also test formation mechanisms.
For 100--1000~km particles, coagulation calculations produce a size 
distribution with power-law slope $q \approx$ 4 
\citep[e.g.,][]{kb2004c,schlicht2011,kb2012, schlicht2013}. Among smaller
objects, the slope has a broader range and depends on a variety of 
parameters.  If 100--1000~km objects on Sedna-like orbits are as common 
as current surveys suggest \citep[e.g.,][]{trujillo2014}, measuring the
slope of the size distribution could constrain the formation mechanism.

\section{SUMMARY}
\label{sec: summary}

We analyze aspects of three mechanisms -- migration, scattering, and 
{\it in situ} formation -- for placing indigenous super-Earth mass planets on 
orbits with large semimajor axes, $a \gtrsim$ 100~AU, around a solar-type star. 
In migration and scattering models, a super-Earth forms within a few Myr at 
$a \approx$ 5--10~AU. Interactions with other protoplanets and a massive disk 
then stabilize the orbit at large $a$.  The {\it in situ} picture requires a 
massive reservoir of solid material where super-Earths form directly.  For all 
three hypotheses, super-Earths at 100--300~AU are much more likely than 
super-Earths beyond 300~AU.

Using a suite of single-annulus coagulation calculations, we investigate the
physical requirements to form super-Earth mass planets at 125--250~AU around 
solar-type stars. In annuli with $M_0 \approx$ 15~\mearth,
super-Earth formation is common. When this mass is in planetesimals with 
\r0\ $\approx$ 1~cm to 1~m, super-Earths form in 1--3~Gyr at 125~AU and in
2--5~Gyr at 250~AU. Although swarms of larger planetesimals (\r0\ $\approx$
10~m to 10~km) can produce super-Earths in $\lesssim$ 4 Gyr at 125~AU, growth
at 250~AU takes longer than the age of the Solar System. Thus, {\it in situ} 
growth of super-Earths at 200--300~AU requires a massive ring of small 
planetesimals.

Collisional damping of the debris from destructive collisions is an essential 
ingredient in super-Earth formation at 125--250~AU. During the first 300~Myr
to 1~Gyr of evolution, the usual pattern of slow growth, runaway growth, and
oligarchic growth leads to the formation of 1000--3000~km protoplanets and a 
swarm of leftover small (0.1--10~km) planetesimals with large relative velocities 
\citep[see also][]{kenyon2002,kb2002b,kb2004a,kb2008,kb2010}. Destructive collisions 
among these leftovers generates a multitude of mm-sized to m-sized fragments.
Damping reduces the relative velocities of the fragments by 1--2 orders of
magnitude, to $e_{k,rel} \lesssim 1$ where $e_{k,rel} = e /r_H$ and 
$r_H = (\mmax / 3\msun)^{1/3}$ is 
the relative size of the Hill sphere of the most massive protoplanet. 
{\it The small fragment velocities enable a second runaway growth phase, where 
protoplanets grow rapidly. When protoplanets have $e_{k,rel} \lesssim$ 1, they 
reach super-Earth sizes in 1--2 Gyr}.

Fragmentation also plays a crucial role in the formation of super-Earths at
125--250~AU. In our calculations, the debris from destructive collisions
has a power-law size distribution, $n(r) \propto r^{-q}$. At 125~AU, setting
$q$ = 4.5--5.5 promotes the formation of super-Earths. When $q$ = 3.5--4.0,
super-Earth formation is very rare. At 250~AU, all $q \approx$ 3.5--5.5 allow
super-Earth formation. In models with $q \approx$ 4.5--5.5, however, super-Earths 
tend to be more massive than in calculations with $q \approx$ 3.5--4.0.
Within the fragmentation algorithm, the variation of ejected mass with impact
energy is also important. In our approach, $\mesc \propto (Q_c / \qdstar)^b$.
Calculations with $b$ = 9/8 tend to produce more massive protoplanets than 
calculations with $b$ = 1.

Improving the calculations requires a better understanding of several parameters 
in the fragmentation algorithm. In our approach, the mass of the debris ejected
during a collision is \mesc\ = $ 0.5 (Q_c / \qdstar)^b$, where $Q_c$ is the
collision energy, \qdstar\ is the collision energy required to eject half of
the mass involved in the collision, and $b$ = 1.00--1.25. For the mass of the
debris, we set $m_L$ = 0.2 \mesc\ and adopt a constant value $q$ for slope of 
the power-law size distribution of the debris.  In a future study, we plan to
investigate the impact of adopting a functional form for the largest object in
the debris, $m_L = m_{L,0} (Q_c / \qdstar)^{-b_L}$ where $b_L \approx$ 0.5--1.0
\citep[e.g.,][]{koba2010b,weid2010}. However, analytic studies and more 
comprehensive numerical simulations of binary collisions would enable more 
informed choices for $m_{L,0}$, $b$, $b_L$, and $q$ as a function of the
collision energy and the bulk properties of the planetesimals.

Observations of debris disks allow interesting tests of the models. Expanding
our single-annulus models to a full multi-annulus calculation will enable 
comparisons between predictions and observations of the time-dependent debris 
disk luminosities. \ALMA\ observations of the dust surface density in the
gaps of debris disks can place limits on super-Earth formation rates.

Discovery of planet X and additional dwarf planets on Sedna-like orbits would
also test the plausibility of {\it in situ} super-Earth formation at 125--250~AU.
From the mass distribution of protostellar disks, indigenous super-Earths with
$a \gtrsim$ 500~AU are unlikely \citep[see also][]{najita2014}. Robust estimates
of the semimajor axis of any planet X directly constrain planet formation theory.
Aside from placing better constraints on an orbit for planet X
\citep{trujillo2014,delafm2014}, orbital elements and (eventually) the size
distribution for new members of the inner Oort cloud can test the predictions
of the coagulation models.

\vskip 6ex

We acknowledge generous allotments of computer time on the NASA `discover' cluster.
We thank the referee, K. Ohtsuki, for a constructive and thorough review.  Advice 
and comments from T. Currie and M. Geller also greatly improved our presentation.  
Portions of this project were supported by the {\it NASA } {\it Astrophysics Theory}
and {\it Origins of Solar Systems} programs through grant NNX10AF35G and the
{\it NASA Outer Planets Program} through grant NNX11AM37G.

\appendix

\section{Appendix}

\orch\ contains many algorithms to follow the formation and evolution of a planetary 
system. With few analytic solutions available, verifying these algorithms requires 
comparisons with other approaches. In previous papers 
\citep[e.g.,][]{kl1998,kb2001,bk2006,kb2008,bk2011a}, we compare our numerical
results with analytic solutions and $n$-body simulations. Here, we describe several
additional tests.

\subsection{Runaway and Oligarchic Growth}

Accurate calculations for runaway and oligarchic growth depend on the spacing factor 
$\delta = m_{k+1} / m_k$ between adjacent mass bins 
\citep[e.g.,][]{oht1988,oht1990,weth1989,weth1990}. 
When bins have fixed masses, adopting large $\delta$ risks severe underestimates 
in the growth times for massive objects. \citet{oht1990} derive the constraints 
on $\delta$ which allow accurate solutions \citep[see al;so][]{fraser2009b}.  
In our approach, mass bins have fixed boundaries but variable masses 
\citep[see also][]{weth1993,kl1998}.  Calculations with $\delta > 1$ always lag 
a `perfect' solution, reaching a fiducial size later in time.  Larger $\delta$'s 
always produce larger lags. 

To understand the origin of the lag, we consider a simplified example.  With 
preset boundaries $m_{k+1/2}$, the average mass of objects in a bin is 
$\bar{m}_k$, where $m_{k+1/2} < \bar{m}_{k+1} < m_{k+3/2}$ and 
$m_{k+3/2} = \delta \cdot m_{k+1/2}$.  When an object in bin $k+n$ merges with an 
object in bin $k$, the mass of the new object is $m_c = \bar{m}_k + \bar{m}_{k+n}$. 
In an ideal calculation with an infinite number of mass bins, the mass of the 
new object is placed in a new bin which is distinct from all other bins. In a
real calculation, however, placing this object in a new bin with index 
$k_c > k+n$ requires $m_c > m_{k+n+1/2}$. To derive a simple estimate of when 
this inequality is satisfied, we set $\bar{m}_{k+n} = \delta^n \bar{m}_k$.  
The constraint on promoting the new object to a new bin then becomes:
\begin{equation}
{ (\delta^n + 1) \bar{m}_k \over \delta^{n+1/2} m_k } > 1 ~ .
\label{eq: pro1}
\end{equation}

Throughout a calculation, $\bar{m}_k$ is distributed uniformly in the range 
$\delta^{-1/2}$--$\delta^{1/2}$~$m_k$. The {\it minimum} $\bar{m}_k$ which
satisfies eq.~(\ref{eq: pro1}) is $\delta^{-1/2} m_k$. We can then write a 
much simpler inequality for promotion to a new bin:
\begin{equation}
\delta^{n+1} - \delta^n - 1 < 0 ~ .
\label{eq: pro2}
\end{equation}
When this inequality is satisfied, promotion to a new bin {\it always} occurs. 
Otherwise, promotion fails some fraction of the time.  This failure is the 
source of the lag in all coagulation calculations with variable mass bins.  
The fraction of failures grows with increasing $\delta$.  Thus, the lag grows 
with larger $\delta$.

Before describing results of our test calculations, it is worth examining 
several examples.  When $n$ = 1, we consider a simple collision between
objects in adjacent mass bins.  Calculations with $\delta \lesssim$ 1.62 
satisfy the inequality in eq.~(\ref{eq: pro2}). In solutions with 
$\delta >$ 1.62, collisions fail to promote the merged object into a 
higher mass bin roughly 20\% of the time.  When merged objects are not 
promoted, the average mass of the largest object in the grid is smaller 
than the mass of a promoted object. Smaller objects have smaller collision 
rates and longer collision times.  Thus, calculations with $\delta >$ 1.62 
always lag an ideal calculation for this simple merger.  

As $n$ increases, we consider collisions between objects with larger and 
larger mass ratios. Automatic promotion to the next bin requires smaller and 
smaller $\delta$. When $n$ = 3, 10, 25, or 62, promotions never fail when 
$\delta \lesssim$ 1.38, 1.19, 1.10, 1.05. For collisions with large $n$, 
smaller $\delta$ yields smaller failure rates. Thus, the lag declines 
with decreasing $\delta$.

Previous discussions of the magnitude of the lag as a function of $\delta$
focus on the evolution of the overall size distribution 
\citep{weth1990,kl1998,morby2009b}. Here, we illustrate the lag 
by following the growth of the largest object. In the first example, we 
consider growth when the cross-section is $A_{ij} = \gamma m_i m_j$ where
$\gamma$ is a constant. Defining the dimensionless time $\eta = \gamma n_0 t$ 
where $n_0$ is the initial number of particles with mass $m_0$, there is an 
analytic solution for the size of the largest object as a function of $n_0$
and $\eta$ \citep{weth1990}.

\begin{figure}
\includegraphics[width=6.5in]{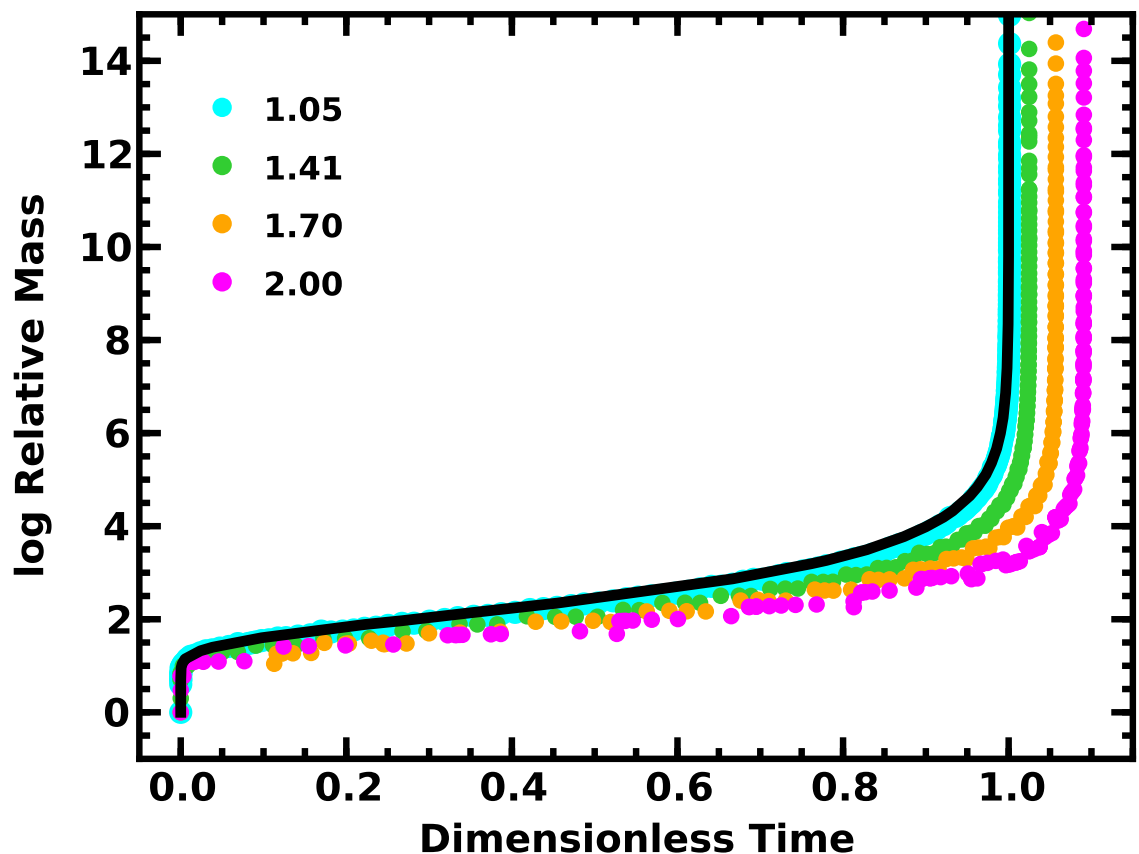}
\vskip 3ex
\caption{%
Mass of the largest object as a function of dimensionless time
$\eta$ for cross-section $A_{ij} \propto m_i m_j$. The solid 
black line plots the analytic solution. Colored points show
results for numerical solutions with different values for the
mass spacing factor $\delta$ as summarized in the legend.
The numerical solutions lag the analytic solution by
9.1\% ($\delta$ = 2.0),
5.7\% ($\delta$ = 1.7),
2.5\% ($\delta$ = 1.4),
2.0\% ($\delta$ = 1.25),
0.3\% ($\delta$ = 1.1), and
0.1\% ($\delta$ = 1.05).
Although calculations with $\delta$ = 1.05 follow the analytic
result almost exactly, solutions with $\delta \le$ 1.4 yield
satisfactory approximations to the exact result.
\label{fig: an1}
}
\end{figure}

Fig.~\ref{fig: an1} compares results for four values of $\delta$ with the
analytic solution. Independent of $\delta$, \rmax\ follows the same evolution:
slow growth for $\eta \approx$ 0.0--0.95 and exponential growth for 
$\eta \approx$ 1. The precise timing of exponential growth depends on 
$\delta$. Models with $\delta$ = 2 reach the runaway when $\eta$ = 1.091,
lagging the exact solution by 9.1\%. When $\delta$ = 1.05, runaway begins 
when $\eta$ = 1.001. With a lag of only 0.1\%, these calculations almost
precisely match the analytic solution. 

Although solutions for $A_{ij} \propto m_i m_j$ provide excellent tests of a 
coagulation code, realistic cross-sections for particles in circumstellar disks 
are not as extreme \citep[see also][]{weth1990,lee2000,mal2001}. 
When the collision rate depends solely on the geometric 
cross-section, $A_{ij} \propto (m_i + m_j)^{2/3}$.  If gravitational focusing 
is important, $A_{ij} \propto (m_i + m_j)^{4/3}$. 
Calculating the evolution for either cross-section yields a more appropriate
assessment of the lag as a function of $\delta$ in actual protostellar disks. 
With no analytic solution available, we adopt calculations with $\delta$ = 
1.10 as the reference. For clarity, we rescale the dimensionless time to
place exponential growth for $\delta$ = 1.10 at $\eta$ = 1.

\begin{figure}
\includegraphics[width=6.5in]{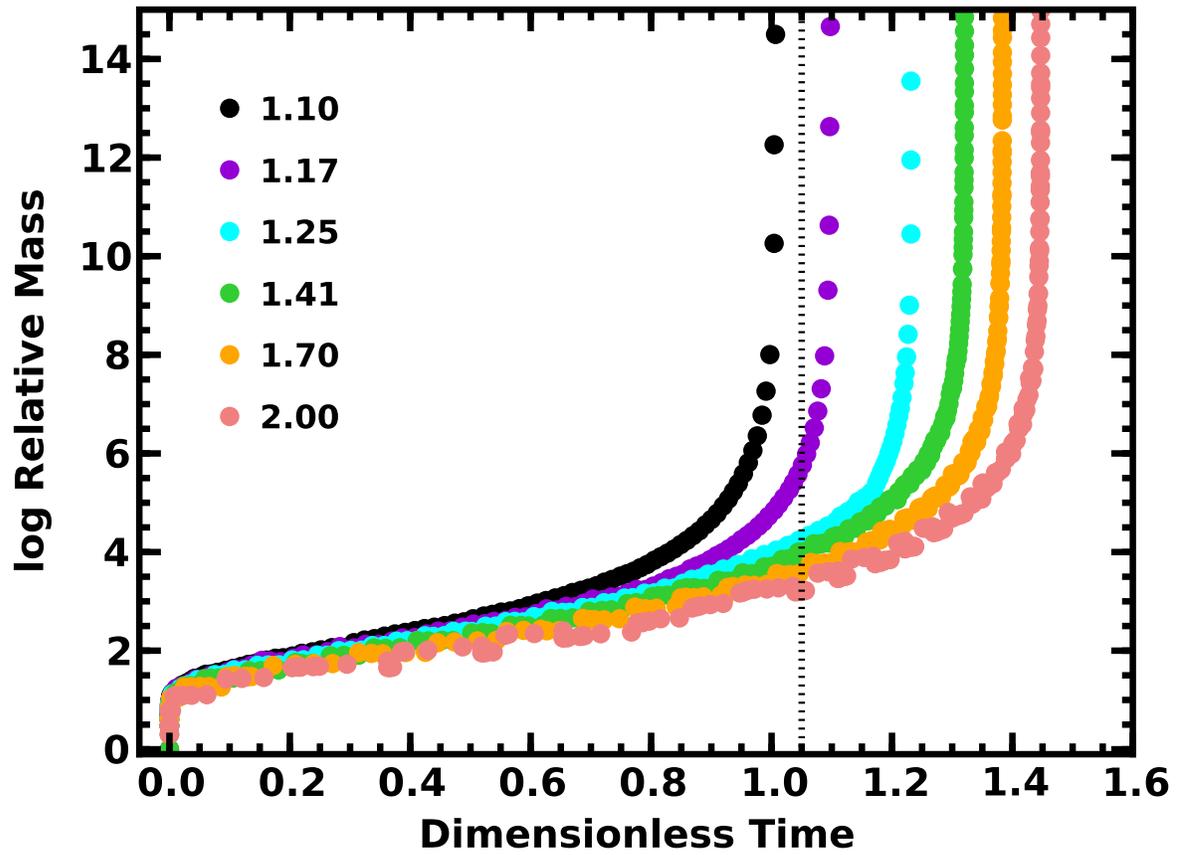}
\vskip 3ex
\caption{
As in Fig. \ref{fig: an1} for $A_{ij} \propto (m_i m_j)^{2/3}$.
The numerical solutions lag the $\delta$ = 1.10 result by
43.7\% ($\delta$ = 2.0),
37.4\% ($\delta$ = 1.7),
31.1\% ($\delta$ = 1.4),
22.5\% ($\delta$ = 1.25), and
9.0\% ($\delta$ = 1.17).
\label{fig: an2}
}
\end{figure}

Fig.~\ref{fig: an2} compares results for six values of $\delta$ with
$A_{ij} \propto (m_i + m_k)^{4/3}$ \citep[see also][]{weth1990}.
All solutions have the same form.  Compared to the analytic model in 
Fig.~\ref{fig: an1}, the largest object grows somewhat faster for 
$\eta$ = 0.0--0.95; close to $\eta$ = 1, runaway growth is somewhat 
less explosive. The timing of the runaway is very sensitive to $\delta$.  
Models with $\delta$ = 2 reach runaway roughly 40\% later in time than 
models with $\delta$ = 1.05.  

\begin{figure}
\includegraphics[width=6.5in]{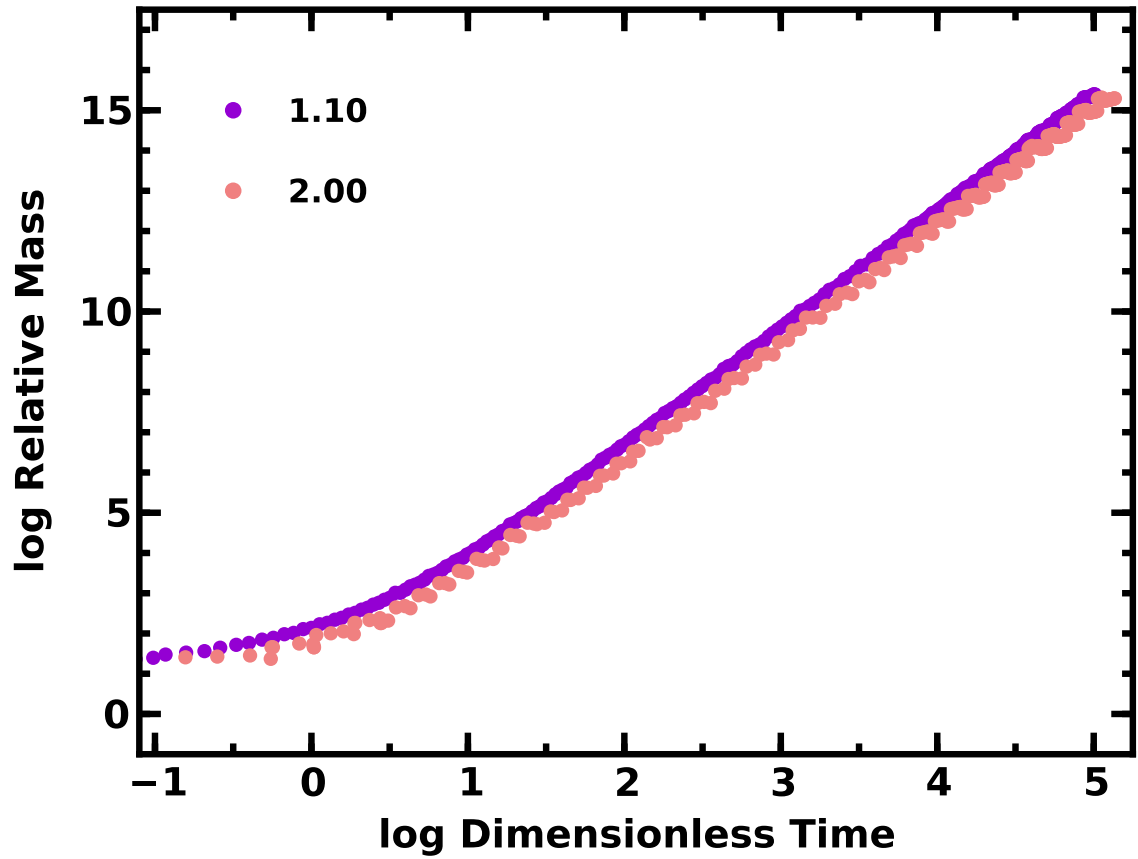}
\vskip 3ex
\caption{
As in Fig. \ref{fig: an2} for $A_{ij} \propto (m_i m_j)^{1/3}$.
The numerical solutions lag the $\delta$ = 1.10 result by
1.9\% ($\delta$ = 2.0),
1.6\% ($\delta$ = 1.7),
1.3\% ($\delta$ = 1.4),
1.0\% ($\delta$ = 1.25), and
0.5\% ($\delta$ = 1.17).
\label{fig: an3}
}
\end{figure}

When $A_{ij} \propto (m_i + m_j)^{2/3}$, growth is insensitive to $\delta$
(Fig.~\ref{fig: an3}). Solutions with $\delta$ = 2 lag calculations with
$\delta$ = 1.1 by less than 2\%. Calculations with $\delta$ = 1.25--1.7
have lags ranging from 0.5--1.5\%. Thus, mass spacing has little impact
on growth with purely geometric cross-sections \citep{weth1990}.

\begin{figure}
\includegraphics[width=6.5in]{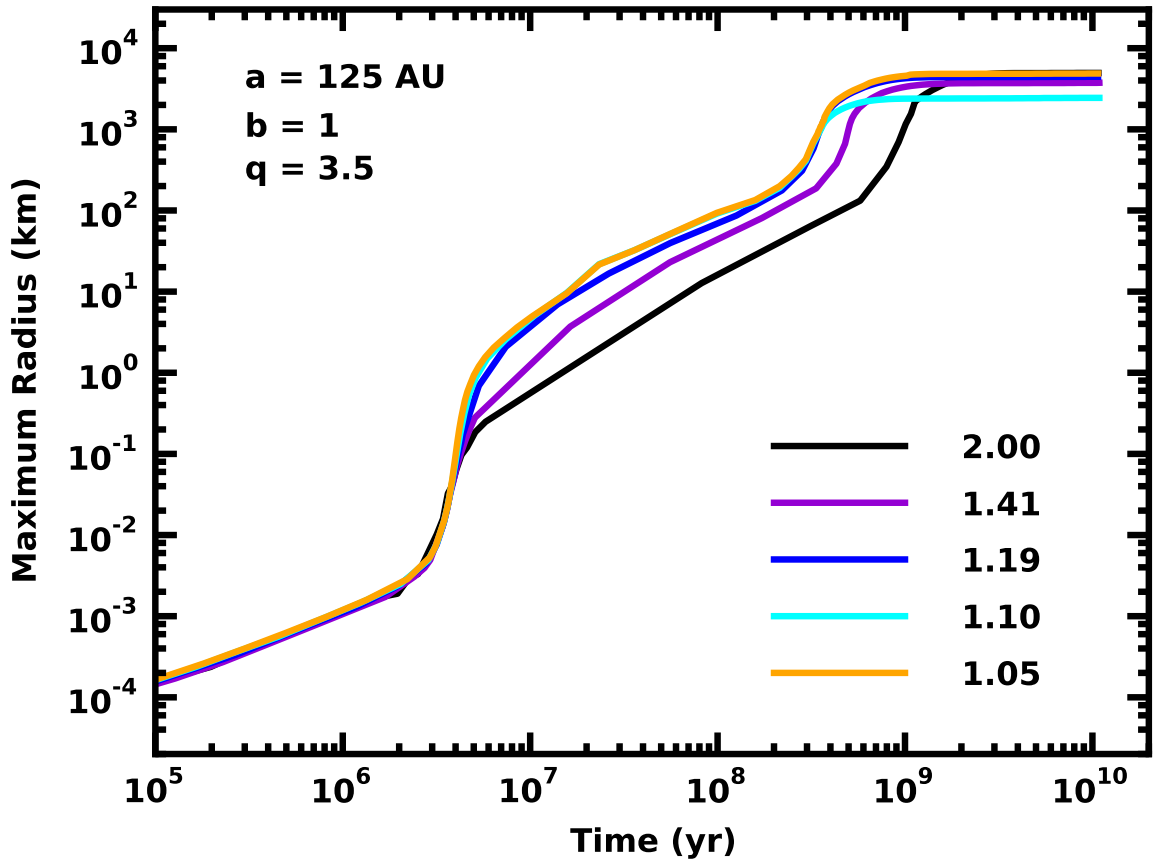}
\vskip 3ex
\caption{
Growth of 1~cm particles at $a$ = 125~AU as a function of
mass spacing parameter $\delta$. Although all calculations
produce objects with similar \rmax, simulations with large
$\delta$ lag those with smaller $\delta$.
\label{fig: growth1}
}
\end{figure}

These examples suggest that $\delta$ is an important factor in models of 
runaway growth but not slow or oligarchic growth \citep[see also][]{weth1990}.
To confirm this point, Fig.~\ref{fig: growth1} compares the evolution of 
\rmax\ for complete evolutionary calculations with different values of 
$\delta$. During an early phase of slow growth (when $t \lesssim$ 1~Myr),
1~cm particles reach maximum sizes of a few meters on time scales that are 
independent of $\delta$. After runaway growth begins, small particles rapidly 
grow to km sizes. Once particles have \rmax\ $\approx$ 0.1~km, calculations 
with larger $\delta$ enter the oligarchic phase before those with smaller 
$\delta$.  Eventually all calculations reach oligarchic growth; the rate of 
growth is then fairly independent of $\delta$. After several Gyr of evolution, 
all models have \rmax\ $\approx$ 3000--7000~km (Fig~\ref{fig: growth1}).

When 0.1~km $\lesssim$ \rmax\ $\lesssim$ 100~km, the growth times in these calculations depend
on $\delta$. For $\delta \lesssim$ 1.2, the time scale to produce 100~km objects 
is $t_{100} \approx$ 500--800~Myr. Although calculations for each $\delta$ 
exhibit a range in $t_{100}$, the median time to reach \rmax\ $\approx$ 100~km 
is independent of $\delta$. For larger $\delta$, the median growth time lags 
the $\delta \lesssim$ 1.2 solutions by 10\%--20\% ($\delta$ = 1.41) to 
25\%--50\% ($\delta$ = 2.0). Thus, calculations with $\delta \lesssim$ 1.2
yield a better estimate of the time scale to produce 100~km objects.

As objects grow larger than 100~km, the growth time depends less on $\delta$
and more on the fragmentation history. For any $\delta$, collisional damping
of small fragments can enable a second runaway growth phase at $t \sim$ 100~Myr 
to 1~Gyr. The timing and magnitude of this phase depends on (i) the amount of 
material available in small objects and (ii) the ability of collisional damping
to reduce particle velocities significantly. Sometimes conditions allow the
production of massive objects with \rmax\ $\gtrsim 10^4$ km. Other simulations
produce objects with \rmax\ $\approx$ 3000--4000~km. 

\subsection{Collisional Damping and Gravitational Stirring}

In our coagulation calculations, we derive changes in particle
velocities from collisional damping, dynamical friction, gas drag, and 
viscous stirring. \citet[][Appendix]{kl1998} and \citet[][\S2]{kb2008} 
describe the algorithms for each of these processes. Collisional
damping is a central issue in this study; thus, we summarize our 
treatment for a single annulus.  We then discuss results for several 
standard examples.

To treat collisional damping in a single annulus, we solve two sets 
of differential equations\footnote{This discussion is adapted from the 
multiannulus approach of eqs. 8--9 from \citet{kb2008}. For a single 
annulus, we drop the subscripts for annulus $i$ and annulus $j$ and
eliminate the term which treats the overlap between adjacent annuli.}:
\begin{equation}
\frac{dh_{in,k}^2}{dt} = \sum_{l=0}^{l=k} \frac{C_{in}}{2}~(m_{l} h_{l}^2 - m_{k} h_{k}^2 - (m_{k} + m_{l}) h_{k}^2)~I_e(\beta_{kl})
\label{eq:dhdtin}
\end{equation}
and
\begin{equation}
\frac{dv_{in,ik}^2}{dt} = \sum_{l=0}^{l=k} \frac{C_{in}}{\beta_{kl}^2}~(m_{l}
v_{l}^2 - m_{k} v_{k}^2 - (m_{k} + m_{l}) v_{k}^2)~I_i(\beta_{kl})
\label{eq:dvdtin}
\end{equation}
where $C_{in} = \alpha_{coll} ~ \epsilon_{kl} ~ \rho_{l} ~
V_{kl} ~ F_{g,kl} ~ (r_{k} + r_{l})^2$,
$\beta_{kl}^2 = (i_{k}^2 + i_{l}^2)/(e_{k}^2 + e_{l}^2)$,
and $\rho_{l}$ is the volume density of planetesimals with mass
$m_{l}$ \citep{weth1993,kl1998}.
In these expressions,
$\alpha_{coll}$ = 0.57-0.855 \citep{weth1993},
$\epsilon_{jk}$ = 1/2 for $k = l$ and 1 for $k \ne l$,
$V_{kl}$ is the relative particle velocity, and 
$F_{g,kl}$ is the gravitational focusing factor.
The integrals $I_e$ and $I_i$ are elliptic integrals described in
previous publications \citep[e.g.,][]{horn1985,weth1993}.

Together with similar equations for coagulation, dynamical friction, 
gas drag, and viscous stirring 
\citep[see][and references therein]{kl1998,kb2002a,kb2008}, we 
solve these equations explicitly using an adaptive time step 
which conserves mass and kinetic energy to machine accuracy. We 
use more accurate and more cpu intensive fourth-order Runge Kutta 
and Richardson extrapolation algorithms to verify the explicit solutions. 
\citet{kl1998} summarize comparisons of our single annulus solutions
for terrestrial planet formation with those of \citet{weth1993}.

To demonstrate our ability to treat collisional damping accurately,
we consider several idealized problems.
In the first test, we follow \citet{lev2007} and consider the 
evolution of 1~cm particles in a single annulus at 30--35~AU. 
The particles have initial eccentricity $e_0$ = 0.1, 
inclination $i_0$ = $e_0/2$, mass density $\rho_s$ = 1~\gcmc, 
and total mass $M_0$ = 10~\mearth.
The calculations include gravitational stirring and collisional
damping, but particles cannot merge or fragment during a collision.

\begin{figure}
\includegraphics[width=6.5in]{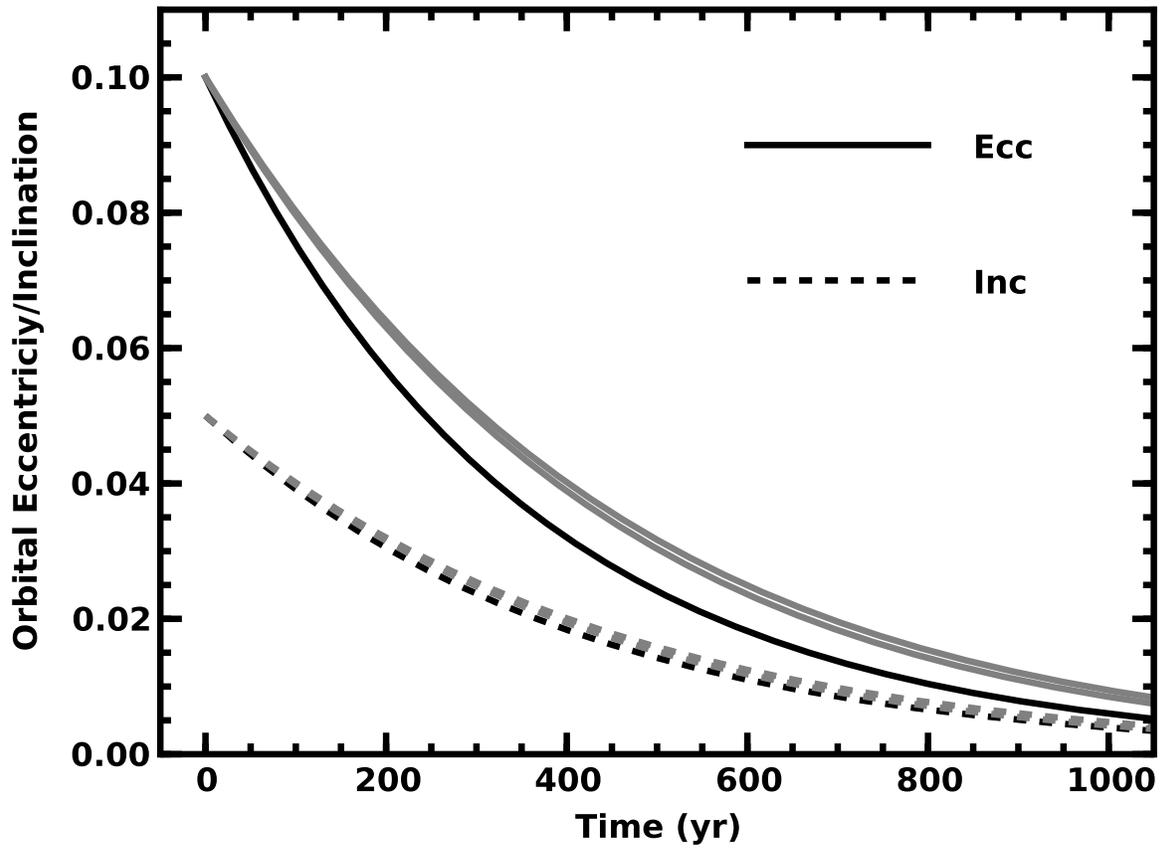}
\vskip 3ex
\caption{
Collisional damping within a ring of small particles at
30--35 AU \citep{lev2007}. The ring contains 10~\mearth\ in
1~cm particles. Heavy lines: damping for inelastic collisions. 
Light lines: damping for elastic collisions with coefficient
of restitution $c_0 = 0.99$ (upper line) and $c_0$ = 0.01 
(lower line).
\label{fig: damp1}
}
\end{figure}

Our results closely match those summarized in \citet{lev2007}
and \citet{morby2009b}. Particle random velocities damp rapidly 
and reach an equilibrium at $e_{eq} = 9.3 \times 10^{-10}$ and 
$i_{eq} = 5.1 \times 10^{-10}$ in 7300~yr (Fig.~\ref{fig: damp1}). 
The e-folding times of $t_e \approx$ 350~yr and $t_i \approx$ 
400~yr are close to the \citet{lev2007} results of 320~yr and 
360~yr. The ratio $t_e/t_i$ = 0.875 is essentially identical to 
their $t_e/t_i$ = 0.889.

To compare this approach to analyses of Saturn's rings
\citep[e.g.,][]{gold1978,bridges1984,sup1995,porco2008},
we modify the damping algorithm. Defining the coefficient 
of restitution as
\begin{equation}
c_R = c_0 v_{rel}^{-k_R} ~ ,
\label{eq: crest}
\end{equation}
with $c_0$ = 0--1 and $k_R$ = 0.1--0.3, the kinetic energy loss
rate is $\delta E_{ij} = A_{ij} v_{ij} (1 - c_R^2) / V_{ij}$
where $v_{ij}$ is the relative velocity of two particles and 
$V_{ij}$ is the volume occupied by the particles. Replacing our 
standard loss term with $\delta E_{ij}$ allows us to compare 
damping rates derived with two different approaches.

The grey lines in Fig.~\ref{fig: damp1} plot results for 
$c_0$ = 0.99 (upper curve) and $c_0$ = 0.01 (lower curve) 
with $k_R$ = 0.15.
In these solutions, the e-folding times for the eccentricity are 
$t_e \approx$ 435 yr ($c_0$ = 0.99) and 425 yr ($c_0$ = 0.01);
for inclination, $t_i \approx$ 435 yr ($c_0$ = 0.99) and 425 yr 
($c_0$ = 0.01). Standard collisional damping in our calculations
proceeds on time scales similar to models with a very small 
coefficient of restitution $c_0 \lesssim 10^{-3}$ 
\citep[see also][and references therein]{kl1999a}.

To test the evolution in an annulus relevant to the calculations 
described in the main text, we examine damping and stirring in a
ring of mono-disperse particles with $\rho_s$ = 1.5~\gcmc\ and 
$M_0$ = 15.79~\mearth\ at 112.5--137.5~AU. For the damping calculations, 
we set $e_0 = 10^2 e_{eq}$ and $i_0 = e_0/2$. To address time scales 
for gravitational stirring, we set $e_0 = 10^{-2} e_{eq}$ and 
$i_0 = e_0 / 2$.

\begin{figure}
\includegraphics[width=6.5in]{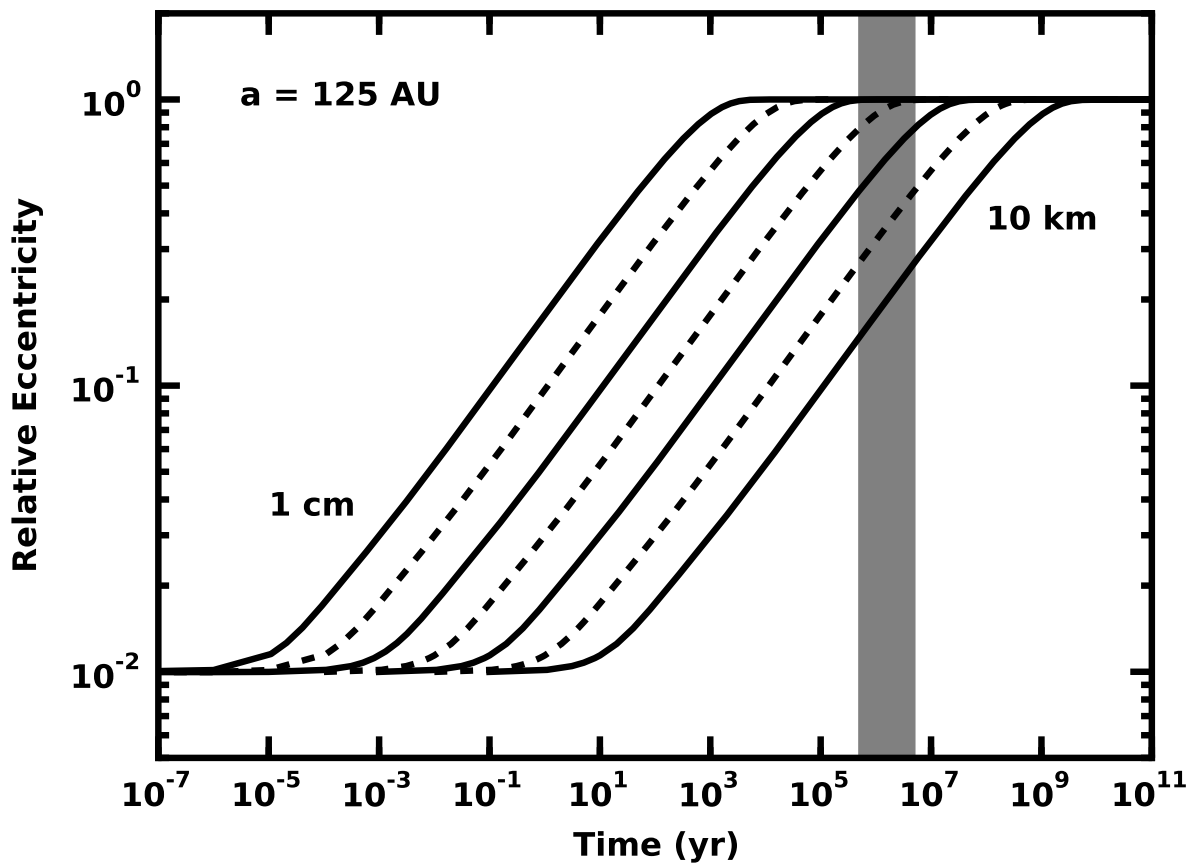}
\vskip 3ex
\caption{
Gravitational stirring within a ring of particles at 112.5--137.5 AU. 
The ring has surface density $\Sigma$ = 0.0215 \gcms\ and contains 
15.8~\mearth\ in mono-disperse particles.
Solid and dashed lines: evolution of the eccentricity $e$ relative 
to the equilibrium eccentricity 
$e_0 = 2.38 \times 10^{-4} (R / {\rm 1~km}) $ for 
1~cm, 10~cm, 1~m, 10~m, 100~m, 1~km, and 10~km particles. 
The time to reach $e^{-1}$ of the equilibrium eccentricity is 
$t_{stir} = 1.67 (R / {\rm 1~km}) $~Myr.
Vertical grey bar: epoch of planetesimal formation and gas dispersal
in a protoplanetary disk 
\citep[e.g.,][and references therein]{kleine2009,dauphas2011b,will2011}.
\label{fig: stir}
}
\end{figure}

\begin{figure}
\includegraphics[width=6.5in]{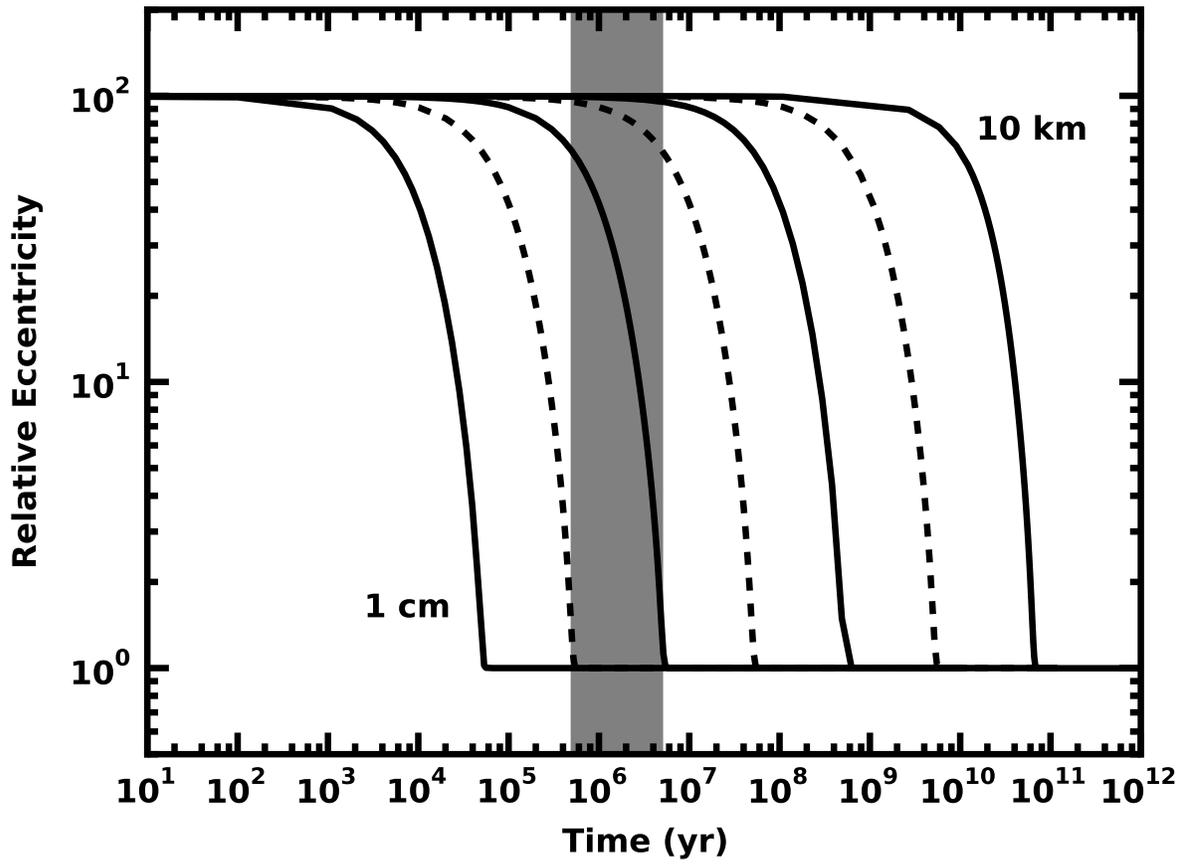}
\vskip 3ex
\caption{
As in Fig.~\ref{fig: stir} for collisional damping.
The $e$-folding time is $t_{damp} = 1.15 \times 10^9 (R / {\rm 1~km})$~yr.
\label{fig: damp2}
}
\end{figure}

In both sets of simulations, particles reach an equilibrium where
the 3D velocity dispersion is roughly 60\% of the escape velocity.
The equilibrium has $h / v$ = 0.49, $i / e$ = 0.55, and 
$e_{eq} = 2.38 \times 10^{-4} (R / {\rm 1~km}) $.
The time scales to reach equilibrium are typically much longer at
125~AU than at 30--35~AU (Figs.~\ref{fig: stir}--\ref{fig: damp2}). 
When gravitational stirring dominates (Fig.~\ref{fig: stir}), 
particles reach equilibrium on time scales ranging from $\sim$ 
1000~yr for 1~cm particles to roughly 1~Gyr for 10~km particles. 
Damping time scales are roughly a factor of 30 longer
(Fig.~\ref{fig: damp2}).

Aside from validating the algorithms for stirring and damping, these
simulations help us to set likely initial conditions for $e$ and $i$ 
at 125--250 AU. For a swarm of particles with velocity dispersion
$c_p$, surface density $\Sigma$, and angular velocity $\Omega$, the
swarm is gravitationally unstable when $c_p \Omega < \pi G \Sigma$
\citep[e.g.,][]{chiang2010}. 
Setting $\Sigma = \Sigma_0 (a / {\rm 1~AU})^{-3/2}$, the swarm is
gravitationally stable when the eccentricity has $e \gtrsim e_s$, 
where
\begin{equation}
e_s \approx 10^{-4} \left ( { \Sigma_0 \over {\rm 30~g~cm^{-2}} } \right ) \left ( {a \over {\rm 100~AU} } \right )^{1/2} ~ .
\label{eq: e-grav}
\end{equation}
For our initial surface density at 125--250~AU, the minimum $e$ for
gravitational stability is $e_s \approx 10^{-4}$. 

Particles with $r \gtrsim$ 0.4~km have $e_{eq}$ larger than $e_s$. 
At 125--250~AU, these particles reach equilibrium on time scales longer 
than the lifetime of the protosolar nebula and other protoplanetary disks 
\citep[e.g.,][and references therein]{dauphas2011b,will2011,clout2014}.  
Thus, we can safely set $e_0$ somewhere between $e_s$ and $e_{eq}$. 

\begin{figure}
\includegraphics[width=6.5in]{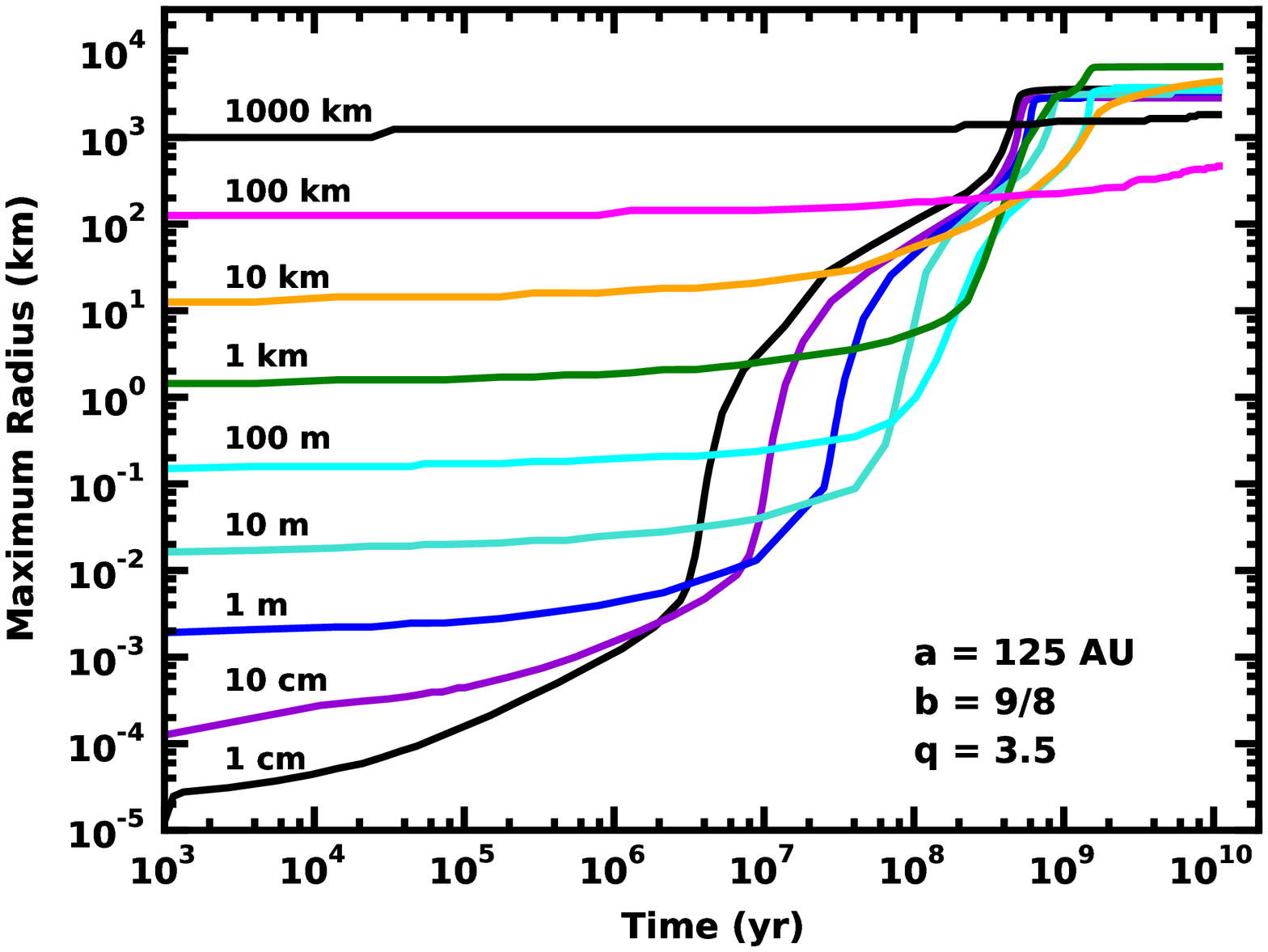}
\vskip 3ex
\caption{%
Growth of the largest object at 125~AU as a function of initial
planetesimal size for calculations with $b$ = 9/8 and $q$ = 3.5.
Although simulations starting with smaller planetesimals produce 
large planets faster than those with larger planetesimals, all 
systems with $M_0 \approx$ 15.8~\mearth\ and \r0\ $\lesssim$ 10~km 
yield planets with \rmax\ $\approx$ 3000--6000~km. In rings with 
\r0\ $\gtrsim$ 100~km, collisions are too infrequent to produce 
planets with \rmax\ $\gtrsim$ 500--2000~km.
\label{fig: rmax-125a}
}
\end{figure}

\begin{figure}
\includegraphics[width=6.5in]{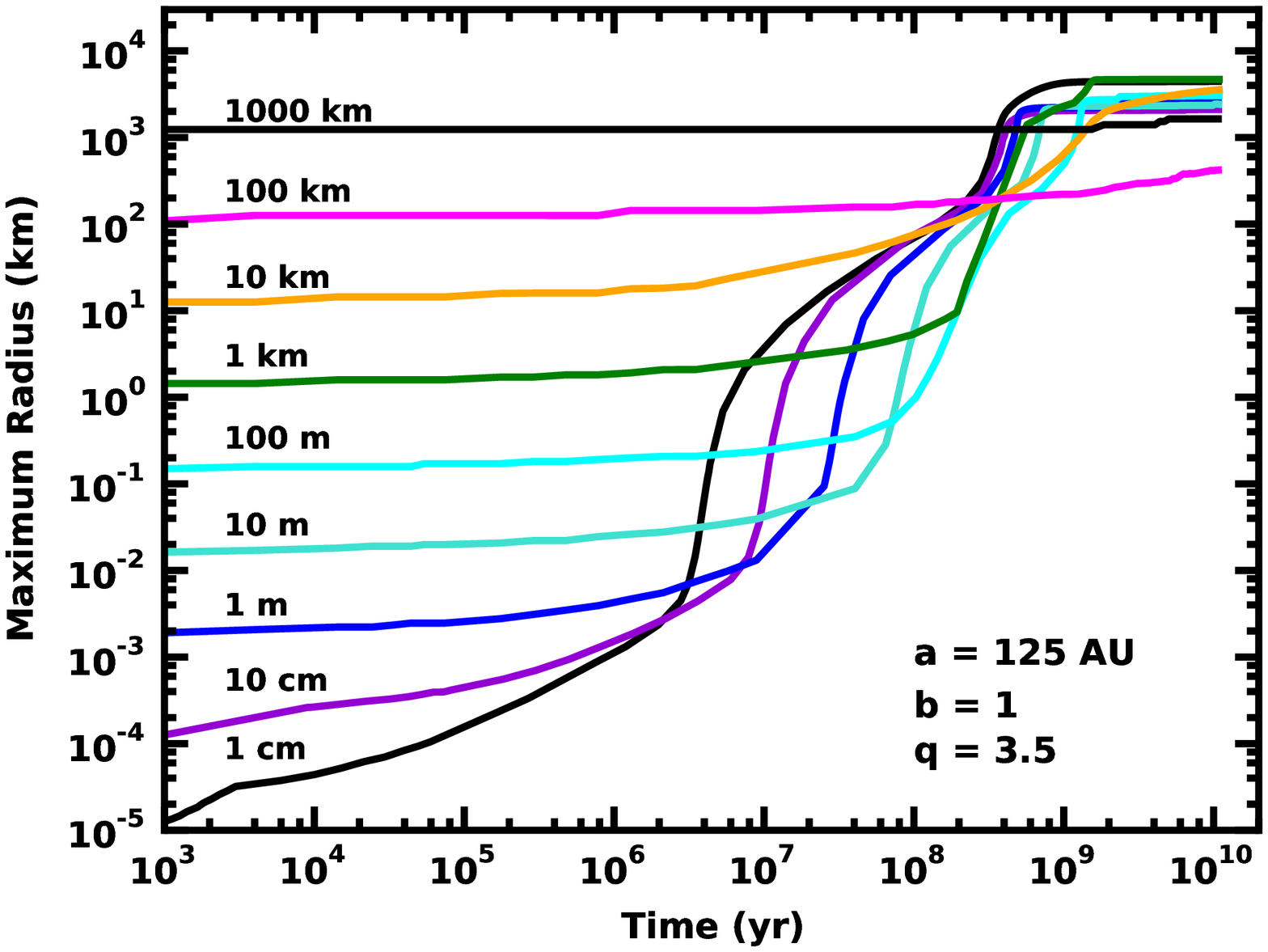}
\vskip 3ex
\caption{%
As in Fig. \ref{fig: rmax-125a} for calculations with
$b$ = 1. 
\label{fig: rmax-125b}
}
\end{figure}

\begin{figure}
\includegraphics[width=6.5in]{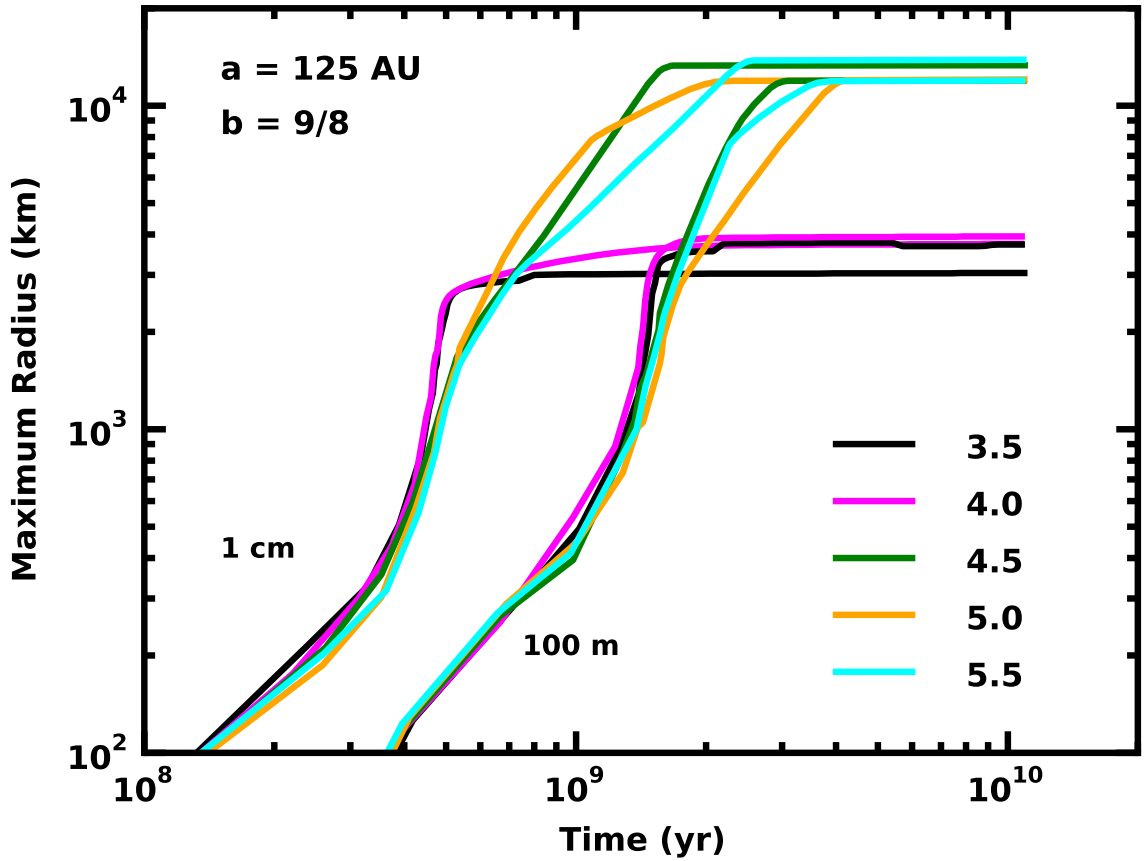}
\vskip 3ex
\caption{%
As in Fig. \ref{fig: rmax-125a} for calculations with
\r0\ = 1~cm or 100~m, $b$ = 9/8, and various $q$
as indicated in the legend. In calculations with 
$q \gtrsim$ 4.5, planets reach super-Earth masses 
in 1--3 Gyr.
\label{fig: rmax-125c}
}
\end{figure}

Although particles with radii $r \approx$ 10--100~m have $e_q < e_s$, 
the time scale to reach this equilibrium is longer than the disk lifetime. 
On these time scales, particles also grow considerably 
(see Figs.~\ref{fig: rmax-125a}--\ref{fig: rmax-125b}).
Thus, we set $e_0 \approx e_s$.

Particles with much smaller radii $r \approx$ 1--100~cm reach very small 
equilibrium $e$ on relatively short time scales, $\sim$ 0.05--5~Myr. 
For our adopted $\Sigma$ and $e_0 = 10^{-4}$ , these particles grow on
similar time scales. To avoid the complications of gravitational 
instabilities, we also set $e_0 = e_s$ for these particles.

In any swarm of particles at 125--250~AU, the likely $e$ and $i$ depend 
on the evolutionary history, which is sensitive to the structure of the 
disk.  In this paper, the adopted $e_0$ and $i_0$ are intermediate 
between the $e_{eq}$ and values expected for a turbulent disk. The
calculations in \S\ref{sec: evol} describe how these choices impact 
our results. The discussion in \S\ref{sec: disc} considers how to
improve these choices.

\bibliography{ms.bbl}

\clearpage

%
\clearpage

\clearpage

\clearpage

\clearpage

\clearpage

\clearpage

\clearpage

\clearpage

\clearpage

\clearpage

\clearpage

\clearpage

\clearpage

\clearpage

\clearpage

\clearpage

\clearpage

\clearpage

\clearpage

\clearpage

\clearpage

\clearpage

\clearpage

\end{document}